\documentclass[seceqn,final]{elsart}

\usepackage{amsmath,amsfonts,graphicx,url}
\usepackage{hyperref}
\usepackage{backrefx}
\usepackage[usenames]{color}

\journal{Physics Reports}
\date{}
\begin{document}
\begin{frontmatter}
\title{Analytical Solutions of Open String Field Theory}
\author{Ehud Fuchs},
\ead{udif@aei.mpg.de}
\author{Michael Kroyter}
\ead{mikroyt@aei.mpg.de}

\address{Max-Planck-Institut f\"ur Gravitationsphysik\\
Albert-Einstein-Institut\\
14476 Golm, Germany \\ \vspace{-180pt}\hspace{282pt}
AEI-2008-056\vspace{200pt}}
\begin{abstract}
In this work we review Schnabl's construction of the tachyon vacuum solution
to bosonic covariant open string field theory and the results
that followed.

We survey the state of the art of string field theory research
preceding this construction focusing on Sen's conjectures and
the results obtained using level truncation methods.

The tachyon vacuum solution can be described in various ways. We
describe its geometric representation using wedge states,
its formal algebraic representation as a pure-gauge solution and
its oscillator representation.
We also describe the analytical proofs of some of Sen's conjectures for
this solution.

The tools used in the context of the vacuum solution
can be adapted to the construction of other solutions,
namely various marginal deformations. We present some of the approaches
used in the construction of these solutions.

The generalization of these ideas to open superstring field theory is
explained in detail.
We start from the exposition of the problems one faces in the construction
of superstring field theory. We then present the cubic and the non-polynomial
versions of superstring field theory and discuss a proposal suggesting their
classical equivalence.
Finally, the bosonic solutions are generalized to this case. In particular,
we focus on the (somewhat surprising) generalization of the tachyon solution
to the case of a theory with no tachyons.
\end{abstract}

\end{frontmatter}

\bibliographystyle{JHEP}

\newcommand{\A}{{\cal A}}
\newcommand{\Q}{{\cal Q}}
\newcommand{\cC}{{\cal C}}
\newcommand{\NN}{{\mathbb N}}
\newcommand{\C}{{\mathbb C}}
\newcommand{\RR}{{\mathbb R}}
\newcommand{\X}{{\cal X}}
\newcommand{\ATFV}{{After the first version of this paper appeared, }}
\newcommand{\aTFV}{{after the first version of this paper appeared, }}
\newcommand{\zz} {s}
\newcommand{\res} {\operatorname{res}}
\newcommand{\arccot} {\operatorname{arccot}}
\newcommand{\al} {\alpha}
\newcommand{\p}{{\cal P}}
\newcommand{\N}{{\cal N}}
\newcommand{\cs}{{\cal S}}
\newcommand{\ep}{{\epsilon}}
\newcommand{\s}{{\sigma}}
\newcommand{\w} {\omega}
\newcommand{\si}{{\sigma}}
\newcommand{\la}{{\lambda}}
\newcommand{\ka}{{\kappa}}
\newcommand{\La}{{\Lambda}}
\newcommand{\K}{{\rm K}}
\newcommand{\cm} {\text{cm}}
\newcommand{\cL}{{\cal L}}
\newcommand{\LL}{{\cal L}}
\newcommand{\BB}{{\cal B}}
\newcommand{\cO}{{\cal O}}
\newcommand{\One} {{\bf 1}} 
\newcommand{\tr} {\operatorname{tr}}
\newcommand{\HH} {{\cal H}}
\newcommand{\II} {{\cal I}}
\newcommand{\csch} {\operatorname{csch}}
\newcommand{\bpz} {\operatorname{bpz}}
\newcommand{\smallfrac}[2] {{\textstyle\frac{#1}{#2}}}
\newcommand{\nodag} {{\phantom{\dag}}}
\newcommand{\dcirc}{\mathop{{}^\circ_\circ}}

\newcommand{\mat}[1] {\begin{pmatrix}#1\end{pmatrix}}
\newcommand{\bra}[1] {\left<#1\right|}
\newcommand{\ket}[1] {\left|#1\right>}
\newcommand{\braket}[2] {\left<#1\vphantom{#2}\right|
                         \left.\!\vphantom{#1}{#2}\right>}
\newcommand{\vev}[1] {\left<#1\right>}

\newcommand{\phbra}[2] {{\vphantom{\bra{V_2}}}_{#1}\!\bra{#2}}
\newcommand{\brao}[1] {\left(#1\right|}
\newcommand{\keto}[1] {\left|#1\right)}
\newcommand{\braketo}[2] {\left(#1\vphantom{#1}\right|
                         \left.\!\vphantom{#1}{#2}\right)}

\newcommand{\brai}[2] {{\vphantom{\bra{#2}}}_{#1}\!\bra{#2}}

\tableofcontents

\section{Introduction}

String field theory is a field theoretical framework for string theory.
As such, it can potentially be used for evaluating scattering
amplitudes using Feynman diagram techniques,
for studying non-perturbative effects of the theory and
for exploring various string vacua by studying classical solutions.

Many attempts have been made to find analytical solutions of
(open covariant bosonic) string field theory, since its introduction by
Witten~\cite{Witten:1986cc}.
It took twenty years until such a solution was constructed by
Schnabl~\cite{Schnabl:2005gv}.
This construction drew much attention and was followed by several
other analytical solutions and other generalizations that improved
significantly our understanding of string field theory.
It is the purpose of this work to review these developments in order to
make them more accessible to the general string theory community.

Previous attempts towards analytical solutions within string field theory
used various tools. Among other approaches one can find:
\begin{itemize}
\item Conformal field theory (CFT)
      methods~\cite{LeClair:1989sp,LeClair:1989sj}:
      Defining the star product and integration in terms of CFT expressions
      enables the usage of many tools and gives a geometric picture to the
      problem, which can then be dealt with in various conformal frames.
      Moreover, many CFT methods are insensitive to the given background and
      are thus universal\footnote{From a technical point of view,
      universality is the property of being dependent on the matter sector
      only through its energy-momentum tensor. We give more details on that
      in the next two sections.}. Universality is an important property of
      solutions that represent tachyon condensation~\cite{Sen:1999xm}.
      Indeed, CFT methods proved essential for the recent progress.

\item Half-string
formalism~\cite{Bordes:1991xh,Rastelli:2001rj,Gross:2001rk,Gross:2001yk,Furuuchi:2001df}:
      A fundamental entity in the construction of string field theory is the
      star product, which can be cast into a form analogous to a matrix
      product, with the two halves of the string playing the roles of the
      two matrix indices. Bogoliubov transformation to the vacuum built upon
      the two string halves simplifies the form of the star product. One
      less favorable feature of this approach
      (and some of the next ones) is that it is based on a given (flat)
      background and is therefore not universal\footnote{Note that the basic
      idea of the half-string formalism can be presented in a universal way.
      The Bogoliubov transformation, on the other hand, refers to a
      specific background.}.
\item Simplifying the non-linear term in the equation of
      motion~\cite{Kostelecky:2000hz}: The equation of motion contains
      a single non-linear piece. As in the case of the Riccati equation,
      simplifying the form of the non-linear term has the potential of
      rendering the equation
      solvable. This approach led to the algebraic construction of the
      surface-state star-projector known as the sliver. The construction
      itself turned out to be equivalent to the half-string approach.
\item Representing the star product in a Moyal
form~\cite{Bars:2001ag,Douglas:2002jm,Bars:2002nu,Erler:2002nr,Bars:2002qt,Bars:2003gu,Bars:2003sm}:
      A natural framework for addressing non-commutative products is the
      Moyal approach, which gained popularity in the string theory community
      following~\cite{Seiberg:1999vs}.
      It was found that the star product can be represented as an infinite
      dimensional tensor product of Moyal pairs.
\item Diagonalizing the star product~\cite{Rastelli:2001hh,Okuyama:2002yr}:
      In the oscillator basis, the star product is defined using some
      infinite dimensional matrices that can (almost) be simultaneously
      diagonalized. This results in the
      continuous basis (aka $\ka$-basis). This approach led to some results
      in the matter sector. The inclusion of the ghosts, however, turns out
      to be more problematic.
\item Vacuum string field
theory~\cite{Rastelli:2000hv,Rastelli:2001jb,Rastelli:2001vb,Rastelli:2001uv,Hata:2001sq,Rastelli:2001wk,Hata:2002xm,Okawa:2002pd}:
      This is not a genuine way to solve string field theory. Instead, the
      form of the theory expanded around the (tachyon) solution, after an
      unknown (and presumably singular) field redefinition, is guessed. This
      approach led to numerous analytical results.
      However, its scope of validity and its relation to the standard string
      field theory were not fully clarified.

\item Identity string field based
solutions~\cite{Takahashi:2001pp,Kluson:2002ex,Takahashi:2002ez,Drukker:2003hh}:
      The star product contains an identity element.
      This state is, however, singular in several respects, e.g.,
      it is not a normalizable state~\cite{Horowitz:1987yz} and in the CFT
      description it represents surface gluing.
      Moreover, it is not killed by $c_0$, which suppose to be a derivation
      of the star product~\cite{Rastelli:2000iu}, implying
      $0\neq c_0\ket{1}=c_0(\ket{1}\star\ket{1})=
      (c_0\ket{1})\star\ket{1}+\ket{1}\star(c_0\ket{1})=2c_0 \ket{1}$, i.e.,
      a contradiction.
      An advantage of the identity based solutions is that they are
      universal.
      On the other hand, while the identity based solutions are formally
      well defined, the problems with the identity string field rendered the
      evaluation of the action of these solutions ambiguous\footnote{\ATFV
      the papers~\cite{Kishimoto:2009nd,Kishimoto:2009hc} appeared, which
      improved the credibility of the identity based solutions.}.
\item Using other surface states to define
      solutions~\cite{Okawa:2003cm,Okawa:2003zc,Yang:2004xz,Drukker:2005hr}:
      In order to bypass the problems with the identity string field, while
      keeping some of the nice properties of the solutions, one can choose
      to work with other star-algebra projectors, such as the butterfly and
      the sliver. Constructing solutions turns out to be more complicated in
      this case than in the case of the identity based solutions.
\end{itemize}

In this work we focus on Schnabl's solution and the developments that
followed.
This construction uses ideas from most of the approaches described above.
Hence, we introduce some of these tools.
While we tried to make this review self-contained, the reader may want to
consult also older reviews focusing on other aspects of string field theory.
The following is a list of related reviews.
\begin{itemize}
\item An introduction to string field theory emphasizing the
      Batalin-Vilkovisky (BV) approach and perturbation theory is the
      classical review by Thorn~\cite{Thorn:1988hm}.
\item Shorter modern general introduction to string field theory can be found
      in~\cite{Rastelli:2005mz,Taylor:2006ye}.
\item The issue of tachyon condensation in string field theory and vacuum
      string field theory is addressed in~\cite{Rastelli:2004cp}.
\item General and extensive reviews of tachyon condensation covering also
      other frameworks for addressing this subject
are~\cite{Ohmori:2001am,DeSmet:2001af,Taylor:2002uv,Taylor:2003gn,Sen:2004nf}.
\item A short review of the early achievements following Sen's conjectures
      is~\cite{Zwiebach:2001nj}.
\item A review that describes many representations of the string field
      algebra with some focus on string field theory as a non-commutative
      field theory is~\cite{Arefeva:2001ps}.
\item The Moyal approach to string field theory is introduced
      in~\cite{Bars:2002yj}.
\item Two papers that stress the algebraic structure of the star product
      are~\cite{Zwiebach:2002fd,Bonora:2003xp}.
\end{itemize}

Let us describe the content of the various sections (by order)
and their interrelations,
in order to answer the perpetual question:
How to use this review?
\begin{enumerate}
\item In the rest of the introduction we briefly describe the advance in
      computing scattering amplitudes in string field theory,
      where impressive progress has been obtained.
      While this is not a result regarding analytical solutions, it is
      important and relevant enough to be shortly described here.
      We refer the reader to the original works for a complete and clear
      exposition of this important subject.
\item This is an introductory section, where string field theory basics are
      described. Researchers familiar with its construction can skip this
      part.
\item Another introductory section, in which Sen's conjectures and their
      study within level truncation are described. It serves as a motivation
      and a historical overview and can also be skipped.
\item Here we present some of the tools needed for the derivation of
      Schnabl's solution.
\item In this section Schnabl's construction of the tachyon vacuum is derived
      using the tools of the previous section. The proofs of Sen's
      conjectures for this solution are then described.
\item This section presents some of the ways by which analytical solutions
      describing marginal deformations were obtained within string field
      theory.
      It relies on some basic concepts from the previous two sections.
\item The recent advance relies mainly on CFT technics. Still, we give an
      oscillator description of Schnabl's solution in this section. It relies
      only on some basic formulas from section~\ref{sec:Schnabl} and
      the oscillator approach is explained therein.
      Other sections do not depend on this one and it can be skipped by
      readers interested in the main root of advance in the field.

\item Here we describe the generalization of the matters presented in previous
      sections to open superstring field theory. The peculiar obstacles for
      the construction of a superstring field theory are described, as well
      as the superstring field theories themselves. This section relies on
      most of the previous ones.
\item An outlook.
\end{enumerate}

\subsection{Scattering amplitudes}
\label{sec:other}

In string field theory scattering amplitudes can be defined even
off-shell.
Off-shell scattering amplitudes can be used as building blocks for
loop amplitudes as well as for the evaluation of the
tachyon potential. The first calculation of off-shell amplitudes was performed
in~\cite{Samuel:1987uu}.
The evaluation of the quartic term in the tachyon potential immediately
followed~\cite{Kostelecky:1988ta}. This calculation was extended to higher levels
and developed in~\cite{Taylor:2002bq,Taylor:2004rh,Forini:2006tn} and in the Moyal
approach~\cite{Bars:2002qt,Bars:2003sm}.

Scattering amplitudes were commonly calculated in the Siegel gauge.
The attractive properties of the gauge choice introduced by Schnabl
changed that.
The Schnabl gauge has also some less attractive properties. It is based on an
operator, which is not hermitian. It is singular in some sense and two
Schwinger parameters should be used to describe a propagator instead of
one~\cite{Schnabl:2005gv}.
Preliminary results in the Schnabl gauge were obtained in~\cite{Fuji:2006me}.
In this paper it was also suggested to use a hermitian version of Schnabl's
operator in order to fix the gauge.

The scattering amplitudes calculated in~\cite{Fuji:2006me} seemed not to obey
some expected symmetries.
A careful study of the subject~\cite{Rastelli:2007gg} showed that the
propagator should be regularized. The results now were shown to obey some of
the symmetries, while it was claimed that the other ones are peculiar to the
Siegel gauge.

Up to this point, only (off-shell) tree amplitudes were considered in the
Schnabl gauge.
A thorough study of scattering amplitudes using a general gauge choice based
on the $b$-ghost, termed linear $b$-gauges, was performed
in~\cite{Kiermaier:2007jg}.
Regularity conditions on the gauge choice were stated and it was found that
the Schnabl gauge is located on the boundary of the regular region in the
space of $b$-coefficients. Nevertheless, no explicit problems with the
results in Schnabl gauge were obtained, unlike in the case of its hermitian
counterpart. Another important discovery of~\cite{Kiermaier:2007jg} was that
Schnabl's condition cannot be imposed at all ghost numbers. This result
does not afflict previous calculations, since it is of importance only when
loop amplitudes are considered, in which string fields of arbitrary ghost
number appear.
General $b$-gauge choices imply different gauge conditions at different ghost
numbers. Still, the question remains, whether the Schnabl gauge can be used
for the evaluation of general loop amplitudes. The disturbing (and novel)
property of this gauge is that it fixes the string mid-point. Thus, it may
seem that moduli space cannot be covered.

This problem was addressed in~\cite{Kiermaier:2008jy}, where the Schnabl
gauge was defined as a limit of regular $b$-gauges. A careful study of the
limit showed that the limiting wedge states should be glued not in the
standard way, but using a slant, which is responsible for a ``hidden boundary
at infinity''. The introduction of slanted wedges enabled the authors to
show that at least at 1-loop level the moduli space is covered
in the Schnabl gauge, despite the apparent problems. This seems
as a strong evidence in favour of the possibility that Schnabl's gauge can
be used to an arbitrary loop level. Moreover, it was claimed that the results
are technically much easier to derive than in the Siegel gauge.
There is still work to be done, though. It would be desirable to
study the general covering of moduli spaces, to calculate explicit scattering
amplitudes and to address the Schnabl gauge using a BV formalism.

\section{String field theory basics}
\label{sec:SFT}

String field theory is an approach to string theory, which is at least
conceptually the most straightforward one. This approach generalizes the
particle physics tools of second quantization to string theory.
In particle physics second quantization amounts to taking the
Schr\"odinger equation of a particle and reinterpreting it as the
equation of motion of a field. It turns out that upon quantization,
such a field describes just a collection of non-interacting particles
of the type of the original one. To this one can add interaction terms
obeying some principles, such as symmetry and renormalizability, in order
to get the full field theory of interacting particles. One added advantage
is that the classical equations of motion of the field can also reveal
non-perturbative effects. Thus, the quantum interaction of particles
and the classical, non-perturbative effects, are related.

In string theory the ``first quantized'' world-sheet approach already
specifies the interaction in the form of the Polyakov path integral.
One may hope then, that the string field theory is uniquely defined by
the world-sheet theory. If this is the case, one may hope to find this
string field theory and then study it both for
understanding and controlling string interactions and as a tool for the study
of non-perturbative effects is string theory.
Indeed, already in 1990, before the big advance in understanding
non-perturbative string theory following the introduction of D-branes by
Polchinski~\cite{Polchinski:1995mt}, a numerical solution to the equations
of motion of string field theory was found~\cite{Kostelecky:1990nt}.
This solution amounts to the closed
string vacuum left after the removal of a D-brane. This, however, was not
understood at the time and the interpretation of the solution
had to wait to the introduction of the celebrated Sen's conjectures, to be
described in section~\ref{sec:Sen}.

There are several perturbative string theories and one should be chosen
in order to construct a string field theory. There are also several
approaches for quantization on the world-sheet, e.g., light-cone. So we
stress at this point that unless otherwise stated, we shall be dealing
with open bosonic strings and will
do so in a covariant formalism, where one can hope to gain some insight
about the general symmetry structure of the theory.

At the linearized level string field theory should give the equation of
motion~\cite{Siegel:1984wx,Siegel:1984xd}
\begin{equation}
\label{LinEOM}
Q_B \Psi=0\,,
\end{equation}
where $\Psi$ is the string field and $Q_B$ is the BRST operator on the
world-sheet of the open bosonic string.
One can think of the string field
as a linear combination with momentum-dependent
coefficients (fields) of states of the first quantized
ghost number one space.
Hence, it is an off-shell generalization of the vertex operator.
In the usual flat space it takes the form,
\begin{equation}
\label{PsiOsc}
\ket{\Psi}=\int d^{26} k \Big(T(k)c_1 + C(k) c_0 + A_\mu(k)\al^\mu_{-1}c_1
+...\Big)\!\ket{k},
\end{equation}
with the usual definition
\begin{equation}
\ket{k}=e^{i k X(0)}\ket{0}.
\end{equation}
The infinity of fields composing it are the momentum-dependent coefficients
of the various modes. Each can be characterized by its level, that is
its eigenvalue with respect to $L_0+1$ (ignoring the momentum part).
Thus, the tachyon field $T$ is of level zero,
while the photon field $A_\mu$ and the (auxiliary) field $C$ are of level
one.
The free equation of motion~(\ref{LinEOM}) should be accompanied by the
free gauge transformation
\begin{equation}
\label{linGauge}
\Psi \rightarrow \Psi + Q_B \Lambda\,.
\end{equation}
Thus, the equation of motion of the string field
corresponds to the condition of being closed and the gauge transformation
identifies string states whose difference is exact.
The gauge string field $\La$ itself\footnote{Throughout this manuscript
we refer to $\La$ as the ``gauge string field''.
This terminology might cause confusion, since in
field theories, such as QED, one usually refers to $A_\mu$ as the ``gauge
field'' and to $\La$ as the ``gauge parameter''. The string field $\Psi$
generalizes $A_\mu$ and also includes such a component field.
Hence, it would have been natural
to refer to $\Psi$ as the gauge string field. Nonetheless, we save this
terminology for $\La$ in order to stress that it is not a parameter, but rather
a collection of infinitely many component fields. We then refer to $\Psi$
as the ``physical string field'', or just as the ``string field'' for short.
Both terminologies appear in the literature, i.e., in some papers the
``gauge string field'' or ``gauge field'' refers to $\La$ as we do here,
while in others it refers to $\Psi$.} consists of an off-shell linear
combinations of states with ghost number zero.
This structure resembles that of differential forms, with $Q_B$ playing the
role of the differential, since it obeys
\begin{equation}
\label{QBnil}
Q_B^2=0\,.
\end{equation}
Note however, that unlike for forms in a finite dimensional space, here there
are ``forms'' of an arbitrary integer degree, from minus infinity to infinity.
The gauge transformation~(\ref{linGauge}) is accordingly reducible at
an infinite order, since a transformation of the form
\begin{equation}
\Lambda^{(n)} \rightarrow \Lambda^{(n)} + Q_B \Lambda^{(n-1)}\,,
\end{equation}
does not contribute to the transformation of $\Lambda^{(n+1)}$
performed in the same way, due to~(\ref{QBnil}). Here, the superscript
represents the ``form-degree''. Since $Q_B$ increases the ghost number by
one, the form-degree can be identified with the (first quantization)
ghost number.
Being a one-form, it is natural to declare that the string field is an odd element.

An appropriate linearized action is~\cite{Neveu:1985sh}
\begin{equation}
\label{actionLin}
S_{\text{lin}}=-\frac{1}{2}\bra{\Psi}Q_B\ket{\Psi},
\end{equation}
where $\bra{\Psi}$ is the BPZ conjugate of $\ket{\Psi}$.
Let us explain the coefficient in front of~(\ref{actionLin}).
The minus sign is needed in order to get the canonical sign for the
component fields. For the
tachyon field $T$, which is the first field in the
expansion~(\ref{PsiOsc}), one gets
\begin{align}
\label{freeTachyonAction}
S=-&\frac{1}{2}\int d^{26}k\, d^{26}k'\,
      T(k)T(k')\bra{k}c_{-1} Q c_1\ket{k'}=\\
  &\frac{1}{2}\int d^{26}k\, d^{26}k'\,
      T(k)T(k')\bra{k}c_{-1} c_0 c_1(1-k'^2)\ket{k'}=
  \int d^{26}k\, \frac{1-k^2}{2}T(k)^2\,.
\nonumber
\end{align}
Here we used the BPZ conjugation property of the $c$ field,
as well as the oscillator expansion of $Q_B$, neglecting in the intermediate
step some terms that do not contribute to the inner product.
We also defined $\al'=1$ and used
\begin{equation}
\bra{0}c_{-1} c_0 c_1\ket{0}=
   \frac{1}{2}\left<\partial^2 c \partial c c \right>=1\,,
\end{equation}
as the normalization convention in the ghost sector.
Given our other convention, that of a mostly positive metric,
the resulting action for the tachyon field $T$ has the canonical form
and we also see that it is indeed a tachyonic field.
For higher level fields one may have to adjust the coefficient in
the definition of the field in order to get a canonical form.
The sign is, however, canonical for all fields.

The inclusion of interactions into this formalism was performed by
Witten~\cite{Witten:1986cc}. The action is given by
\begin{equation}
\label{action}
S=-\int\Big(\frac{1}{2} \Psi\star Q \Psi+
   \frac{g_o}{3} \Psi\star \Psi\star\Psi \Big)\,.
\end{equation}
Here $g_o$ is the open string coupling constant that can (at least classically)
be set to unity by a field rescaling. We shall set it to unity unless
otherwise stated.
From~(\ref{action}), one can derive the equation of motion\footnote{The
derivation is somewhat formal at this stage.
In order to be able to truly derive it, one should first specify the meaning of
the operations involved and conclude that a variant of the
fundamental lemma of the calculus of variations holds in this case.
All that, in fact, works out well.},
\begin{equation}
\label{eom}
Q \Psi+\Psi\star\Psi = 0\,.
\end{equation}

The integration, the star product and the derivation $Q$ are (bi-)linear
operations over the space of (arbitrary ghost number) string fields $\A$
\begin{equation}
\int:\A\rightarrow \C\,,\qquad
\star:\A\times\A\rightarrow\A\,,\qquad
Q:\A\rightarrow \A\,.
\end{equation}
Witten assumed that these entities obey some
algebraic relations, described below. He claimed that the
emergent (noncommutative geometry) structure is the natural
one for describing string field theory.
Regardless of this structure, the first requirement of interacting
string field theory is that in the limit $g_o\rightarrow 0$, it reduces
to the non-interacting theory~(\ref{actionLin}).
This is the case if one identifies\footnote{The definition of
$\int \Psi_1\star\Psi_2\star\Psi_3$ is more involved. We return to this
issue shortly.},
\begin{equation}
\label{IntPsiStarPsi}
Q=Q_B\,,\qquad \int \Psi_1\star\Psi_2=\braket{\Psi_1}{\Psi_2}.
\end{equation}
Henceforth, $Q$ will represent the BRST operator $Q_B$.
Other realizations of $Q$ (denoted $\Q$),
obeying the algebraic structure, are also of importance.
A particular example is obtained
when the theory~(\ref{action}) is expanded around a solution.
We describe this case in section~\ref{sec:Sen}.

Other then linearity, Witten's axioms state that
the star product is associative and $Q$ is
an odd derivation with respect to it, namely\footnote{Throughout this paper
$A$ in $(-1)^A$ represents the grading of $A$.}
\begin{equation}
Q^2=0\,,\qquad Q(A_1\star A_2)=
   (Q A_1)\star A_2+(-1)^{A_1}A_1\star(Q A_2)\,,
\qquad \int Q A=0\,.
\end{equation}
Here, $A_{1,2}$ are string fields of an arbitrary ghost number.
The star product preserves the grading and the ghost number.
Another axiom states that under integration the string fields
behave like differential forms,
\begin{equation}
\label{intPsiPsi}
 \int A_1\star A_2=(-1)^{A_1 A_2}\int A_2\star A_1\,.
\end{equation}
These axioms are identical to those obeyed by differential forms,
except the fact that in the case of differential forms~(\ref{intPsiPsi}) holds even
without integration\footnote{It turns out that
in the bosonic case the integral is
non-zero only for a string field of ghost number three. Thus, for the bosonic
string the sign factor $(-1)^{A_1 A_2}$ equals unity.
However, if the string fields themselves are allowed to carry
Grassmann grading, the sign can become important.
This is the case when the BV formalism is used in order to handle the
gauge symmetry.
Moreover, in the case of Berkovits' version of superstring field theory,
to be introduced in section~\ref{sec:SUSYtheories},
the total ghost number should
equal two and the minus sign is important.}.
The action~(\ref{action}) can thus be considered as a generalization
of the Chern-Simons action.

From the analogy with the Chern-Simons action we recognize that the
action~(\ref{action}) is invariant under the infinitesimal
gauge transformation
\begin{equation}
\label{infGauge}
\delta\Psi=Q\Lambda+\Psi\star\Lambda-\Lambda\star\Psi\,,
\end{equation}
where $\Lambda$ is a ghost number zero string field.
This infinitesimal gauge transformation can be exponentiated to give
the finite gauge transformation,
\begin{equation}
\label{finGauge}
\Psi\rightarrow e^{-\La}(Q+\Psi)e^\La\,,
\end{equation}
where multiplication, exponentiation, as well as any other function
of string fields, should be performed by expanding the relevant function
in a series, while keeping the order of string fields intact and using
the star product for the evaluation of all products.
The $1$ that is obtained at the zeroth order is realized by the identity string
field, described below.
No ambiguities can emerge, since the star product is the only way by
which string fields can be multiplied.
We continue to write it explicitly in the
introductory sections, but it will be left implicit in the remaining of
the paper, except in cases where its inclusion improves
readability.

The gauge transformation~(\ref{infGauge}) is reducible,
since a variation of the form
\begin{equation}
\delta\Lambda=Q\Lambda_{-1}+\Psi\star\Lambda_{-1}-\Lambda_{-1}\star\Psi\,,
\end{equation}
induces no change for $\Psi$ in~(\ref{infGauge}), provided that $\Psi$ obeys
the equation of motion~(\ref{eom}), i.e., it is on-shell.
Similarly,
\begin{equation}
\label{infGaugeinf}
\delta\Lambda_{-n}=Q\Lambda_{-(n+1)}+\Psi\star\Lambda_{-(n+1)}-
   \Lambda_{-(n+1)}\star\Psi\,,
\end{equation}
will induce no change in $\Lambda_{-n+1}$. This follows from
\begin{equation}
\delta^2 A=(Q\Psi+\Psi\star\Psi)\star A-A\star(Q\Psi+\Psi\star\Psi)\,,
\end{equation}
where again $A$ is a string field of an arbitrary degree.
Thus, the (non-linear) gauge generator
$\delta$ is nilpotent only on-shell~\cite{Marcus:1986uc},
in which case the gauge system is infinitely reducible.
A sensible way to deal with reducible gauge systems, as well as with
gauge systems that close only on-shell, is given by the BV
formalism~\cite{ZinnJustin:1974mc,Batalin:1981jr,Batalin:1984jr,Batalin:1984ss,Voronov:1982cp}
(and the reviews~\cite{Henneaux:1992ig,Gomis:1994he}).

The BV formalism was implied for quantizing string field
theory in~\cite{Bochicchio:1986zj,Bochicchio:1986bd,Thorn:1986qj}\footnote{A
BRST approach to string field theory was implied
in~\cite{Maeno:1990tz,Erler:2004hv,Erler:2004sy} using the
``mid-point light-cone gauge''.}.
It turns out that the infinity of ghost fields that should be added
can be identified with string fields of (first-quantized) ghost
number $n_g<1$, while the infinity of anti-fields can be identified
with string fields of ghost number $n_g>1$. The identification is
such that the ghost numbers of a field and its anti-field sum up
to 3, i.e, the anti-field of the classical (ghost number one)
string field carries ghost number two, the anti-field of the first
(ghost number zero) ghost carries ghost number three and so on.
As usual, the coefficient fields carry alternating parity. When
this alternating parity is combined with the parity of the
various ghost-number string states, it results in string fields
of a fixed, odd, parity. These string fields can be summed to give
(an odd) string field, whose ghost number is not constrained,
\begin{equation}
\label{BVPsi}
\Psi=\sum_{n=-\infty}^\infty \Psi_n\,.
\end{equation}
The classical master equation is obeyed by an action, which is
identical in form to~(\ref{action}), only with the string field being
given by~(\ref{BVPsi}), rather than by $\Psi_1$.
This action can be gauge fixed by the introduction of a gauge fixing
fermion. A common gauge choice is the Siegel gauge
\begin{equation}
\label{SiegelGauge}
b_0 \Psi=0\,.
\end{equation}
This is an extension of the Feynman gauge for photons, since this
condition results in the Feynman gauge for the massless vector
component of the string field, upon level truncation\footnote{One
can consider also other gauges, such as the analogue of the Landau gauge.
A study of gauges that interpolate between the Feynman-Siegel and
the Landau gauges was carried out
in~\cite{Asano:2006hk,Asano:2006hm}. Related
discussion was given in~\cite{Feng:2006dr}. Note, however, that there
is more than one possible extension of, e.g., the Feynman gauge, for
the string field. Another such extension, the Schnabl gauge, played
important role in the recent developments.}.
In this gauge, the equation of motion~(\ref{eom}) reduces to
\begin{equation}
\label{EOM}
c_0 L_0\Psi+\Psi\star\Psi=0\,.
\end{equation}

In order to preserve associativity, Witten suggested
that the star product should implement the geometric picture of
gluing the right half of the first string with the left half
of the second string, while integration should amounts to gluing
together both halves of the same string around the middle. This
suggestion is consistent with~(\ref{IntPsiStarPsi}), since in this
case one gets just a gluing of two strings with their orientation
reversed, i.e., the BPZ inner product~\cite{Belavin:1984vu}.
With these identifications of the star product, the integral and the
derivation, it was shown by Giddings et. al. that the Veneziano
amplitude is reproduced from the action~(\ref{action}) and that
at higher order in $g_o$ it reduces to
the Polyakov path integral with correct covering of moduli space
\cite{Giddings:1986wp,Giddings:1986bp,Giddings:1986iy}.
A gap in the proof of moduli space covering was filled up by
Zwiebach~\cite{Zwiebach:1990az}.

For practical calculations it is useful to represent the star product
using either CFT expressions as derived
in~\cite{LeClair:1989sp,LeClair:1989sj} or using oscillator expressions
in the flat background as described
in~\cite{Gross:1987ia,Gross:1987fk,Gross:1987pp,Samuel:1986wp,Ohta:1986wn,Cremmer:1986if}.
The former method has the advantage of being universal. That is, it
does not depend on a choice of a background, as long as it is
composed of a $bc$ system coupled to a $c=26$ BCFT. Indeed, most of the
recent developments in the field were achieved using this formulation.
Nevertheless, the
latter method has also proved useful, especially, but not only,
for describing phenomena peculiar to a given background.
We now turn to describe the CFT approach.
We describe the oscillator method and
the recent related results in section~\ref{sec:oscillator}.

\begin{figure}[mbth]
\begin{center}
\includegraphics[width=7cm]{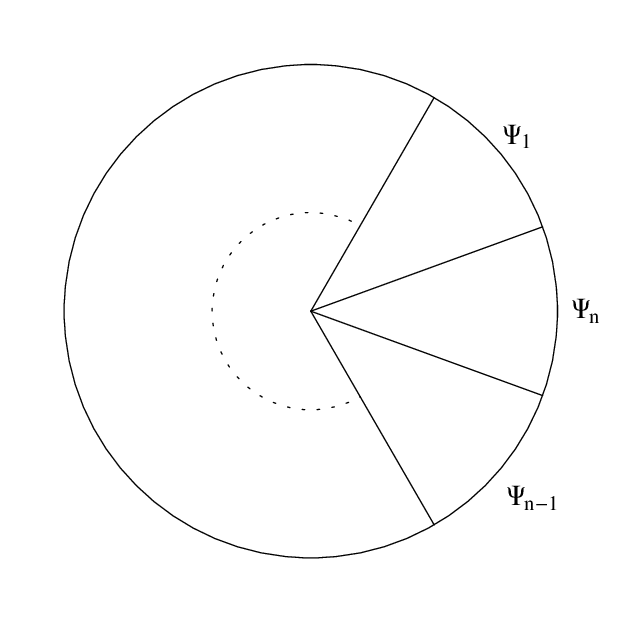}
\end{center}
\caption{The N-vertex represented on the unit disk. Each string state is
transformed to a $\frac{2\pi}{N}$ wedge and the edges of these wedges
are glued to each other.}
\label{fig:Nvertex}
\end{figure}

In the seminal papers~\cite{LeClair:1989sp,LeClair:1989sj},
LeClair, Peskin and Preitschopf defined more general string vertices
by CFT expectation values on the disk, as
\begin{equation}
\label{CFTvertex}
\int \Psi_1\star\ldots\star\Psi_n=
  \vev{f^{(n)}_1\circ\Psi_1(0),\ldots,f^{(n)}_n\circ\Psi_n(0)}.
\end{equation}
Here, the string fields on the left hand side are interpreted as
operators acting on the vacuum. In the right hand side
a given set of conformal transformations acts on these
operators, mapping them from the upper half plane
to various points on the disk, in some given canonical representation
of the disk.
The brackets represent the CFT expectation value of the theory living
on the disk. This expectation value turns the data encoded in the
insertions (the choice of the operators to be inserted $\Psi_k$ and the
choice of mappings that implement the insertions $f_k^{(n)}$) into a
number. The formalism is general. In order to reproduce Witten's
theory the conformal transformations should be fixed, such that one
gets the advocated gluing. If we choose the unit disk as the canonical
representative, as in figure~\ref{fig:Nvertex},
the conformal transformations take the form
\begin{equation}
\label{fDef}
f^{(n)}_k=\Big(\frac{1+i \xi}{1-i \xi}\Big)^{\frac{2}{n}}
    e^{\frac{2\pi i k}{n}}\,.
\end{equation}
Note that the local coordinate patch in the upper half plane, i.e.,
the top half-unit-disk, is
transformed under these maps to wedges with angles of $\frac{2\pi}{n}$ and
these wedges are glued together such that the right half of the
$k^{\text{th}}$ string is glued to the left half of the
$(k+1)^{\text{th}}$ string.
In this representation the symmetry of the vertex under cyclic
permutation is manifest.
Other coordinate systems can be more adequate for specific problems.
We return to this point in section~\ref{sec:primer}.

The CFT representation makes it clear that the $n$-vertices are
multi-linear functions. In particular, the two-vertex forms a
bi-linear map, rather than a hermitian inner product, as one might
have expected~\cite{Witten:1986qs}.
It is possible to impose a reality condition on the string
field,
\begin{equation}
\label{RealPsi}
\Psi^*=\Psi\,.
\end{equation}
The conjugation in the above equation represents a hermitian conjugation
followed by the inversion of the orientation of the string. In terms of the
operators producing the string state this conjugation corresponds to the
composition of hermitian and BPZ conjugations.
When this reality condition is imposed, the two-vertex forms also a hermitian
inner product.

The CFT methods of~\cite{LeClair:1989sp,LeClair:1989sj} can be used not
only for defining string vertices, but also for defining string fields
of a special class, called surface states, which will be very useful in the
following sections. Surface states are generalizations of the
one-vertex, in which the function $f_1^{(1)}$ of~(\ref{fDef}) is replaced by
another (almost) arbitrary function $f$. Thus, there is a 1-1 correspondence
between the set of permissible functions $f$ and the set of surface states.
Given $f$, we define the surface state $\bra{S}$ (or $\bra{S_f}$ if several
such states are involved) by declaring that its contraction with an
arbitrary test state,
\begin{equation}
\Psi=\cO_\Psi(0)\ket{0},
\end{equation}
obeys,
\begin{equation}
\label{SPsiSS} 
\braket{S}{\Psi}=\left<f\circ \cO_\Psi\right>.
\end{equation}
This definition might seem to be unrelated to any particular surface,
as opposed to what one might expect from the term ``surface state''. Nonetheless,
a particular surface certainly enters this definition, as we have to
specify the representation of the disk chosen for the evaluation of the
CFT expectation value. One might think that this choice is immaterial,
since in a CFT, expectation values are by definition invariant under
reparametrization of the disk (i.e., under conformal transformations).
Indeed, the representation of the
disk can be modified. However, this leads also to a change of the
function $f$, which is in fact necessary for the invariance of
expectation values. Hence, the function $f$ and the choice of the
representation for the disk hold the same information.
Surface states can, therefore, be defined either by fixing the disk
(e.g., to the upper half plane) and specifying $f$, or by fixing $f$
while specifying the surface that should be used
for the evaluation of the expectation value. It is also possible
to specify both, a surface $\Sigma$ and a conformal transformation $f$.
While such representations are somewhat degenerate, they might be very
useful in some cases. One can write $f_\Sigma$ and $\bra{S_\Sigma}$,
in order to stress that everything depends on the choice of the surface
$\Sigma$. Similarly, the expectation value~(\ref{SPsiSS}) can be written as,
\begin{equation}
\braket{S_\Sigma}{\Psi}=\left<f_\Sigma\circ \cO_\Psi\right>_\Sigma\,.
\end{equation}

\begin{figure}
\begin{center}
\input{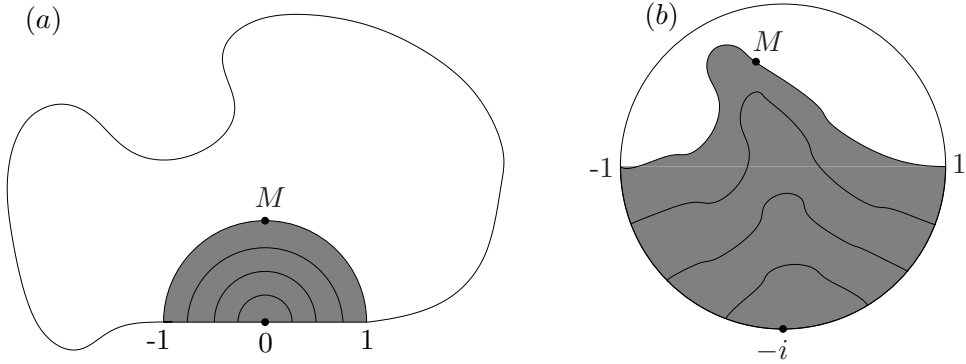}
\end{center}
\caption{Two possible representations of a surface state. In $(a)$, the local
  coordinate patch takes its canonical form, while in $(b)$, the whole surface
  takes a canonical form (a unit disk here, but other canonical
  representations, e.g., the upper half plane, are possible). It is generally
  impossible to find a canonical transformation in which both take their
  canonical form. The only exception is that of the surface state
  representing the $SL(2)$ invariant vacuum $\bra{0}$. It is, however,
  possible to fix three points of the local coordinate patch. These are most
  naturally chosen to be the two points of the boundary of the local
  coordinate patch that touch the second patch and the puncture at which the
  test state is to be inserted, i.e., the point $\pm 1$ and $0$ in $(a)$ can
  be sent to $\pm 1$ and $-i$ in $(b)$ (this choice can become singular for
  some singular surface states, e.g., the identity string field). This
  assignment completely fixes the local coordinates. In particular, it
  fixes the mid-point $M$. Then, the points $\pm 1$ and $M$ fix a
  local coordinate system also on the other patch. One can refer to the two
  as the local coordinate patch of the test state and the local coordinate
  patch of the state itself.
  Unfortunately, in much of the literature both are called ``the local coordinate
  patch'', which can lead to confusion. Note, that in the representation
  $(a)$, a general surface state might not be represented as a simple surface
  in the complex plane, as it might involve a multiple cover of parts of the
  complex plane. This will result in cuts in this representation. This is not
  a problem, as long as the representation $(b)$ is well defined.}
\label{fig:SurState}
\end{figure}

A nice geometric interpretation of surface states (see figure~\ref{fig:SurState}~$(a)$)
can be obtained by choosing
$f(z)=z$~\cite{Rastelli:2001vb}.
In the BPZ picture, the inner product
corresponds to the gluing of two states, which can be thought of as a state and a
test state. The test state is created by the insertion of the operator
$\cO_\Psi$ at the origin, while the state itself is created by an insertion at infinity.
The regions around $z=0$ and $z=\infty$ are the local coordinate patches of the two states.
These two local coordinate patches meet on the half-circle $|z|=1$
and we think of the string wave functions on the curve $|z|=1$ as
evolving from the insertions at $z=0,\infty$ towards this curve.
The inner product is then the overlap of the wave functions of the two strings at
$|z|=1$. On the other hand, in the representation of surface states
discussed above, the state is not created
using an insertion at infinity. Rather, it is created by choosing a non-standard
form for the local coordinate patch of the surface state itself.
Of course, the surface state can also be represented in the usual manner using the
insertion at infinity and both representations should lead to the same wave function
at $|z|=1$, since the test state $\Psi$ in~(\ref{SPsiSS}) is arbitrary.
We conclude that a surface state is a state for which the insertion that creates
the state can be traded for a surface deformation. This can only be the case if
the operator that creates the surface state is an operator that generates a
finite surface deformation\footnote{So far we are considering genuine surface state,
without any further insertions. The generalization to
``surface states with insertions'', which is in fact a degenerated
representation of all string fields, would be described below.}.

This geometric picture leads also to an algebraic representation for surface states.
It is known that infinitesimal deformations of a surface are obtained by
the Virasoro generators. Hence, finite deformations are obtained by an
exponentiation of those generators. Thus, we conclude that
a general surface state can be written as,
\begin{equation}
\label{surfS}
\ket{S}=\exp\big(\sum_{n=2}^\infty s_n L_{-n}\big)\ket{0},
\end{equation}
where the $s_n$ are the coefficients defining the surface state.
The simplest surface state is the vacuum, for which the conformal map is the
identity map $f(z)=z$. Using the standard local coordinate patch, while evaluating the
CFT expectation value on the upper half plane, is the same as a contraction
with the vacuum state. Hence, we conclude that for $\ket{0}$, $s_n=0\ \forall n$.
The relation between the coefficients $s_n$ and the conformal transformation
$f$ in the general case was described in~\cite{Schnabl:2002gg}.

The form~(\ref{surfS}) for surface states implies that in the oscillator representation
of flat space background, a surface state is a squeezed state, i.e., its
matter part (we focus on one dimension for simplicity) can be represented by,
\begin{equation}
\ket{S}=e^{\frac{1}{2} a_n^\dag S_{nm} a_m^\dag}\ket{0}.
\end{equation}
The matrix $S_{nm}$ can be deduced from the conformal transformation $f$ or
from the coefficients $s_n$.
The converse, however, does not necessarily hold, namely there are many
squeezed states, which are not surface states.
A simple criterion for deciding whether a given (matter)
squeezed state is (the matter part of) a surface state was introduced
in~\cite{Fuchs:2002zz}. Another criterion was later introduced that
stressed integrability as the main character of surface
states~\cite{Boyarsky:2002yh} (integrability within this context was
further studied in~\cite{Bonora:2002un,Boyarsky:2003kr}).
The two criteria were shown to be
equivalent in~\cite{Fuchs:2004xj}, where it was also shown that
the first column of the defining matrix of a squeezed state
is equivalent to a unique surface state. Related study of the ghost
sector was performed
in~\cite{Maccaferri:2003rz,Kling:2003sb,Ihl:2003fw,Fuchs:2004xj}.

An important property of surface states is that a
given surface defines a surface state for an arbitrary BCFT. Thus, it is
possible to discuss, e.g., the sliver (a particular surface state) on a
D25-brane, as well as the sliver on the D0 brane. It is also possible to
study separately the matter and ghost parts of a surface state and further
separate the surface state into various sectors. Note, however, that
string fields that are defined by different surface states in different
sectors might be inconsistent~\cite{Schnabl:2002gg}. One normally thinks
of ``balanced'', e.g., ``geometric'' surface states, i.e., the Virasoro operators
in~(\ref{surfS}) are total Virasoro operators.

Another important consequence of the representation~(\ref{surfS}) is that
surface states are BRST closed,
\begin{equation}
\label{QS=0}
Q\ket{S}=0\,,
\end{equation}
since the BRST charge commutes by construction with all the Virasoro
generators. An important special case is
\begin{equation}
\label{Q1=0}
Q\ket{1}=0\,,
\end{equation}
where $\ket{1}$ is the identity string field, which is a surface state by
definition, since it is generated by the conformal transformation $f_1^{(1)}$
defining the one-vertex. The relation~(\ref{Q1=0}) is important for the
consistency of Witten's axioms, since the identity string field spans a
one-dimensional subspace isomorphic to the complex numbers inside the string
field algebra and ``constants'' should be annihilated by any derivation, $Q$
in particular.

\section{Sen's conjectures}
\label{sec:Sen}

An important advance in string field theory followed the realization that it
is the ideal framework for addressing Sen's
conjectures~\cite{Sen:1999mh,Sen:1999xm}.
Sen's work addressed the fate of the tachyon that is living on the bosonic
D-brane\footnote{Similar conjectures hold also for the open string tachyon
on a non-BPS D-brane and the tachyon living on the D-\=D
pair~\cite{Sen:1998ii,Sen:1998sm,Sen:1998tt,Sen:1998ex,Sen:1999mg}.}.
According to the conjectures, the tachyon describes the instability of the
D-brane.
An effective tachyon potential should have a local maximum around zero,
where the D-brane exists. This potential should also have a
local minimum\footnote{The tachyon potential in the supersymmetric cases is
bounded from below and the potential is symmetric around the origin. The
bosonic potential is not symmetric and CFT analysis suggests that it is
unbounded below. The ``local minimum'' of the tachyon potential at the
non-perturbative vacuum, might well be a saddle point.}.
A solution in which the tachyon acquires this
value as a vev (other fields also acquire vevs) represents the absence of
the original D-brane\footnote{The actual time-dependent process
of tachyon condensation, was described by Sen in the
``rolling tachyon'' papers~\cite{Sen:2002nu,Sen:2002in}.}.

This picture suggests the following properties:
\begin{enumerate}
\item The depth of the local minimum equals the tension of the original
      D-brane. This reflects the energy difference between the solutions
      with and without the D-brane.
\item	Other solutions exist, representing lower dimensional D-branes. These
      solutions	are lumps from the perspective of the tachyon field.      
\item There are no perturbative states around the tachyon solution, since
      perturbative states in open string field theory represent open string
      degrees of freedom and there are no open strings when the D-brane is
      absent.
\end{enumerate}
\begin{figure}[tbh]
\begin{center}
\includegraphics[width=7cm]{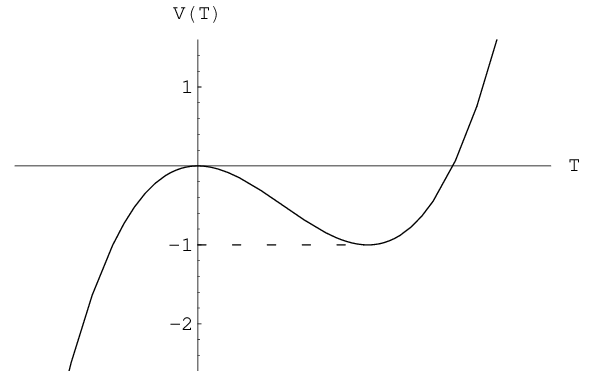}
\end{center}
\caption{The tachyon potential in units of D-brane tension.}
\label{fig:potential}
\end{figure}

The first of Sen's conjectures states the following.
Fix the value of the tachyon field to a constant and solve the equations
of motion of all the other components of the string field.
The action as a function of the tachyon field should be similar
to the one depicted in figure~\ref{fig:potential}. The local minimum
of the potential represents the solution describing the condensation of
the D-brane. Thus, the action of the solution should equal minus the
tension of the D-brane (with proper normalization of the space-time volume).

The second conjecture is related to the study of properties of solutions
that are generally space-time dependent.
For a solution to represent a lump of, e.g.,
co-dimension one, it has to reduce to the vacuum solution when one moves
in this one dimension away from the core, which should be thin in some sense.
Analogously to the first and third conjectures, the solution should have
the correct tension and carry the appropriate perturbative spectrum in
order to describe a D-brane of the appropriate dimension.

In order to understand the third conjecture we have to examine
the form of the action~(\ref{action}) when expanded around the solution.
The expansion is obtained by substituting
\begin{equation}
	\Psi\rightarrow\Psi_0+\Psi\,,
\end{equation}
into the action. As the original action is cubic it is also cubic with
respect to the shifted string field. The zero order term is the constant that is used
for testing the first conjecture. However, as a constant it does not contribute
to the equation of motion and so can be neglected.
The first order term is absent by definition, since we expand around a solution, 
which is an extremum of the action.
The third order term cannot be changed by a shift of the field. Thus, theories
that one gets by expansion around different backgrounds differ only in the form
of their kinetic, i.e., second order, term. Direct evaluation gives,
\begin{equation}
\label{actionAroundSol}
S=-\int\Big(\frac{1}{2} \Psi\star \Q \Psi+
   \frac{1}{3} \Psi\star \Psi\star\Psi \Big).
\end{equation}
The new kinetic operator is defined by its action on a string field with arbitrary
ghost number $A$ by,
\begin{equation}
\label{QQdef}
  \Q A=Q A+\Psi_0\star A-(-)^A A\star \Psi_0\,.
\end{equation}
This form of the kinetic operator is not fixed by the ghost number one string field
$\Psi$ that appears in the action. Moreover, as far as the action is concerned,
we could have written the kinetic operator as $\Q \Psi=Q\Psi+2\Psi\star\Psi_0$.
We have to choose the symmetric
form $\Q\Psi=Q\Psi+\Psi\star\Psi_0+\Psi_0\star\Psi$ in order to be able to write the
equation of motion around the solution as,
\begin{equation}
	\Q\Psi+\Psi\star\Psi=0\,.
\end{equation}
The dependence on the parity of the form $-(-)^A$ appearing in~(\ref{QQdef}) implies
the derivation property of the new kinetic operator
\begin{equation}
	\Q\Q A=0\quad\Longleftrightarrow\quad Q\Psi_0+\Psi_0\star\Psi_0=0\,.
\end{equation}
This implies that the gauge symmetry written in terms of $\Q$ takes exactly
the same form as when written in terms of the original $Q$
in~(\ref{infGauge}) and~(\ref{infGaugeinf}).

Sen's conjectures were addressed using several formalisms, such as
effective theory analysis~\cite{Sen:1999md,Garousi:2000tr,Bergshoeff:2000dq,Kluson:2000iy,Sen:2002an} and toy models such as p-adic string (field)
theory~\cite{Ghoshal:2000dd,Moeller:2002vx,Yang:2002nm,Moeller:2003gg}
(developed in~\cite{Freund:1987kt,Brekke:1988dg})\footnote{We refer the
reader to some of the reviews mentioned in the introduction for details
regarding developments using other formalisms.}.
However, as the conjectures involve a zero-momentum tachyon, which is quite
far from being on-shell, it should be clear that a field theoretical
approach is advantageous.
Other than the cubic string field theory, which is the main framework
described in this review, there exists also BSFT~\cite{Witten:1992qy,Witten:1992cr,Shatashvili:1993kk,Shatashvili:1993ps}\footnote{BSFT
stands for either ``background-independent string field theory'' or for
``boundary string field theory'', as it involves the evaluation of
correlation functions in a theory perturbed by the insertion of (open string)
operators at the boundary.}.
Analytical solutions within the framework of
BSFT~\cite{Gerasimov:2000zp,Kutasov:2000qp,Ghoshal:2000gt} have been derived
with much less effort than those of cubic string field theory.
The reason is that, in this formalism, one can consistently truncate all
the other fields and study the action of the tachyon field.
However, it is not quite clear how should the massive (non-renormalizable)
string modes be dealt within BSFT.
While the analysis of tachyon condensation seems to be consistent
within BSFT, it is desirable to have such a description also in the framework
of cubic string field theory, which is more adequate for the study of
other problems.

It turned out that within the framework of cubic string field theory the
problem is technically much harder, since the tachyon field does not
decouple. The main tool for studying the solution was the numerical approach
called ``level truncation'', to which we turn next.

\subsection{Level truncation}
\label{sec:LevelTrunc}

Level truncation was first introduced quite some time before
Sen's conjectures by Kostelecky and Samuel~\cite{Kostelecky:1990nt},
following a preliminary study~\cite{Kostelecky:1988ta}.
These authors tried to address the problem of dealing with an
infinite number of coefficient fields by truncating the string field
and the action to fields whose level is at most $l$.
Then, they assumed (without a formal proof\footnote{In fact,
in~\cite{Kostelecky:1988ta} they write
``Our truncation is {\it not} systematic''.
In~\cite{Kostelecky:1990nt}, on the other hand, the authors do give some
justification and write ``The approach is systematic''. The real
``proof'' of the power of this approach is ``experimental'':
it works very well.})
that by taking
the limit $l \rightarrow \infty$ the correct string field theory
results are obtained. They carried out their
analysis up to level four and indeed, they managed to identify
convergence to a solution
with energy lower than that of the trivial solution.
They realized that the solution is non-perturbative with respect
to the open string coupling constant $g_o$.
They even managed to notice that there is a decrease
in the amount of open string degrees of freedom around the
non-perturbative vacuum, although they did not have a reason to
suspect that none are left.

At that time, neither Sen's
conjectures nor D-branes~\cite{Polchinski:1995mt} were known.
Thus, while some papers did improve much the analysis of the
level-truncation results~\cite{Kostelecky:1995qk}, it was not until
Sen's conjectures that this tool (and string field theory in general)
gained popularity.

String field theory has a huge amount of
gauge freedom~(\ref{infGauge}). Level truncation is usually
performed with a fixed gauge. The Siegel gauge was almost exclusively
chosen. Gauge fixing achieves two goals. First,
the number of fields at any given level is reduced. The second
goal is related to the fact that level
truncation breaks the gauge symmetry, which
mixes fields at all levels. The gauge symmetry is
restored in the infinite limit level.
Gauge transformations induce flat directions
in the potential. At a fixed level these flat directions are not
exactly flat and flatness is achieved only asymptotically as the level
goes to infinity. These almost flat directions have many shallow
minima. Thus, without gauge fixing one finds many more solutions
than one would like to find, as many of them approximate solutions
that lie on the same gauge orbit. However, since gauge symmetry is
only approximate in level truncation, it is generally hard to
disentangle the pure-gauge solutions from the real ones.
Gauge fixing removes this degeneracy.

On the other hand, gauge fixing can also induce problems, since
gauge choices are generally not globally well defined. This goes
under the name of Gribov ambiguities in gauge theories.
The range of validity of the Siegel gauge was studied by Ellwood
and Taylor~\cite{Ellwood:2001ne}.
They found out that both the trivial and the non-perturbative
vacua remain in the range where the Siegel gauge is valid, as the
level is increased. Outside this region they found a complicated
pattern for the boundaries of Siegel gauge validity that
bifurcates as the level is increased.

One can make level truncation more efficient by using symmetries.
As mentioned by Sen in~\cite{Sen:1999xm},
a set of fields that does not include the tachyon
and that enter at least quadratically in the action, can be consistently
set to zero when looking for the tachyon vacuum. The equations of motion
of the fields that were set to zero are satisfied, as they are all at least
linear with respect to these fields. This simple observation can be used
in many ways to reduce the amount of fields to be considered.
First, one can set to zero all states that carry momentum,
since the condensation is with respect to the zero-momentum tachyon field
and the action carried no momentum. A related observation is that the tachyon
field is a Lorentz scalar as is also the action. At least two non-scalars
are needed in order to produce a scalar (the action). Thus, only scalars
should be kept in the search for the tachyon vacuum.
One can also set to zero all the fields with a non-unity ghost
number\footnote{While the classical action is a functional
only of a ghost-number-one string fields, the BV treatment of the gauge
symmetry introduces a string field of arbitrary ghost number (even before
fixing the gauge).
However, the tachyon field is a classical ghost-number-one field
and the fields with other ghost numbers indeed enter the action at least
quadratically. Thus, it is consistent to consider only ghost-number-one
string fields regardless of the question of gauge fixing.}.
Further, it is enough to consider only string fields that live in the
``universal subspace'' spanned by the ghost fields and the matter Virasoro
generators\footnote{The BRST charge $Q$ is composed of these
operators, so the quadratic term does not mix universal and non-universal
states. For the cubic term we note that it is defined as a correlator of
the three states acted upon by conformal maps. Conformal maps can be
defined by exponentials of (total) Virasoro generators, which do not
change the universality property and the expectation value of two
universal states with one non-universal state vanishes.}~\cite{Sen:1999xm}.
It was shown in~\cite{Rastelli:2000iu} that ghost-number one universal
states (named ${\cal H}_{univ}^{(1)}$)
are spanned by applying all possible combinations of
matter Virasoro operators and ghost Virasoro operators on the vacuum
$c_1\ket{0}$\footnote{A natural further restriction is to use only
total Virasoro generators. Acting on $c_1\ket{0}$ this gives a set of
states to which other states in ${\cal H}_{univ}^{(1)}$ do couple
linearly. Thus, it cannot be used for further restricting the tachyon
vacuum. However, the set of total Virasoro generators acting on the
ghost-number zero vacuum $\ket{0}$ forms an important subalgebra of the
star product, namely the algebra of surface states described above.}.
This provided a simple systematic enumeration of the
states relevant for tachyon condensation.
Other symmetries with respect to which the tachyon state and the action
are scalars include twist symmetry and the $SU(1,1)$
symmetry of~\cite{Zwiebach:2000vc}
(see also~\cite{Siegel:1985tw,Hata:2000bj}). The twist generator is
represented in ${\cal H}_{univ}^{(1)}$ as $(-1)^{L_0+1}$. Thus, the
reduction to twist even states corresponds to using only even levels.
The $SU(1,1)$ symmetry can be applied when one works in the Siegel
gauge. Then, it acts on the ghost sector with the generators
\begin{equation}
\label{SU11}
G=\sum_{n=1}^\infty (c_{-n}b_n-b_{-n}c_n)\,,\qquad
X=\sum_{n=1}^\infty n c_{-n}c_n\,,\qquad
Y=\sum_{n=1}^\infty \frac{1}{n} b_{-n}b_n\,.
\end{equation}
Here $G$ is the ghost number, while the other generators induce rotations
between the ghost and the anti-ghost modes.

Other methods used for restricting the solutions in level truncation are the
use of the residual BRST symmetry~\cite{Hata:2000bj} and the use of 
conservation laws~\cite{Rastelli:2000iu}. Conservation laws imply relations
among various coefficients of the string field~\cite{Schnabl:2000wt}.
More importantly, they allowed for the automatization of the calculations,
which enabled going to an extremely high precision.
Finally, in~\cite{Gaiotto:2002uk} it was found numerically that some
coefficients of the string field take universal values and these values
were evaluated analytically. It was suggested there that the universality
property follows from (an infinity of)
linear relations that the solution should obey
and an example of such a relation was derived.
Nonetheless, no full understanding of this universality property was achieved
and some problems with it were found in~\cite{Gaiotto:2002wy}.

\subsubsection{Level truncation: Example}
\label{sec:LevelTruncExample}

In the most common case of a flat 26-dimensional space, the string field
can be expanded as,
\begin{equation}
\begin{aligned}
\label{PsiOscLE}
\ket{\Psi}&=\int d^{26} k \Big(T(k)c_1 + \tilde T(k)c_0 c_1\\
&+ C(k) + \tilde C(k) c_0 + A_\mu(k)\al^\mu_{-1}c_1 +
  \tilde A_\mu(k)\al^\mu_{-1} c_0 c_1\\
&+ \beta (k) b_{-2} c_1+\gamma (k)c_{-2} c_1
 +\beta_\mu (k)\al^\mu_{-1}+\gamma_\mu (k)\al^\mu_{-1}c_{-1}c_1\\
 &+\cC(k) c_{-1}+ B_\mu (k)\al^\mu_{-2}c_1
  +B_{\mu\nu} (k)\al^\mu_{-1}\al^\nu_{-1}c_1+\tilde{(\cdots)}+\ldots
\Big)\!\ket{k}.
\end{aligned}
\end{equation}
Here, we took into account the fact that upon using the BV formalism the
string field is no longer restricted to ghost number one as
in~(\ref{PsiOsc}). 
The first line includes the level zero fields, the second line includes the
level one fields and the next level fields appear at the last two lines.
All fields ``appear twice'', with the $c_0$, where they carry a tilde and
without it (at level two the ``tilded'' fields appear collectively inside
brackets). The ``same'' fields with and without a tilde have opposite
statistics. Imposing the Siegel gauge allows one to discard the ``tilded''
fields, reducing the degrees of freedom by a half.

As explained above, the
fact that we are looking for the tachyon solution implies that one can
restrict the string fields back to ghost number one, excluding the fields
in the third line. Disregarding all fields other than the scalars drops the
$A_\mu$ field, the last remaining level one field (all odd level fields
drop), as well as the $B_\mu$ field and all components but the trace of the
$B_{\mu\nu}$ field. Hence, up to level two, the string field can be written
as,
\begin{equation}
\ket{\Psi}=\int d^{26} k \Big(T(k)c_1 +\cC(k) c_{-1}
 +B (k)\eta_{\mu\nu}\al^\mu_{-1}\al^\nu_{-1}c_1\Big)\!\ket{k}.
\end{equation}
We should still have to check that these fields are all $SU(1,1)$ singlets.
Direct inspection shows that they are all annihilated by the $G$, $X$ and
$Y$ operators of~(\ref{SU11}).

Setting the fields to be proportional to
$\delta(k)$ is the next step\footnote{This step is modified when lump
solutions are studied (see section~\ref{sec:LevelTrunc3} below).}. This
amounts to a Fourier transform to
space-time fields, which are kept constant. The volume of the space-time,
i.e., the volume of the D-brane, is kept finite.
The last thing to check is the reality
condition. Imposing it shows that the three component fields are real.

A shorter way to get to the form of the level truncated string field is to
use the universality, mentioned above, i.e., to write\footnote{The second
term could also be written as $\frac{\cC}{2}L_{-1}^g L_{-1}^g c_1\ket{0}$.
In this case explicit evaluation is simpler in terms of the ghost fields
themselves. Hence, we stick to the representation~(\ref{LevTruncExample}).},
\begin{equation}
\label{LevTruncExample}
\ket{\Psi}=(T+\cC c_{-1}b_{-1}+B L_{-2}^{m})c_1\ket{0}.
\end{equation}

To get the form of the tachyon potential one has to plug this expansion into
the action, divide by the volume and change the sign\footnote{Recall that for
static cases, the action density is minus the potential ($L=T-V$).}. 
Keeping the total level of the action not larger than four, i.e., keeping
all the kinetic terms, but only interaction terms that involve the field $T$
(this is denoted in the literature as level (2,4) truncation), the tachyon
potential reads~\cite{Rastelli:2000iu},
\begin{align}
V_{(2,4)}&=-\frac{1}{2}T^2-\frac{1}{2}\cC^2+\frac{13}{2}B^2\\
\nonumber
 &+K^3\Big(\frac{1}{3}T^3+\frac{11}{27}T^2 \cC-\frac{65}{27}T^2 B
  +\frac{19}{243}T \cC^2 +\frac{7553}{729}T B^2
  -\frac{1430}{729}T\cC B\Big).
\end{align}
We see here a general characteristic of the cubic terms. They include the
constant $K^3$, where we defined, 
\begin{align}
\label{K}
K=\frac{3\sqrt{3}}{4}\,.
\end{align} 
One could also consider (2,6) level truncation, in which case
some terms should be added to the potential,
\begin{equation}
V_{(2,6)}=V_{(2,4)}+K^3\Big(\frac{1}{243}\cC^3-\frac{272363}{19683}B^3
 -\frac{1235}{6561}\cC^2 B+\frac{83083}{19683}\cC B^2\Big).
\end{equation}

Let us demonstrate how these results are derived, starting with the
quadratic terms, which can be written as the following CFT expectation value,
\begin{equation}
V_2=-S_2=\frac{1}{2}\int \Psi Q\Psi=\frac{1}{2}\vev{(I\circ \Psi)(0)Q\Psi(0)}.
\end{equation}
The conformal transformation $I=-\frac{1}{z}$ is used to send the first
insertion to infinity, i.e., it is the BPZ conjugation.
It is instructive to note that this transformation
cannot generate a $c_0$ mode from any other mode. Thus, the use of the
Siegel gauge implies that neither $\Psi$ nor $I\circ \Psi$ carry this
mode, which is imperative for a non-zero result. The only object that
can supply this mode is the BRST charge
\begin{equation}
Q=c_0 L_0+\mbox{(terms with no $c_0$ mode)}\,.
\end{equation}
Thus, we can write,
\begin{equation}
V_2=\frac{1}{2}\vev{(I\circ \Psi)(0)(c_0 L_0\Psi(0))}.
\end{equation}
The BPZ transformation can be explicitly written in
terms of modes as,
\begin{equation}
\phi_n \rightarrow (-1)^{n+h}\phi_{-n}\,,
\end{equation}
where $\phi$ is any world-sheet field and $h$ is its dimension.
Specifically,
\begin{equation}
L^m_n \rightarrow (-1)^n L^m_{-n}\,,\qquad
b_n \rightarrow (-1)^n b_{-n}\,,\qquad
c_n \rightarrow -(-1)^n c_{-n}\,.
\end{equation}
Also, it is important to note that the BPZ transformation, while
inverting the ordering of the modes does not change
their formal Grassmann ordering.

It is now easy to see that there
can be no mixing of modes at the quadratic level and all that is
left is to evaluate the coefficients. The $L_0$ can be immediately eliminated
in favour of the conformal weight of the state, which is just the level minus
one. The rules above give for the coefficient of $\cC^2$,
\begin{equation}
\frac{1}{2}\bra{0} c_{-1} b_1 c_1 c_0 c_{-1} b_{-1}c_1\ket{0}=-\frac{1}{2}\,,
\end{equation}
while the coefficient of $B^2$ includes also a matter contribution that is
given in terms of the conformal anomaly of the matter Virasoro algebra,
\begin{equation}
\frac{1}{2}\bra{0} c_{-1} c_0 c_1 \ket{0}_g \bra{0} L_2^m L_{-2}^m \ket{0}_m=
\frac{1}{2}\cdot\frac{26}{12}(2^3-2)=\frac{13}{2}\,.
\end{equation}

Consider now the cubic terms, starting with the simplest one, the
$T^3$ term. Its coefficient is given in the CFT language by,
\begin{equation}
V_3^{TTT}=\frac{1}{3}\vev{(f_1\circ \Psi_T) (f_2\circ \Psi_T) (f_3\circ \Psi_T)}.
\end{equation}
The three conformal transformations $f_n$ are obtained by sending the upper half
to the unit disk using,
\begin{equation}
\label{UHPtoDisk}
w=\frac{1+i\xi}{1-i \xi}\,,
\end{equation}
then rescaling $w$ and relocating it to the three points of the
``rotated Mercedes-Benz logo'',
\begin{equation}
w\rightarrow e^{\frac{2\pi i n}{3}}w^{\frac{2}{3}}\,,
\end{equation}
and finally sending it back to the upper half plane using the inverse
of~(\ref{UHPtoDisk})\footnote{We could have decided to avoid the last
step and evaluate the expectation values in the unit disk coordinates.
In this case the maps are given by~(\ref{fDef}).},
\begin{equation}
\xi=i\frac{1-w}{1+w}\,.
\end{equation}
The composition of these three transformations for the three values of $n$
gives the conformal functions $f_n$.
The only relevant information about these $f_n$ is,
\begin{equation}
\begin{aligned}
\label{fns}
f_3(0)=& 0\,,\qquad f_1(0)=\sqrt{3}\,,\qquad f_2(0)=-\sqrt{3}\\
f'_3(0)=& \frac{2}{3}\,, \qquad f'_{1,2}(0)=\frac{8}{3}\,.
\end{aligned}
\end{equation}
We can now evaluate the CFT expectation value,
\begin{equation}
\vev{c(-\sqrt{3})c(0)c(\sqrt{3})}=2\cdot 3^{\frac{3}{2}}\,,
\end{equation}
and the conformal factor,
\begin{equation}
\Big(f_1'(0)\Big)^{-1} \Big(f_2'(0)\Big)^{-1} \Big(f_3'(0)\Big)^{-1}=
\Big(\frac{8}{3}\Big)^{-1} \Big(\frac{2}{3}\Big)^{-1} \Big(\frac{8}{3}\Big)^{-1}
=\frac{27}{128}\,.
\end{equation}
Assembling the ingredients, the final expression takes the form,
\begin{equation}
V_3^{TTT}=\frac{1}{3} \cdot 2\cdot 3^{\frac{3}{2}} \cdot \frac{27}{128}=\frac{K^3}{3}\,,
\end{equation}
as stated above.

In order to evaluate the other coefficients we use the conservation
laws for the ghosts and for the matter Virasoro operators.
Consider, for concreteness, the coefficient of $T\cC B$, which gets
contributions from six terms in the expansion. Cyclic permutations
can be used in order to trade the six terms for two terms with opposite
ordering multiplied by a symmetry factor of 3,
\begin{align}
V_3^{T\cC B}=\frac{1}{3}\cdot 3 &\cdot \bra{V_3}\big(c_1\ket{0}\big)^{(1)}\cdot\\
\nonumber
  &\cdot\Big[\big(c_{-1}\ket{0}\big)^{(2)} \big(L_{-2}^m c_1\ket{0}\big)^{(3)}+
	  \big(L_{-2}^m c_1\ket{0}\big)^{(2)} \big(c_{-1}\ket{0}\big)^{(3)} \Big].
\end{align}
Here, $\bra{V_3}$ is the three-vertex and the superscripts represent the three
spaces on which it acts. The conservation laws should now be used in order to
trade the $c_{-1}$ and $L_{-2}^m$ for other generators that would lead to the
expression that we evaluated before.
The $L_{-2}^m$ conservation law (after substituting $c=26$) reads,
\begin{equation}
\bra{V_3}\Big(\big(L_{-2}^m+26\cdot\frac{5}{54}+\ldots\big)^{(2)}+
  \big(\ldots\big)^{(1)}+\big(\ldots\big)^{(3)}\Big)=0\,.
\end{equation}
The entries that we omitted are matter Virasoro operators $L_n^m$ for
$n\geq 0$, in the various spaces, that do not contribute in the case at hand.
Thus, in our case, we only have to replace the $L_{-2}^m$ by $-\frac{65}{27}$.
The $c$-ghost conservation law is as simple,
\begin{equation}
\bra{V_3}\Big(\big(c_{-1}-\frac{11}{27}c_1+\ldots\big)^{(2)}+
  \big(\ldots\big)^{(1)}+\big(\ldots\big)^{(3)}\Big)=0\,.
\end{equation}
Here, the ellipses stand for $c_n$ with $n\geq 1$, in the two other spaces
and $n\geq 2$ in the space that we call $2$. These modes all drop out.
Hence, in our case, we only have to replace the $c_{-1}$ by $\frac{11}{27}c_1$.
The two orderings contribute the same and we are left with,
\begin{equation}
V_3^{T\cC B}=\frac{1}{3}\cdot 3\cdot 2\cdot\bra{V_3}\big(c_1\ket{0}\big)^{(1)}
  \big(\frac{11}{27}c_1\ket{0}\big)^{(2)} \big(-\frac{65}{27} c_1\ket{0}\big)^{(3)}
	=-\frac{1430}{729}\vev{c c c}
\end{equation}
The remaining vev is the tachyon vev that we evaluated before and substitution
leads to the quoted result. The evaluation of the other terms is similar.

\subsubsection{Sen's first conjecture}
\label{sec:LevelTrunc1}

The restrictions on the solutions and the automatization of the evaluation
enabled getting to high level in the level truncation scheme, which
resulted in extremely impressive results. The precision in the
evaluation of the D-brane tension got far beyond the
initial two percents of~\cite{Sen:1999nx} and even beyond the
$0.1\%$ of~\cite{Moeller:2000xv}.
On the other hand, it seemed that these accurate calculations introduced an
overshooting of the D-brane tension~\cite{GaiottoRastelli}
when the level in raised above $l=14$.
Taylor addressed this problem~\cite{Taylor:2002fy},
by evaluating the tachyon potential using the methods he developed earlier
in~\cite{Taylor:2002bq}, as a function of the level.
A technical problem in this approach is that
the radius of convergence of the terms in the expansion is finite, due
to a pole at a negative value of the tachyon field. This pole is related
to the boundary of validity of the Siegel gauge mentioned above
and is irrelevant to the
problem at hand, since the tachyon vacuum is located at a positive value
of the tachyon field. In order to overcome this technical problem,
the Pad\'e-approximation was used.
Then, the tension as a function of the level was numerically fit to
a polynomial in $l^{-1}$.
The results suggested that the
overshooting problem was merely an artifact.
The (common) expectation that the approach to the correct tension would be
monotonic with respect to the level was too naive.
In fact, the form of the approach graph is gauge dependent and a
non-monotonic approach is quite generic.
In the Siegel gauge it was found that after the solution overshots,
it starts to return to the correct value around $l\approx 26$.
Using a fit with the variable $l^{-1}$,
Gaiotto and Rastelli~\cite{Gaiotto:2002wy} managed to show
that their results also imply that around $l\approx 28$ the tension
returns towards the correct value.
Their final result as the level $l \rightarrow \infty$ gave
the desired result with an impressive accuracy of $\,3\cdot 10^{-3}\%\,$.
The analysis was based on level truncation up to $l=18$ with most of the
restrictions described above taken into account. This resulted in more
than 2000 fields to consider and over $10^{10}$ interaction terms.
If it was not for the restriction of the coefficient fields,
the amount of terms would have been larger by quite a few orders
of magnitude and the evaluation, even with the strongest computers, would
have been impossible to perform.

\subsubsection{Sen's third conjecture}
\label{sec:LevelTrunc2}

Sen's second conjecture states that the cohomology, at least at ghost
number one, of the kinetic operator around the non-perturbative vacuum
should vanish. As mentioned above, a hint in this direction appeared
already in~\cite{Kostelecky:1988ta},
although it was not understood at the time.
A preliminary examination of the kinetic term following Sen's work
was performed in~\cite{Hata:2001rd}. The analysis there
indicated that the kinetic term seems to vanish, at least
for a few low-level scalars. A more systematic study was
later performed by Ellwood and Taylor~\cite{Ellwood:2001py}.
They studied the exactness of the kinetic operator at ghost number one
in level truncation up to level six by calculating the projection
to the space of exact states of all scalar closed states.
There are several problems with such a numerical study.
First, when level truncated, the kinetic
operator does not square to zero. Still, the nilpotence of $\Q$ is
improved as the level is increased and so this should not be a problem
of principle. Second, in order to calculate the projection, an inner
product in the space of string fields should be introduced and there is
no canonical choice for such an inner product. This problem was handled
by considering several inner products and showing that the results do
not have a strong dependence on the choice. Another issue is that,
at a given level $l$, one cannot trust the analysis regarding states
whose mass squared is much higher than $l$. The authors considered
only the states below a cutoff, $m^2<l-1$.
It was found out that at level 6 the projection to the exact space
of an arbitrary closed state below the cutoff was of length of more
than $99\%$ of the state's norm squared,
regardless of the inner product chosen.
For a generic (that is not closed) state, the projection gave only
about $35\%$ of the original norm squared.

A more elegant approach for investigating the triviality of
the spectrum was given in~\cite{Ellwood:2001ig}.
While this paper also used level truncation,
its main tool is general and was used also for proving
Sen's conjecture using Schnabl's solution, as we describe in
section~\ref{sec:Sen3}. The idea is to find a ghost number $-1$
state $A$ obeying\footnote{An operator (in an operator algebra
with derivation and identity) with such a property
is called a contracting homotopy.}
\begin{equation}
\label{QA1}
\Q A=\ket{1},
\end{equation}
where $\ket{1}$ is the identity string field
obeying
\begin{equation}
\ket{1}\star\Psi=\Psi\star\ket{1}=\Psi\qquad \forall \Psi\,,
\end{equation}
and $\Q$ is the kinetic operator around the tachyon vacuum $\Psi_0$
as defined in~(\ref{QQdef}).
The existence of $A$ is equivalent to a strong version
of Sen's conjecture. According to Sen, the cohomology of
$\Q$ should vanish at ghost number one, while the existence of $A$
implies that it is zero for all ghost
numbers\footnote{In~\cite{Giusto:2003wc}
it was claimed that the other cohomology
groups are non-empty. This result contradicts the result
of~\cite{Ellwood:2001ig}. We return to this point in~\ref{sec:Sen3}.}.
The proof that the existence of $A$ implies the vanishing of the
cohomology is straightforward.
Suppose that $\Psi$ is closed, i.e., $\Q \Psi=0$. Then,
\begin{equation}
\Q (A\star \Psi)=(\Q A)\star\Psi - A\star (\Q\Psi)=
   \ket{1}\star\Psi-0=\Psi\,,
\end{equation}
and it follows that $\Psi$ is exact. Thus, the cohomology is empty.
Suppose now that the cohomology is empty. The identity string
field is closed, since
\begin{equation}
\Q\ket{1}=Q\ket{1}+\Psi\star\ket{1}-\ket{1}\star\Psi=0\,,
\end{equation}
where the first term vanishes trivially~(\ref{Q1=0}),
while the two other terms cancel each other.
Triviality of the cohomology implies that the identity is also exact,
that is, a state $A$ obeying~(\ref{QA1}) exists.

Since $A$ depends
on the solution $\Psi_0$, which was only known in level truncation,
the state $A$ was also found using level truncation. First, a
measure of being close to having the desired property is introduced
in the space of string fields at a given truncation level,
\begin{equation}
\ep=\frac{|\Q_l A_l-\ket{1}_l|}{|\ket{1}_l|}\,.
\end{equation}
Here $A_l$ is an arbitrary state at level $l$, the norm is
arbitrary, but following~\cite{Ellwood:2001py} it is assumed
that different norms will give similar results. Also, while
the norm of the state $\ket{1}$ diverges for many natural norms,
the norm of $\ket{1}_l$ is finite,
since the Hilbert space at any finite $l$ is finite dimensional.
Then, the minimum of $\ep$ is found at the level-$l$ space.
With this procedure it seems that $\ep\rightarrow 0$ as the level
is increased (it gets to about $2\%$ at $l=9$). However, an
examination of $A_l$ reveals that it does not converge.

The resolution of the above problem comes from noting that $A$ is
defined only up to a gauge transformation, since the transformation
\begin{equation}
A\rightarrow A+\Q B\,,
\end{equation}
leaves~(\ref{QA1}) intact.
This symmetry is broken by level truncation and the gauge orbit
is replaced by many isolated local minima. For every level, $A_l$
lands arbitrary at different isolated minima. This is the reason
for not having a well defined limit. Repeating the analysis for a
Siegel-gauge-fixed $A$ resulted in $\ep\rightarrow 0$
(about $3\%$ at $l=9$) as well as convergence of $A_l$ to a well
defined limit.

\subsubsection{Sen's second conjecture}
\label{sec:LevelTrunc3}

The examination of the conjecture concerning lump solutions
is more involved than the other two. The construction of
lumps involves the evaluation of non-universal terms\footnote{Lumps are
manifestly background dependent.} also in the cubic part of the potential.
This introduces not only space-time dependence but also
non-locality, due to the exponentiation of momenta in the interaction term.
Level truncation might become problematic in such a case, since now it
involves also a truncation of the exponential of momenta to a finite order in its
Taylor expansion and the infinite derivative theory might differ from the
limit of a finite derivative theory when the number of derivatives approaches
infinity~\cite{Moeller:2002vx}.
Still, the study of lump solutions using level
truncation produced appealing evidence in favour of Sen's
conjectures~\cite{Harvey:2000tv,deMelloKoch:2000ie,Moeller:2000jy},
although the agreement was much better for lumps of a small co-dimension
than for lumps with a large co-dimension. It was claimed that the exact
result is the same, only the approach as a function of the level is slower
for the large co-dimension lumps.

The simplest case is the truncation of the action to level zero, i.e.,
keeping only the tachyon field
\begin{equation}
\Psi=T(x) c_1\ket{0}\,,
\end{equation}
as we did for the non-interacting theory
(in $k$-space~(\ref{freeTachyonAction})).
The truncated string field action~(\ref{action}) reads,
\begin{equation}
S=-\int d^{26}x \Big(\,\frac{1}{2}\partial_\mu T(x) \partial^\mu T(x)
  -\frac{1}{2}T(x)^2+\frac{1}{3}K^3 \tilde T(x)^3\Big),
\end{equation}
where we defined,
\begin{align}
\label{tildT}
\tilde T(x)=& \; e^{\log K \partial_\mu \partial^\mu}T(x)\,.
\end{align}
Since it is complicated to evaluate the derivatives in the definition of
$\tilde T(x)$ exactly, one is led to expanding the exponent in~(\ref{tildT}).
Level truncation now turns into a double expansion in the
parameters\footnote{While $K$ is a fixed number~(\ref{K}), it is treated
as a free parameter for the purpose of the expansion.}
$\log K$ and $l$.

When the directions in which the lump is localized are compactified,
the double expansion becomes a single expansion again. This happens
because the compactified momenta are quantized. The oscillators contribute
to the level as before, while the momenta contribute to the level
a factor of $\al' k^2$.
From the uncompactified point of view such solutions represent an infinite
array of equally spaced lumps. Another drawback of this approach is that
the radius is fixed and the analysis should be repeated for different radii.
On the other hand the existence of a single expansion parameter simplifies
the analysis. The authors of~\cite{Moeller:2000jy} considered a co-dimension
one lump for simplicity (and because the approach to the correct result was
already known to be the fastest in this case).
A possible subtlety of their approach is the appearance of null-states
when the radius is rational in $\al'$ units.
The authors avoided this complication by using only non-rational values for
the compactification radius.
Their result for the lump tension agreed with expectation within less than
a percent at level 3. They also found a lump profile that does not change
much with changes of the radius. Another check for the identification of
their solution as a lump was the numerical study of the cohomology around the
solution.
The consistency checks of~\cite{Hata:2000bj} were performed for these lump
solutions in~\cite{Mukhopadhyay:2001pq} with reasonably good results.

\subsubsection{Sen's conjectures in the supersymmetric case}
\label{sec:LevelTruncSUSY}

Level truncation was also used in the supersymmetric case\footnote{For
conventions and definitions of the theories involved see
sections~\ref{sec:SUSYintro},~\ref{sec:SUSYtheories}.}.
The standard open superstring theory does not include a tachyon. However,
one can consider the theory that lives on a non-BPS D-brane or a D-\= D pair
with the appropriate conjectures as stated above.

The study of these conjectures using superstring field theory was
initiated by Berkovits~\cite{Berkovits:2000zj}.
In this paper he used his version of superstring field
theory~\cite{Berkovits:1995ab}, which he generalized in order to include
also the GSO($-$) sector living on the non-BPS D-brane, which supports
a tachyon. The action was truncated and only the tachyon field was kept.
Despite the non-polynomiality of the action there are only two terms that
contribute within this truncation. The resulting tachyon potential is quartic
and its two symmetrically located local minima correspond to about $60\%$
of the predicted value.
The results of this paper were extended in~\cite{Berkovits:2000hf}, where
it was shown how to work with non-BPS D-branes as well as with the
D-\= D system. It was also shown that expanding the action to an arbitrary
level gives rise to a finite number of terms, despite the non-polynomiality
of the action. The minimum of the tachyon potential at level $l=3/2$
was found to give about $85\%$ of the brane tension.
Adding one more level brought the number to about
$90\%$~\cite{DeSmet:2000dp,Iqbal:2000st}.

The tachyon potential in Witten's superstring field
theory~\cite{Witten:1986qs} was studied in~\cite{DeSmet:2000je}.
It was found that at least at the few lowest levels the potential is
unbounded from both sides and develops singularities before any local
minimum is obtained. Since there are also other problems with this theory
as described in section~\ref{sec:SUSYtheories}, the study
of the other superstring field theories seems to be more promising.

In the modified version of Witten's
theory~\cite{Preitschopf:1989fc,Arefeva:1989cp,Arefeva:1989cm}
the tachyon potential does not suffer from the above problems,
at least when the non-chiral choice of $Y_{-2}$ is being
made~\cite{Aref'eva:2000mb}. In this paper truncation to level
$1/2$ was shown to reproduce already $97.5\%$ of the tension.
Truncation to level 2 gives, in the Siegel gauge, the remarkable
result of $99.96\%$ of the tension~\cite{Ohmori:2003vq}.
This result is, however, gauge dependent and other possible
gauge choices give less impressive
results~\cite{Aref'eva:2000mb,Ohmori:2003vq}.

Lump solutions of superstring field theory were constructed
in~\cite{Berkovits:2000hf,Ohmori:2001sx}. As in the bosonic case,
one gets for these lumps good agreement with the conjectured result.
We are not aware of a work addressing Sen's third conjecture,
regarding cohomology, in the supersymmetric case.

\section{Algebraic primer}
\label{sec:primer}

One of the main hurdles in studying string field theory used to be
the technical difficulty of working with the string field algebra.
This algebra is greatly simplified by working in an appropriate
coordinate system and considering the relevant basis of states.

\begin{figure}[tbh]
\begin{center}
\input{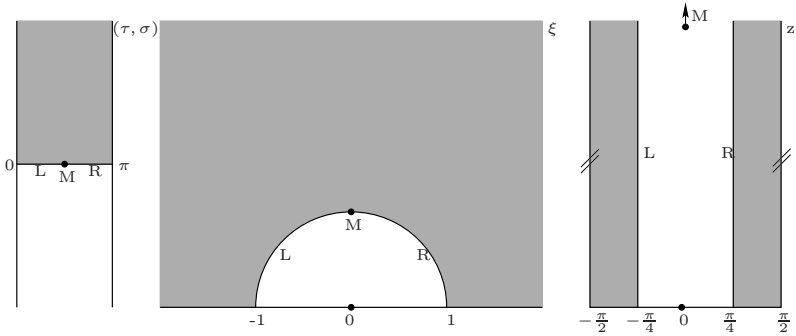}
\end{center}
\caption{Different coordinate systems. In each coordinate system
the Right, Left and Middle of the string are marked.
The last coordinate system is a cylinder with circumference
$\pi$ (the two marked lines should be identified) and the
mid-point is at $i\infty$. The grayed out region is the
coordinate patch where no operator insertions are allowed.
}
\label{fig:coordinates}
\end{figure}
Figure~\ref{fig:coordinates} shows three possible coordinate
systems\footnote{Note that in the literature one finds that
left-right is defined either with respect to infinity as was
done by Schnabl~\cite{Schnabl:2005gv} or with respect to
the origin as we do here, following Okawa~\cite{Okawa:2006vm}.}.
The ($\tau,\sigma$) system gives an intuitive view of a string,
where $\sigma\in[0,\pi]$ is its width and $\tau\in(-\infty,\infty)$
represents the world-sheet time.
The upper half plane, related to the previous one by
$\xi=-\exp(\tau-i\sigma)$,
is the canonical coordinate system for calculating CFT correlators.
The semi-infinite cylinder $C_\pi$ defined by $z=\arctan(\xi)$
turns out to be the most convenient coordinate system for string
field theory, as the star product is greatly simplified there.

We also need to transform the operators that we work with to
the new coordinate system
\begin{equation}
\tilde\cO(z) = \tan\circ\cO(\xi)\,.
\end{equation}
For primary fields such conformal transformations have
the simple form
\begin{equation}
\tilde\cO^h(z) = f\circ\cO^h(\xi) = f'(z)^h\cO^h(f(z))\,,
\end{equation}
where in our case $f(z)=\tan(z)=\xi$.
We are also interested in the transformation of the mode
expansion coefficients
\begin{equation}
\cO^h(\xi) = \sum\cO^h_n\xi^{-n-h}\,,
\end{equation}
which are not primary fields.
Still, computing their transformation is straightforward
\begin{align}
\tilde\cO^h_n = f\circ\cO^h_n
   &= \oint\frac{dz}{2\pi i} z^{n+h-1}\tilde\cO^h(z)
    = \oint\frac{dz}{2\pi i} z^{n+h-1}f'(z)^h\cO^h(f(z)) \nonumber\\
   &= \sum_{m=-\infty}^\infty \cO^h_m
	\oint\frac{dz}{2\pi i} z^{n+h-1}f'(z)^h f(z)^{-m-h}\,.
\end{align}
It is also useful to make a change of variables inside the integral
\begin{equation}
\label{xiTransform}
\tilde\cO^h_n = \sum_{m=-\infty}^\infty \cO^h_m
	\oint\frac{d\xi}{2\pi i} \big(f^{-1}(\xi)\big)^{n+h-1}
	   \big(f'(z(\xi))\big)^{h-1} \xi^{-m-h}\,.
\end{equation}
Usually one looks at transformations that are regular at the origin,
then using $SL(2)$ transformations one can get to the
canonical form
\begin{equation}
\label{fInit}
f(0)=0\,, \qquad f'(0)=1\,.
\end{equation}

We use the above relations to calculate some modes of the energy
momentum tensor
\begin{align}
\label{LLm1}
\LL_{-1} &= \sum L_m \oint\frac{d\xi}{2\pi i} (1+\xi^2)\xi^{-m-2} = L_{-1}+L_1
    \,\\
\label{LL0}
\LL_{0} &= \sum L_m \oint\frac{d\xi}{2\pi i} \arctan(\xi)(1+\xi^2)\xi^{-m-2}
    = L_0 -2\sum_{n=1}^\infty \frac{(-1)^n}{4n^2-1}L_{2n}\,.
\end{align}
These transformed Virasoro generators obey the usual Virasoro algebra
\begin{equation}
[\LL_n,\LL_m]=(n-m)\LL_{n+m}\,.
\end{equation}
This is so because the Virasoro algebra comes out from the OPE relations
near the origin where $\tan(\xi)\sim\xi$.

BPZ conjugation on the other hand is not of the form~(\ref{fInit}).
BPZ conjugation amounts to transforming the world-sheet
$(\tau,\sigma)\rightarrow(-\tau,-\sigma)$. Therefore, it can be seen
as mapping an incoming string into an outgoing string.
On the upper half plane the BPZ transformation is
\begin{equation}
I(\xi)=-\frac{1}{\xi}\,.
\end{equation}
Using~(\ref{xiTransform}) we get the transformation rule
\begin{equation}
\label{BPZgen}
{\cO^h_n}^\flat \equiv I\circ\cO^h_n = \sum_m\cO^h_m\oint\frac{d\xi}{2\pi i}
    (-1)^{h+n}\xi^{-n-m-1} = (-1)^{h+n}\cO^h_{-n}\,.
\end{equation}
Here, an extra minus sign came from the change of orientation of the
integration contour from a contour around $\xi=\infty$ to a contour
around $\xi=0$.
For the energy momentum tensor its reality implies that the hermitian
conjugation of its modes obeys
\begin{equation}
\label{Ldag}
L^\dag_n=L_{-n}\,.
\end{equation}
From~(\ref{BPZgen}) and~(\ref{Ldag}) we read the relation between BPZ and hermitian
conjugations for the Virasoro modes,
\begin{equation}
\label{BPZHerm}
L^\dag_n=(-1)^n L_n^\flat\,.
\end{equation}

On the cylinder the BPZ transformation becomes
\begin{equation}
I(z) = \arctan(-\frac{1}{\tan{z}}) = z+\frac{\pi}{2}\,.
\end{equation}
The simple form of the BPZ transformation on the cylinder does not
translate to a simple relation between modes and their BPZ conjugates.
The problem with applying~(\ref{xiTransform}) in this case is that
to deform the contour from $z=\pi/2$ to $z=0$ we need to go through
the cut. Hence, the map between modes
and their BPZ conjugation on the cylinder is generally quite
cumbersome.

There are two equivalent ways to expand the BPZ conjugate operators.
One is to apply the BPZ conjugation on the UHP operators
in~(\ref{xiTransform}).
The other option is to use the composed transformation $f(I(\xi))$
directly. Specifically, for the Virasoro generators on the cylinder
we get
\begin{equation}
\label{Ldagger}
\LL^\flat_n = -\oint\frac{d\xi}{2\pi i}(1+\xi^2)
       \big(-\arccot(\xi)\big)^{n+1}T(\xi)\,.
\end{equation}
Here, an extra minus sign comes from orientation change as in~(\ref{BPZgen}).
Hermitian conjugate Virasoro operators transform as,
\begin{equation}
\label{LHerm}
\LL^\dag_n = \oint\frac{d\xi}{2\pi i}(1+\xi^2)
       \big(\arccot(\xi)\big)^{n+1}T(\xi)\,.
\end{equation}
Comparing~(\ref{Ldagger}) and~(\ref{LHerm}), we see that the Hermitian and
BPZ conjugation obey~(\ref{BPZHerm}) also in this
coordinate system. In particular,
\begin{equation}
\LL_0^\dag=\LL_0^\flat\,.
\end{equation}
This can also be seen by noticing that the expansion of $\LL_0$ in terms
of the usual generators includes only even modes~(\ref{LL0}).

It is easy to see that
\begin{equation}
\LL_{-1}^\dagger = \LL_{-1} = K_1\,,
\end{equation}
where $K_1$ is one of the mid-point preserving reparametrization
operators~\cite{Witten:1986qs},
\begin{equation}
\label{Kn}
K_n=L_n-(-1)^n L_{-n}\,.
\end{equation}
This operator plays a dominant role in string field theory.
Another important combination comes from the commutation relation
\begin{equation}
\label{SL2Algebra}
\left[\LL_0,\LL_0^\dagger\right] =
    \oint\frac{d\xi}{2\pi i}(1+\xi^2)(\arctan \xi+\arccot \xi)T(\xi) =
    \LL_0+\LL_0^\dagger\,.
\end{equation}
The integrand can be written using a step function,
\begin{equation}
\arctan \xi+\arccot \xi = \frac{\pi}{2}\epsilon(\Re(\xi)) = \frac{\pi}{2}
\left\{\begin{array}{r}-1\quad\Re(\xi)<0 \\ 1\quad\Re
 (\xi)>0\end{array}\right.\,.
\end{equation}
The calculation of the integral is a bit subtle, since $\arccot(\xi)$ has
a branch cut in the range $\xi=(-i,i)$, while $\arctan(\xi)$ has a branch
cut on the rest of the imaginary axis.
Therefore, the contour can cross the imaginary axis only at the points
$-i,i$. 
This is not evident when writing the integrand as a step function.
The step function does suggest that we should split the contour
of integration to its left ($C_L$) and right ($C_R$) parts.
For each part we have the integrand of $K_1$, it is therefore
natural to name the two parts $K_1^L$ and $K_1^R$.
This gives the relation
\begin{equation}
\label{L0L0gadK1}
\hat\LL_0 \equiv \LL_0+\LL_0^\dagger = \frac{\pi}{2}(-K_1^L+K_1^R)\,.
\end{equation}
$\hat\LL_0$ has an interesting property of increasing the
$\LL_0$-level of states.
This can be seen from the commutation relation
\begin{equation}
[\LL_0, \hat\LL_0] = \hat\LL_0\,.
\end{equation}
For the standard Virasoro operators, only $L_{-1}$ has such a
behaviour with respect to $L_0$.
For $\LL_0$ we have two such states, $\hat\LL_0$ and $\LL_{-1}$
(or linear combination of these like $K_1^L, K_1^R$).

\begin{figure}[tbh]
\begin{center}
\input{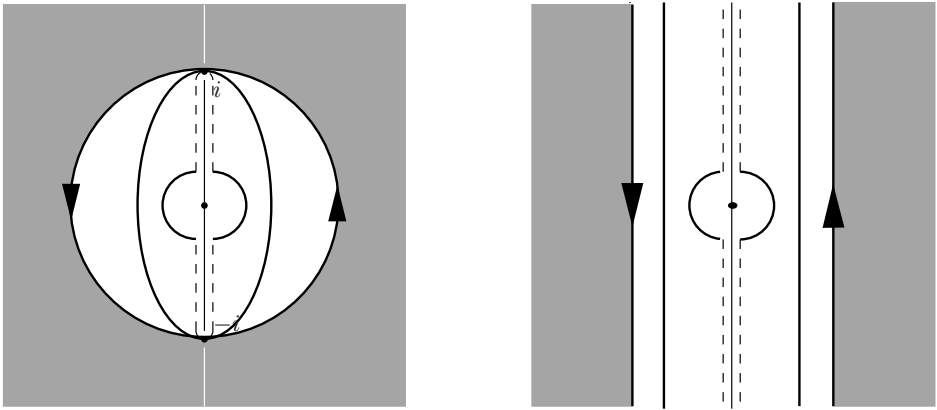}
\end{center}
\caption{Possible choices for the contour in~(\ref{SL2Algebra}).
The natural choice in the upper half plane is the unit circle,
which transforms to two infinite straight lines at $\pm\pi/4$
in the cylinder coordinates.
We can squeeze the circle in the upper half plane in a way that
keeps the lines in the cylinder straight.
More generally, we can deform the contour as we like, as long as
we do not move the points $\pm i$, which are fixed by the existence
of cuts in the rest of the imaginary axis. In the cylinder
coordinates, this translates to arbitrary curves connecting
$\pm i \infty$.}
\label{fig:StepFunction}
\end{figure}

Notice that the definition of $K_1^L$ and $K_1^R$ is restricted
by the cut structure.
This is illustrated in figure~\ref{fig:StepFunction}.
These operators have ``semi-derivation'' properties with respect
to the star product
\begin{align}
K_1^L(\Psi_1 \Psi_2) = (K_1^L\Psi_1) \Psi_2\,,\\
K_1^R(\Psi_1 \Psi_2) = \Psi_1 (K_1^R\Psi_2)\,.
\end{align}
Similar relations hold for the $b(z)$ ghost
\begin{align}
\label{Bhat}
\hat\BB_0 \equiv \BB_0+\BB_0^\dagger &= \frac{\pi}{2}(-B_1^L+B_1^R)\,,\\
\label{Bleft}
B_1^L(\Psi_1 \Psi_2) &= (B_1^L\Psi_1)\Psi_2\,,\\
\label{Bright}
B_1^R(\Psi_1\Psi_2) &= (-1)^{\Psi_1}\Psi_1(B_1^R\Psi_2)\,.
\end{align}

The commutation relation~(\ref{SL2Algebra}) defines an $SL(2)$ algebra.
This relation only holds if the underlying CFT has vanishing central
charge. Turning on a central charge in the CFT will add an infinite
central charge to~(\ref{SL2Algebra}).
A way to regularize this central charge was given in~\cite{Fuchs:2006an}.

The advantage of the cylinder coordinates is that it allows for
a simple geometrical interpretation of the star product.
The most basic example, is the multiplication of two vacuum states.
In the upper half plane, one has to cut out the coordinate
patch from the two states (see figure~\ref{fig:coordinates}).
Then, the left edge of one of the remaining half unit circles
needs to be sewed to the right edge of the other.
Finally, the coordinate patch will have to be reintroduced,
leading to a non-trivial world-sheet.
In the cylinder coordinate things are much simpler.
Cutting the coordinate patch leaves us with two half-infinite strips
(see again figure~\ref{fig:coordinates}).
Sewing the two half-infinite strips and the coordinate patch, which is
also a half-infinite strip gives a cylinder with circumference
$\frac{3\pi}{2}$.

\begin{figure}
\center{\includegraphics[width=10cm]{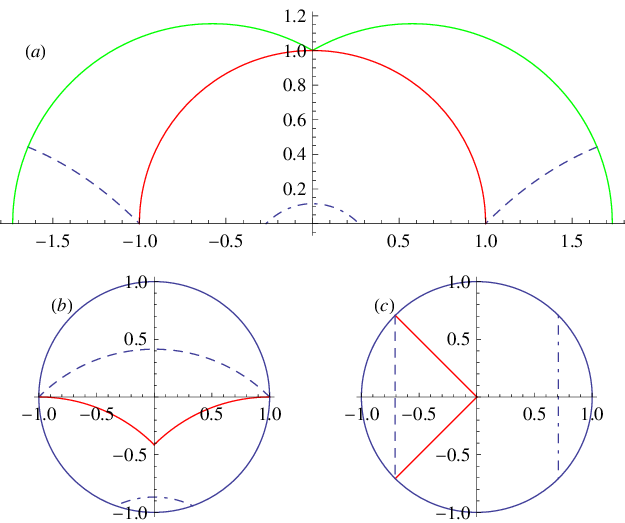}}
\caption{Three pictures for wedge states.
In $(c)$ we have one of the canonical representations of a wedge state.
The local coordinate patch (of the test state) is the part of the disk
to the right of the red curve. The local coordinate patch is always
``in the size of'' the vacuum.
Hence, we see that the wedge state here is $W_{\frac{1}{3}}$, since it is
one third the size of the vacuum $W_1$. Alternatively, we could have
interpreted the left part as the local coordinate patch. Then we would have
concluded that the wedge state is $W_3$. In this representation we can see
the two limits of wedge states, $W_0$ -- the identity string field, whose
local coordinate patch fills all of the disk, except for a cut on the
negative part of the real axis and $W_\infty$ -- the sliver, for which the
local coordinate patch takes the form of an infinitesimal sliver.
The representation $(b)$ is the analogue of $(b)$ of
figure~\ref{fig:SurState}. Here, the local coordinate patch is the part of the
disk below the red curve. Note that the two parts of the curve still meet at
an angle of $\frac{3\pi}{2}$, since the transformation is conformal -- it is
a M\"obius transformation. To get to the representation $(a)$ of
figure~\ref{fig:SurState}, one should first send the local coordinate patch to
half a disk, using $z\rightarrow z^{\frac{2}{3}}$. Then, a M\"obius
transformation gives us~$(a)$. The $z^{\frac{2}{3}}$ transformation is
singular and hence non-conformal at $z=0$. This is what enables
``flattening'' the angle between the two parts of the curve. On the other
hand, the transformation is also singular on the negative real axis, where it
introduces a cut, at the end of which a conical singularity appears. To
``undo'' the cut, the two green curves in~$(a)$ should be identified.
Had we tried to represent a $W_{>1}$ state, we would have had in addition to
an identification of curves in the plane also a multi-cover of it.
In all three representations we drew two curves, a dashed one outside the
local coordinate patch and a dot-dashed one inside it. These two curves are
straight lines in the representation $(c)$, but take different forms in
the other two. In particular, in $(a)$, the curve is ``split'' to two halves,
which are glued on the green lines.
The most useful representation of wedge states in the recent developments
proved to be none of these ones, but the cylinder coordinates.
More on the various geometrical representations of wedge states can be found,
e.g., in section 3.3 of~\cite{Rastelli:2001vb}. 
}
\label{fig:WedgeReps}
\end{figure}

Regardless of the choice of a coordinate system, the result of multiplying
two vacuum states is the wedge state $\ket{3}$. Wedge
states~\cite{Rastelli:2000iu,Furuuchi:2001df,Rastelli:2001vb,Schnabl:2002gg}
are a set of surface states satisfying the Abelian algebra
\begin{equation}
\ket{r}\star\ket{s} = \ket{r+s-1} \qquad r,s \ge 1 ,
\end{equation}
where the wedge state $\ket{2}$ is the vacuum state
$\ket{0}$\footnote{The equality $\ket{0}=\ket{2}$ may be a source for
confusion. This should not be the case, since the wedge state $\ket{r=0}$
does not exist.}.
The state $\ket{3}$ is the result of multiplying two vacuum states, as
discussed in the previous paragraph, and $\ket{1}$ is the identity
string field.
In the cylinder coordinates\footnote{Strictly speaking there are many
``cylinder coordinates'', parametrized by the circumference (the distance
along the $x$-axis between the two lines that are identified) of the cylinder.
Thus, each wedge state is naturally defined in a different coordinate system
and all those coordinate systems are collectively referred to as ``cylinder
coordinates''. A wedge state can be defined in the ``standard'' (that is
the one that is natural for the vacuum $\ket{0}$) cylinder coordinates
using an operator insertion, which corresponds to rescaling (see below).},
a wedge state $\ket{r}$ corresponds to
a cylinder with circumference $\frac{r \pi}{2}$, of which the coordinate patch
of the test state occupies a (canonical) strip of width $\frac{\pi}{2}$.
Note, that while (integer) powers of $\ket{0}$ generate the surface states
$\ket{r}$ with $r\in \NN$ (the formal definition $\ket{2}^0=\ket{1}$ is
consistent with all the structures introduced so far), there are no obstructions
that prevent us from similarly defining them for any $1\leq r\in \RR$.
A different convention used for the wedge states is
\begin{equation}
\label{SSalg}
W_\alpha\equiv\ket{\alpha+1} \quad\Rightarrow\quad
W_\alpha W_\beta = W_{\alpha+\beta}\,.
\end{equation}
We shall be using both conventions interchangeably.
There are several geometrical representations of the wedge states. We
present some of them in figure~\ref{fig:WedgeReps}.

\begin{figure}[tbh]
\begin{center}
\input{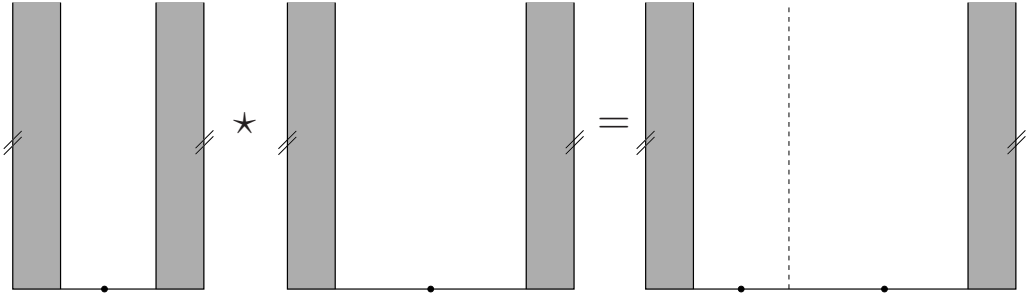}
\end{center}
\caption{Multiplication of two wedge states results in a wedge state
with the combined width of the two original states. The local coordinate
patches are removed and the glued surface gets a new local coordinate patch.
Operator insertions are mapped accordingly.
One can have many operator insertions and these can be anywhere on
the coordinate patch. We mark one such insertion for each of the surface
states. The resulting surface state has two insertions located at
non-generic points. One can use the OPE in order to replace these two
insertions by a single, symmetrically located, insertion.}
\label{fig:WedgeProduct}
\end{figure}

While wedge states form a very simple subalgebra,
they are of limited
use by themselves. Specifically, they cannot describe the solution
to the equation of motion since they have ghost number zero.
To overcome this limitation, we consider wedge states with insertions.
The product of two such states takes in the cylinder coordinate the form,
\begin{align}
\label{OperatorMultiply}
\cO(z^r_1)\ket{r}\star
\cO(z^s_1)\ket{s} =
\cO(z^r_1+\frac{\pi(s-1)}{4})
\cO(z^s_1-\frac{\pi(r-1)}{4})
\ket{r+s-1} .
\end{align}
This wedge state multiplication is illustrated in
figure~\ref{fig:WedgeProduct}.

Using wedge states with insertions, we can describe any string field
we want. Actually, since we are allowing an arbitrary insertion,
we have more than one description per field.
The simplest example to this redundancy is given by representing the
wedge states themselves as\footnote{This representation is in accord
with the fact that they are surface states, in light of~(\ref{surfS}).}
\begin{equation}
\ket{n} = U_n^\dag\ket{0} = \Big(\frac{2}{n}\Big)^{\LL^\dag_0}\ket{0} ,
\end{equation}
where $\LL^\dag_0$ is the hermitian conjugate of $\LL_0$ as defined
in~(\ref{LHerm}).
This expression can be understood from looking at
the inner product of the wedge state with some test state
\begin{equation}
\bra{0}\cO_h(z)\ket{n} =
\bra{0}\cO_h(z)U_n^\dag\ket{0} =
\Big(\frac{2}{n}\Big)^h\bra{0}\cO_h\big(\frac{2z}{n}\big)\ket{0} ,
\end{equation}
where we used the fact that $\LL_0^\dag$ is the dilatation operator for
bra states and assumed in the last step that $\cO_h$ is a primary field
of conformal weight $h$.
This algebraic result matches the geometrical picture of starting
from an operator inserted on a cylinder of width $\frac{n \pi}{2}$ and
scaling it to the canonical cylinder of width $\pi$.
Equivalently, wedges with insertions are written as
\begin{equation}
\cO(z_n)\ket{n} = U_n^\dagger U_n^\nodag \cO(z_n)\ket{0} .
\end{equation}

We can use the fact that $\LL_0$ and $\LL_0^\dag$ form an $SL(2)$
algebra~(\ref{SL2Algebra}) to get the following identities using basic
group theory techniques
(such as using an explicit two-dimensional representation)
\begin{align}
U_r \LL_0^\dag U_r^{-1} &= \frac{2-r}{r}\LL_0 + \frac{2}{r}\LL_0^\dag\,,\\
U_r^{-1} \LL_0^\dag U_r &= \frac{r-2}{2}\LL_0 + \frac{r}{2}\LL_0^\dag\,,\\
U_r^\dag \LL_0 U_r^{\dag-1} &= \frac{r}{2}\LL_0 +
 \frac{r-2}{r}\LL_0^\dag\,,\\
U_r^{\dag-1} \LL_0 U_r^\dag &= \frac{2}{r}\LL_0 + \frac{2-r}{2}\LL_0^\dag\,,
\end{align}
and also
\begin{align}
U_r U_s &= U_{\frac{rs}{2}}\,,\\
U_r U_s^\dag &= U_{2+\frac{2}{r}(s-2)}^\dag
U_{2+\frac{2}{s}(r-2)}^\nodag\,,\\
\label{eL0L0UdagU}
e^{\beta(\LL_0+\LL_0^\dag)} &= U_{2-2\beta}^\dag U_{2-2\beta}^\nodag\,.
\end{align}

The building blocks that we will be working with are wedge states with
local operator insertions as in~(\ref{OperatorMultiply}) and
line integrals of the $T(z)$ and $b(z)$ fields,
\begin{align}
K \equiv K_1^L =-\int_{-i\infty}^{i\infty} T(z)\, \frac{dz}{2\pi i} \,,\qquad
B \equiv B_1^L =-\int_{-i\infty}^{i\infty} b(z)\, \frac{dz}{2\pi i} \,.
\end{align}
These line integral can be freely moved as long as the end-points are
fixed and no other operator insertions are crossed, as explained in
figure~\ref{fig:StepFunction}.

The $K$ operator can be used to generate wedge states from the
identity state\footnote{This follows from the easily derived commutation
relation $[K_1,\cL_0+\cL_0^\dag]=0$, together with the
relations~(\ref{L0L0gadK1}) and~(\ref{eL0L0UdagU}) and the fact that
$K_1$~(\ref{Kn}) annihilates the ($SL(2)$ invariant)
vacuum~\cite{Schnabl:2002gg,Schnabl:2005gv}.},
\begin{equation}
\label{KWedge}
e^{\frac{\pi}{2}(n-1)K}\ket{1} = \ket{n} .
\end{equation}
From this relation we learn that
\begin{equation}
\label{KdevWedge}
\frac{\pi}{2} K W_n=\partial_n W_n\,.
\end{equation}
This implies that $K$ acts as a derivation with respect to strip length,
since the factor of $\frac{\pi}{2}$ is the ratio between the length
of the wedge and its number.

\newpage
\section{Schnabl's solution}
\label{sec:Schnabl}

It took almost twenty years from the time Witten introduced the equation
of motion of string field theory~\cite{Witten:1986cc} until Schnabl
found its analytical solution~\cite{Schnabl:2005gv}.
Prior candidate solutions were either numerical, singular in some sense
or too abstract.
The litmus test for a solution is its ability to reproduce the $D$-brane
tension according to Sen's conjecture. There were some suggestions
for analytical solutions before Schnabl, but none of them could pass
this test.
Some candidate solutions turned out to have zero action, and
were reinterpreted as pure-gauge solutions.
                                                                                
Still, the simplest form of Schnabl's solution is as a formal
pure-gauge solution~\cite{Okawa:2006vm}.
We will see that the gauge transformation is
singular, and that the solution is physical.
From the form of the finite gauge transformation~(\ref{finGauge}),
we see that a pure-gauge solution can be written as,
\begin{equation}
\label{PureGauge}
\Psi=e^{-\La}Qe^\La\,.
\end{equation}
Reparametrizing $\La$ we can rewrite the above as
\begin{equation}
\label{GaugeSol}
\Psi = \Gamma^{-1}(\Lambda)Q\Gamma(\Lambda)\,.
\end{equation}
The $\La$ in~(\ref{GaugeSol}) differs from the $\La$ in~(\ref{PureGauge}) and
$\Gamma(\Lambda)$ is some known function of the form
\begin{equation}
\Gamma(\Lambda) = 1 + \Lambda + \cO(\Lambda^2)\,.
\end{equation}
The choice of $\Gamma$ dictates the relation between the gauge string fields
that appear in~(\ref{PureGauge}) and~(\ref{GaugeSol}).
We refer to the choice of $\Gamma$ as a scheme choice.
While the scheme choice has no physical significance, it might lead to
simplified expressions for $\La$, as well as to a simplifications or
complications of the equation of motion and of the reality
condition. Hence, in practice, it is worthwhile to examine various schemes.

Schnabl's solution is most easily represented in the
left scheme\footnote{The left scheme is the scheme in which the $Q\La$ term
in $\Psi$ is always to the left of $\La^n$, as is explicit
in~(\ref{PsinLeft}) below.},
\begin{equation}
\Gamma(\La)=\frac{1}{1-\la\La}\,,
\end{equation}
where the gauge string field $\La$ takes the simple form\footnote{An
interesting property of $\La$ is that its $\LL_0$ eigenvalue
vanishes.}~\cite{Okawa:2006vm},
\begin{equation}
\label{LambdaSolution}
\Lambda = B c(0)\ket{0},
\end{equation}
and $\la$ is a parameter. It turns out that the solution is
a pure-gauge one for $|\la|<1$, while for $\la=1$, it is the
desired tachyon vacuum. The solution does not converge
for other (real) values of
$\la$~\cite{Schnabl:2005gv}\footnote{This picture of a family
of pure-gauge solutions for $\la<1$ that turn into a physical solution
for $\la=1$ and are not defined elsewhere,
is supported by a numerical study,
in which it was found that the energy density approaches a step
function as a function of $\la$ as the truncation level is
increased~\cite{Takahashi:2007du}.}. The solution itself takes then the form,
\begin{equation}
\label{ScPsi}
\Psi = (1-\la\La)Q\frac{1}{1-\la\La}=Q\La\frac{\la}{1-\la\La}\,,
\end{equation}
where in the last equality we integrated by parts\footnote{Recall that the
identity string field is annihilated by $Q$~(\ref{Q1=0}).}.
It is useful to expand $\Psi$ in powers of $\la$,
\begin{equation}
\label{PsiExpand}
\Psi=\sum_{n=1}^\infty \lambda^n \Psi_n\,.
\end{equation}
This gives,
\begin{equation}
\label{PsinLeft}
\Psi_n = (Q\Lambda)\Lambda^{n-1}\,.
\end{equation}

Written as in~(\ref{ScPsi}), $\Psi$ naturally satisfies the equation
of motion.
Unfortunately, it seems that the action vanishes.
This can be seen from
\begin{equation}
\label{ZeroAction}
\partial_\lambda S(\Psi) = \vev{\partial_\lambda\Psi
    (Q\Psi+\Psi^2)},
\end{equation}
which vanishes upon using the equation of motion. Since the action is clearly
zero for $\lambda=0$, it remains zero for any value of $\lambda$.
This is not surprising, since
the above is equivalent to the variational method by which one
gets the equation of motion in the first place.
                                                                                
It turns out that the argument for the vanishing of the action fails,
since the action has a finite radius of convergence in $\lambda$.
One clue for this singularity comes from looking at the gauge condition.
Schnabl's solution does not obey the Siegel gauge~(\ref{SiegelGauge}),
instead it obeys
\begin{equation}
\BB_0\Psi = 0\,.
\end{equation}
This is the Schnabl gauge condition.
Despite the fact that $\Psi$ obeys the gauge condition,
it is (by construction) a pure-gauge state.
This shows that Schnabl's gauge does not fix the gauge completely.
This is a bit strange, in light of the similarity between the
Siegel and the Schnabl gauge choices.
It may suggest that the pure-gauge states obeying the condition
are of a somewhat singular nature.
Specifically, $\Psi_1$ is an exact state obeying
the gauge condition,
\begin{equation}
\BB_0\Psi_1 = \BB_0 Q\Lambda = (\LL_0-Q\BB_0)\Lambda = 0\,.
\end{equation}
It is instructive to try to transform $\Psi_1$ to the Siegel gauge,
using the one-parameter family of transformations~\cite{Fuchs:2007gw}
(see also~\cite{Kiermaier:2007jg,Kiermaier:2008jy}),
\begin{equation}
\BB_0^s = U_s \BB_0 U_s^{-1}\,,\qquad \La_s=U_s \La\,,\qquad
U_s = s^{-L_0}\,,
\end{equation}
where $s=1$ corresponds to Schnabl's gauge and $s=0$ gives the Siegel
gauge.
The limit $s\rightarrow 0$ is well defined when $U_s$ acts on physical
states, but $\La_s$ is singular in the limit.

Next, we want to simplify the expanded expression for $\Psi$.
We start with
\begin{equation}
\Lambda^n = B c_1\ket{0} \star B c_1\ket{0} \star \cdots \star B c_1\ket{0}.
\end{equation}
This can be viewed as a cylinder of width $\frac{(n+1)\pi}{2}$ with
alternating $B$ line integrals and $c$ insertions.
As mentioned above, the $B$ integral can be freely moved over
the cylinder as long as the end points are fixed at $\pm i \infty$
and it does not cross any $c$ operators.
We can use the (anti)-commutation relation\footnote{Here and elsewhere
in the paper, the brackets represent the graded commutator, i.e., it
is the anti-commutator for two odd objects and the commutator otherwise.}
\begin{equation}
\label{Bc1}
[B,c_1] = 1\,,
\end{equation}
in order to move the $B$ line integrals across the $c$ insertions.
This results in the $B$ hitting another
$B$ that annihilates it.
Hence, we are left with
\begin{equation}
\Lambda^n = \ket{0}^{n-1} \star B c\ket{0}
    = \ket{n} \star B c\ket{0}.
\end{equation}
Next, we calculate
\begin{equation}
\label{Psi1}
Q\Lambda = (K-B Q)c(0)\ket{0} = (1-B c)K c(0)\ket{0} = (cBKc)(0)\ket{0},
\end{equation}
where we used
\begin{equation}
c\partial c=cKc\,,
\end{equation}
which follows from~(\ref{KdevWedge}).
We conclude that
\begin{equation}
\label{Psi_n}
\Psi_n=(Q\Lambda)\Lambda^{n-1} =
 c\ket{0} \star BK\ket{n-1} \star c\ket{0}=\frac{d}{dn}\psi_{n-1}\,,
\end{equation}
where we defined
\begin{equation}
\label{psi_n}
\psi_n \equiv \frac{2}{\pi} c\ket{0} \star B \ket{n}\star c\ket{0} =
\frac{2}{\pi} c\ket{0} \star B e^{\frac{\pi}{2}(n-1)K}\ket{1}\star c\ket{0}.
\end{equation}
Here, we used~(\ref{KWedge}) in order to relate the two ways of writing
$\psi_n$ in~(\ref{psi_n}) and~(\ref{KdevWedge}) was used for trading $K$
in~(\ref{Psi_n}) for the derivative.
Note, that this expression is somewhat formal for the case $n=0$
($\Psi_1$). Carefully taking the limit, one gets
\begin{equation}
\psi_0=\frac{2}{\pi}(cBc)(0)\ket{0},\qquad \psi'_0=(cBKc)(0)\ket{0},
\end{equation}
in agreement with~(\ref{Psi1}).

\begin{figure}[tbh]
\begin{center}
\input{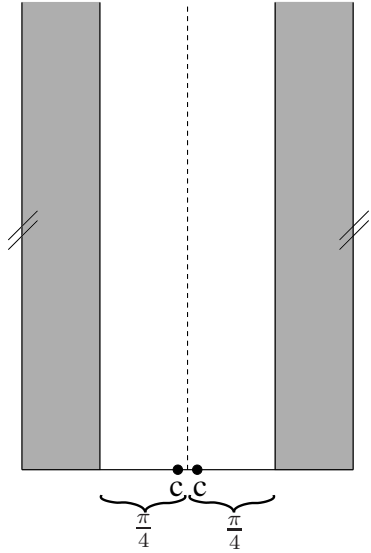}
\end{center}
\caption{Schnabl's solution: The left (length $\frac{\pi}{4}$)
strip length is represented in the split-string formulation by
$W_{\frac{1}{2}}$. To the right of it is the $c$ insertion.
The dashed line represents the $B,K$ line integrals as well as
the varying strip length insertions.
The $\Psi_1$ part contributes zero length, so the $c$ insertions
almost touch, but the $B,K$ in between them make this part also
non-trivial. The rest of $\Psi$ contributes larger and larger
strip lengths, up to infinity. A (gray) coordinate patch is
appended to the resulting surface in the usual way.}
\label{fig:SchSol}
\end{figure}
Summing~(\ref{PsiExpand}) with $\la=1$, we can write
\begin{equation}
\label{SchPsi0}
\Psi=c\ket{0}\star\sum_{n=0}^\infty
 \big(B K e^{\frac{\pi}{2}(n-1)K}\big)\ket{1}
    \star c\ket{0}= c\ket{0}\star \frac{B K e^{-\frac{\pi}{2}K}}
  {1-e^{\frac{\pi}{2}K}}\ket{1}
    \star c\ket{0}.
\end{equation}
Here, the factor of $e^{-\frac{\pi}{2}K}$ removes a piece of
strip. The result can most easily be understood in terms of the
split-string formalism~\cite{Okawa:2006vm,Erler:2006hw,Erler:2006ww}.
Examining the strip from left to right, one encounters
a $c$ insertion after a length of $\frac{\pi}{4}$, which is
half the length of the vacuum strip (local coordinate patch not included),
since the $c$ is inserted in the middle on the vacuum state.
Then, there is a varying strip length, from zero to infinity,
which is summed over, with the $BK$ line integral insertions.
Finally, there is again a $c$ insertion followed by a strip of
length $\frac{\pi}{4}$. Thus, we can write
\begin{equation}
\label{SchPsi}
\Psi=W_{\frac{1}{2}}c\frac{BK}{1-W_1}c W_{\frac{1}{2}}\,,
\end{equation}
where free operator insertions (namely $c,B,K$) represent
these operators acting on the identity string field,
i.e., inserted on a zero-sized strip. Now, the only product appearing in
the equation is the star product\footnote{The string field $c\ket{0}$
for example, is written in this notation as
$W_{\frac{1}{2}}c W_{\frac{1}{2}}$.
In the literature one usually finds the notation
$F\equiv W_{\frac{1}{2}}$.}.
The $B$ and $K$ insertions can be freely moved and so the expression containing
them in the numerator with the denominator that represent a varying
strip length is well-defined. We present Schnabl's solution schematically
in figure~\ref{fig:SchSol}.

\subsection{Sen's first conjecture}
\label{sec:Sen1}

We already showed in~(\ref{ZeroAction}) that the action for $\Psi_\lambda$
seems to vanish. Actually we showed that the action vanishes order by
order in $\lambda$. This suggests that $\Psi_\lambda$ can be regularized
by expanding it up to some finite order $N$ in $\lambda$.
Such a regularization will not affect terms up to $\lambda^N$ in
the action. Yet, at higher orders some of the terms needed
for~(\ref{ZeroAction}) to hold will be missing, allowing for
a non-vanishing action.

Concentrating on the kinetic term in the action, we have of the order of
$N^2$ terms with powers of $\lambda$ larger than $N$.
Therefore, for the proposed regularization to hold, it should
be the case that
\begin{equation}
\label{KineticTerm}
\vev{\psi'_m Q \psi'_n}\stackrel{?}{=}\cO(N^{-3}) \qquad m+n>N\,.
\end{equation}
Using CFT methods we can calculate the correlator
\begin{align}
\label{uglyCor}
\vev{\psi_m Q \psi_n} =
\frac{1}{\pi^3}\bigg( &
  \Big(\cos\big(\frac{(m-n)\pi}{m\!+n\!+2}\big)+1\Big)
  \Big((m+n+2)
    \sin\big(\frac{2\pi}{m+n+2}\big)-\pi\Big) + \nonumber \\
& \sin^2\!\big(\frac{\pi }{m+n+2}\big) \Big(
  (m+n+2)(m-n)\sin\big(\frac{(m-n)\pi}{m+n+2}\big)
\nonumber \\
& \quad
  - 2\pi(m+n+1) + 2\pi m n \cos\big(\frac{(m-n)\pi}{m+n+2}\big)
\Big)\bigg)\,.
\end{align}
In the limit where both $n$ and $m$ are large, this correlator becomes
a constant. The terms in~(\ref{KineticTerm}) behave in this limit
(due to the two derivatives) like $N^{-2}$, so~(\ref{KineticTerm})
does not hold and the summation produces a finite term.
This demonstrates that, at least for calculating the action, cutting
the series at finite $N$ is not a consistent regularization.

We can fix this regularization by rewriting the sum using the Bernoulli
numbers. The Bernoulli numbers are defined by the generating function
\begin{equation}
\frac{x}{e^x-1} = \sum_{k=0}^\infty \frac{B_k}{k!} x^k\,.
\end{equation}
The middle term in the solution~(\ref{SchPsi0})
(without the $B$ insertion) can thus be written as,
\begin{align}
\sum_{n=0}^\infty
K e^{\frac{\pi}{2}(n-1)K}=\frac{K e^{-\frac{\pi}{2}K}}{1-e^{\frac{\pi}{2}K}}=
\sum_{n=0}^{N-1} 
 K e^{\frac{\pi}{2}(n-1)K} - \frac{2}{\pi}e^{\frac{\pi}{2}(N-1) K}
  \sum_{k=0}^\infty \frac{B_k}{k!}\big(\frac{\pi}{2}K\big)^k \,,
\end{align}
where $N$ is arbitrary.
If we cut-off the Bernoulli series at $M$~\cite{Erler:2007xt},
we get a regularized form for the solution~(\ref{SchPsi0}),
\begin{equation}
\label{RegPsi}
\Psi^{(N,M)} = \sum_{n=0}^{N-1} \psi'_n -
\sum_{k=0}^M \frac{B_k}{k!}\psi_{N}^{(k)}\,,
\end{equation}
where $\psi_{N}^{(k)}$ is the $n^{th}$ derivative of $\psi_{N}$ with respect
to $N$.
The large $N$ behaviour of the kinetic term for this solution is
\begin{equation}
\vev{\Psi^{(N,M)} Q \Psi^{(N,M)}} = \vev{\Psi Q\Psi} + \cO(N^{-M-1})\,,
\end{equation}
where $\vev{\Psi Q\Psi}$ is the result without regularization artifacts.
Similarly, for the cubic term one gets
\begin{equation}
\vev{\Psi^{(N,M)}\Psi^{(N,M)}\Psi^{(N,M)}} = \vev{\Psi\Psi\Psi}
 + \cO(N^{-M-1})\,.
\end{equation}
Therefore, it is enough to set $M=0$, meaning that one needs only the
first term in the Bernoulli series, $B_0=1$.
Hence, a properly regularized form of the solution is\footnote{It may happen
that more terms are needed. This is the case for the ``tachyon'' solution in
superstring field theory~\cite{Erler:2007xt}. One should also note that
this is not the only way to regularize the solution. The simplest alternative
is level truncation. The Bernoulli term (sometimes called
``the phantom piece'') has a vanishing inner product with all Fock space
states in the limit $N\rightarrow \infty$. Thus, level truncation is a
proper regularization of the solution. However, it is usually not adequate
for the derivation of analytical results. The need of the phantom piece was
a source of confusion, exactly since this term, which is
imperative for proving Sen's conjecture, has a vanishing inner product
with all Fock space states. \ATFV the paper~\cite{Erler:2009uj} appeared, in
which a new, ``phantom-less'' version of the tachyon solution was
presented.},
\begin{equation}
\label{SchSolBerReg}
\Psi = \lim_{N\rightarrow \infty}\Big(\sum_{n=0}^{N-1} \psi'_n-\psi_N\Big)\,.
\end{equation}

To calculate the kinetic term, we replace $n$ with $x N$ and $m$ with
$y N$. For large $N$ we get
\begin{equation}
\label{xyLimit}
\vev{\psi'_x Q\psi'_y} =
\frac{8\pi x y}{(x+y)^6}
\left(\pi  x y \cos \big(\frac{\pi(x-y)}{x+y}\big)+(y^2-x^2)
   \sin \big(\frac{\pi  (x-y)}{x+y}\big)\right),
\end{equation}
where the derivatives now are with respect to $x$ and $y$ and we write $\psi_x$
instead of $\psi_{Nx}$.
Naively, one could think that it is possible to replace the summation
with an integral. Substituting~(\ref{SchSolBerReg}) gives,
\begin{align}
\label{WrongKinetic}
\nonumber
\vev{\Psi Q\Psi} & =
\int_0^1 dx dy \vev{\psi'_x Q\psi'_y} -
\int_0^1 dx \vev{\psi'_x Q\psi_1} -
\int_0^1 dy \vev{\psi_1 Q\psi'_y} +
\vev{\psi_1 Q\psi_1} \\
& = \vev{\psi_0 Q\psi_0} = 0  \qquad\qquad\qquad\qquad\mbox{(wrong)} \,.
\end{align}
The first two equalities are straightforward. The last equality holds, since
the relevant primitive of~(\ref{xyLimit}), i.e., the one that
agrees with~(\ref{uglyCor}) at infinity, is zero on the lines $x=0$ and $y=0$.
We are getting the wrong results, since when both $x$ and $y$ are small,
the approximation~(\ref{xyLimit}) does not hold and
the summation cannot be replaced by an integral.
Therefore, we calculate the sum explicitly for the case $x+y<1$.
The sum over any constant $x+y$ corresponds to a sum over terms
with a $\lambda^{(x+y)N}$ coefficient.
These sums vanish by the virtue of the equation of motion as discussed earlier.

Therefore, to calculate the action for the kinetic term, we only
need to subtract the integral over $x+y<1$ in~(\ref{WrongKinetic})
\begin{equation}
\vev{\Psi Q\Psi} = 
 - \int_0^1 dx \int_0^{1-x} dy \vev{\psi'_x Q\psi'_y} = -\frac{3}{\pi^2}\,.
\end{equation}
From the equation of motion we know that
\begin{equation}
\vev{\Psi Q\Psi} = -\vev{\Psi\Psi\Psi} \,.
\end{equation}
One might want to check explicitly that this equation indeed
holds~\cite{Okawa:2006vm,Fuchs:2006hw}.
It is a non-trivial check of the regularization.
Specifically, without the Bernoulli term, the equation of motion
does not hold, when contracted with the solution itself.

The energy density equals minus the action per unit volume.
\begin{equation}
E = -\frac{S}{V_{26}}
  = \frac{1}{g_o^2 V_{26}}\big(\frac{1}{2}\vev{\Psi Q\Psi}
    +\frac{1}{3}\vev{\Psi\Psi\Psi}\big)
  = \frac{\vev{\Psi Q\Psi}}{6g_o^2 V_{26}}
  = -\frac{1}{2\pi^2 g_o^2} \,.
\end{equation}
This is exactly the tension of the D25-brane~\cite{Okawa:2002pd}.
The volume factor comes from integrating over the zero-modes and was set
to unity before.
Lower dimensional D-branes do not have a zero-mode for the directions
with the Dirichlet boundary condition, giving the correct volume
factor for these D-branes.

\subsection{Sen's third conjecture}
\label{sec:Sen3}

As we described in~\ref{sec:LevelTrunc2}, the cohomology of the
kinetic operator around the tachyon vacuum is empty, provided that
a string field $A$ exists obeying~(\ref{QA1}).
An exact form of such an operator was found by
Ellwood and Schnabl~\cite{Ellwood:2006ba}
using an analogy to the numerical solution found in the Siegel
gauge in~\cite{Ellwood:2001ig}. The string field $A$ takes the form,
\begin{equation}
\label{A}
A=-\frac{2}{\pi}B\int_0^1 W_r dr\,.
\end{equation}

To show that this string field obeys~(\ref{QA1}),
we recall~(\ref{QQdef}) and evaluate the terms separately.
First\footnote{Conventions here can be confusing, since $W_1=\ket{0}$,
namely the perturbative vacuum,
while $W_0=\ket{1}$, namely the identity string field.},
\begin{equation}
Q A=-\frac{2}{\pi}[Q,B]\int_0^1 W_r dr=
   -\int_0^1 \partial_r W_r dr=-W_1+W_0=\ket{1}-\ket{0}.
\end{equation}
Then, we substitute~(\ref{SchPsi}) to get,
\begin{equation}
\begin{aligned}
\Psi A=& -\frac{2}{\pi}W_{\frac{1}{2}}c\frac{BK}{1-W_1}c B
      \int_{1/2}^{3/2} W_r dr=
  -\frac{2}{\pi}W_{\frac{1}{2}}c BK \int_{1/2}^\infty W_r dr=\\ &
  -W_{\frac{1}{2}}c B \int_{1/2}^\infty \partial_r W_r dr=
W_{\frac{1}{2}}c B (W_{\frac{1}{2}}-W_\infty)=(cB)(0)\ket{0},
\end{aligned}
\end{equation}
where in the first equality we used the wedge state
algebra~(\ref{SSalg}) and in the second we commuted the ghosts,
expanded the denominator and merged it with the integral.
Then, we used the fact that $K$ acts as a derivative with respect to
wedge length~(\ref{KdevWedge}),
performed the integral and neglected the boundary term coming from
an infinitely long strip, since its inner product with an arbitrary
Fock space goes like $N^{-3}$.
Similarly one gets
\begin{equation}
A\Psi=(Bc)(0)\ket{0}\,.
\end{equation}
All in all one can write
\begin{equation}
\Q A=Q A+\Psi A+A \Psi=\ket{1}-\ket{0}+[B,c](0)\ket{0}=\ket{1},
\end{equation}
where~(\ref{Bc1}) was used.
This ends the proof.

As mentioned in~\ref{sec:LevelTrunc2}, the existence of $A$
proves that the cohomology is empty not only for ghost number one
string fields, but for string fields of all ghost numbers.
Nonetheless, the opposite was concluded in a numerical study 
performed in the Siegel gauge~\cite{Giusto:2003wc}.
Imbimbo, one of the authors of that work, repeated the numerical
analysis, this time in the Schnabl gauge and found again, that
while it seems that the cohomology at ghost number one is empty,
there seem to be non-empty cohomologies at other ghost
numbers~\cite{Imbimbo:2006tz}\footnote{Similar conclusions for
the ghost number one space were obtained in~\cite{Kwon:2007mh}.}.
This result contradicts the derivation presented above and one
has to think how can they be reconciled.
One option is that the numerical results get modified as higher
levels are included and the analytical prove presented here holds.
While this is always a possibility with numerical analysis,
the results of~\cite{Imbimbo:2006tz} seem to be
quite robust. Another option is that we were too hasty in
neglecting the contribution of the terms based on
infinitely long strips. This is related to the notorious problem
of defining the correct space of string fields. Since it is
not clear how should this space be defined\footnote{There is no
canonical positive definite norm, for example, to be used for
defining the space of string fields.
We only have some criteria, such as demanding this
space to form a star-algebra and the existence of physical solutions
within it. See also footnote~\ref{foot:hilbert}.},
it may well be the case that either the formal steps
in the Ellwood-Schnabl construction or the numerical analysis
of Imbimbo, fail somehow in the relevant space.
This question certainly deserves a deeper study.

\subsection{Generalizations of the solution}
\label{sec:generalization}

String field theory has a huge amount of gauge symmetry.
In particular, there are many solutions equivalent to the one found
by Schnabl. One may wonder what is the specific property that
makes this solution simpler than the other gauge equivalent ones.
Deeper understanding of this point can serve as a key for finding
solutions with other physical content.

One of the features that enabled the construction of Schnabl's solution,
was the simplification in the form of the equation of motion in the
$z$ conformal frame. In~\cite{Rastelli:2006ap}, Rastelli and Zwiebach
tried to pin down the properties of this conformal frame that made
the string field equation of motion tractable. Instead of studying
the equation of motion~(\ref{EOM}), a simplified ghost-number-zero
toy model was used~\cite{Gaiotto:2002uk}. The equation of motion
was postulated to be,
\begin{equation}
\label{toyEOM}
(L_0-1)\Psi+\Psi^2=0\,.
\end{equation}
Next, they postulated that in different conformal frames the
equation of motion should be a generalization of~(\ref{toyEOM}), namely
\begin{equation}
\label{toyRZ}
(L-1)\Psi+\Psi^2=0\,.
\end{equation}
Here, $L$ represents the zero mode of the energy momentum tensor
in the given conformal frame.
Then, they required that $L$ and its hermitian conjugate $L^\dag$
obey a generalization of~(\ref{SL2Algebra}),
\begin{equation}
\label{RZLL}
[L^\dag,L]=s(L+L^\dag)\,.
\end{equation}
The parameter $s$ introduced here seems to be spurious, since it is
possible to absorb it by a rescaling of $L$. However, the requirement
that $L$ is the zero-mode of the energy momentum tensor in a peculiar
conformal frame fixes this constant. Note, that a rescaling of the
conformal transformation does not change the value of $s$, since
$L$ depends on $f$ only through the combination $\frac{f}{f'}$,
\begin{equation}
L=\oint \frac{d\xi}{2\pi i}\frac{f(\xi)}{f'(\xi)}T(\xi)\,.
\end{equation}

It was then suggested that the property that makes the cylinder
coordinates adequate is their relation to the sliver projector.
Star-algebra projectors~\cite{Rastelli:2001rj,Gaiotto:2002kf},
i.e., ghost number zero string fields obeying the equation
\begin{equation}
\Psi^2=\Psi\,,
\end{equation}
were used in string field theory in various contexts.
As always, one can classify projectors according to their rank.
The identity string field is a projector with a maximal rank.
An important role in string field theoretical research
was played by rank one projectors, the most familiar
ones being the sliver~\cite{Rastelli:2000iu,Kostelecky:2000hz}
and the butterfly~\cite{Gaiotto:2001ji,Schnabl:2002ff,Gaiotto:2002kf}.

The association of conformal frames to projectors is via the relation
\begin{equation}
\Psi=\lim_{\gamma\rightarrow \infty}e^{-\gamma L}\ket{0}.
\end{equation}
If the limit exists, it defines a projector. In order to be adequate
for defining analytical solutions, it should obey some regularity
conditions, which basically imply that a left-right factorization of
$L$ is permissible. Projectors obeying these conditions were named
``special projectors'' and their conformal frames named ``special
conformal frames''.
With these conditions at hand, it was shown that the sliver is the only
projector with $s=1$ in~(\ref{RZLL}), while for $s=2$ one has two
possible projectors, namely the butterfly and a new one, named
the ``moth''. The solutions of~(\ref{toyRZ}) were found for $s>1$.
These solutions look like generalizations of Schnabl's (toy)
solution, in which the Bernoulli numbers were generalized is some way.
Still, Schnabl's original solution seems to be the simplest one,
both algebraically and because it is based on a simple conformal
transformation in which the star product is realized simply by
gluing (without any other operations).

Another issue left out in~\cite{Rastelli:2006ap} is the ghost
number one case, as opposed to the toy model.
This physical case was studied along similar lines
in~\cite{Okawa:2006sn}.
It was found that, while special projectors give rise to simple
solutions, other projectors can also be used in principle.
The map that sends solutions from one conformal
frame to another was constructed.
In particular generalizations of the wedge state subalgebra leading to
an arbitrary projector was found and it was established that (unlike
in the toy model) the formal form of the solution is the same for all
conformal frames.

Classification of the resulting solutions led to the conclusion that
they all reside in a proper subspace of the universal space, in which
they are supposed to be found.
This enabled the authors to check whether the numerical solutions
in the Siegel gauge can be related to their construction. It turns
out that the vacuum solution found in the Siegel gauge does not
belong to the restricted space and so cannot be obtained from a 
projector in the way described in~\cite{Okawa:2006sn}.
Thus, while it is still believed that the numerical solution in the
Siegel gauge is related to Schnabl's solution (and its generalizations
described here) by gauge transformations, these gauge
transformations cannot be obtained using reparametrization.

Another generalization of Schnabl's solution was considered
in~\cite{Ishida:2008jc}, where a B-field was included\footnote{The
treatment of a B-field background in string theory in the
language of non-commutative field theory was developed
in~\cite{Seiberg:1999vs}.
The implementation of this idea within string field theory was studied
in~\cite{Sugino:1999qd,Kawano:1999fw,Witten:2000nz,Schnabl:2000cp,Furuuchi:2002zy,Maccaferri:2005mq}.}.
It was shown that Schnabl's solution (in its CFT form) is not modified.
This is an expected result due to the universal form of the solution and
Sen's observation regarding the universality of the tachyon
vacuum~\cite{Sen:1999xm}.

\section{Marginal deformations}
\label{sec:Marginal}

A boundary conformal field theory can be deformed by adding to it
a (boundary) term of the form\footnote{See~\cite{Recknagel:1998ih}
for a general study of boundary deformations.}
\begin{equation}
\delta S_{BCFT}=\lambda\int V dx\,.
\end{equation}
If $V$ is a weight-one vertex-operator, then the resulting theory
will remain conformal, at least to leading order in $\lambda$.
If the theory remains conformal also for finite $\lambda$, one refers
to $V$ as being ``exactly marginal''.
The new theory can be described as a solution of the string field theory
of the original BCFT. This solution should be given by
\begin{equation}
\label{marg1stOrder}
\Psi=\lambda \Psi_1+{\cal O}(\lambda^2)\,,\qquad \Psi_1=cV(0)\ket{0}.
\end{equation}
The first order term $\Psi_1$ trivially satisfies the linearized equation
of motion~(\ref{LinEOM})
\begin{equation}
\label{QPsi1}
Q\Psi_1=0\,,
\end{equation}
since it is nothing but the unintegrated vertex operator associated with the
integrated vertex operator $V$\footnote{A more accurate description of vertex
operators is given in section~\ref{sec:SUSY}.}.

It was suggested in~\cite{Sen:1990hh,Sen:1990na,Sen:1992pw} that for
exactly marginal deformations, the appropriate higher order terms should
exist. These higher order terms could, in principle, be evaluated
by expanding the string field as in~(\ref{PsiExpand})
and solving the recursion relations
\begin{equation}
\label{MarginalRecursion}
Q\Psi_n =-\sum_{k=1}^{n-1}\Psi_k\Psi_{n-k}\,,
\end{equation}
with the initial condition~(\ref{marg1stOrder}).
Note, however, that while the perturbed BCFT is a unique theory,
the solution to the recursion equations is not unique.
Far from that, the huge gauge symmetry of string field theory is
hidden in them. Thus, for a finite value of $\lambda$ the relation
is that a single BCFT corresponds to a gauge equivalence class
of string field theory solutions.
Moreover, one can add to $\Psi_n$ in~(\ref{MarginalRecursion}) any
solution of~(\ref{QPsi1}). This means that an arbitrary marginal
deformation can be added to the solution at any order. Adding
the same marginal deformation that one starts with, results in
a reparametrization of $\la$, while adding another one
results in a solution whose physical content is different for
finite values of $\la$. Thus, one should be cautious when
constructing and interpreting these solutions.

The first investigation (using level truncation) of marginal
deformations as solutions to string field theory
was performed in~\cite{Sen:2000hx}. Following this work other papers
appeared~\cite{Zwiebach:2000dk,Iqbal:2000qg,Takahashi:2002ez,Kluson:2002hr,Kluson:2003xu,Katsumata:2004cc,Sen:2004cq,Yang:2005iu,Kishimoto:2005bs}.
The first analytical solutions of marginal deformations
appeared in~\cite{Kiermaier:2007ba,Schnabl:2007az},
where the solution describing general marginal deformations
with regular OPE's was given. The OPE whose regularity
we refer to is that of $V$ with itself.
Thus, the simple analytical solutions given could not describe
more general deformations (of the correct dimensions)
with OPE of the form,
\begin{equation}
V(z)V(0)\sim \frac{1}{z^2}+\frac{\tilde V}{z}\,.
\end{equation}
It is known that the presence of a non-zero $\tilde V$ in the OPE above
results in a perturbation, which is not exactly
marginal~\cite{Recknagel:1998ih}.
The case in which $\tilde V$ is absent,
\begin{equation}
\label{VVsing}
V(z)V(0)\sim \frac{1}{z^2}\,,
\end{equation}
might correspond to an exactly marginal deformation.
This case was also considered in~\cite{Kiermaier:2007ba}
and it was found that one has to add some counter-terms to the
expressions that appear in the regular case. The first two
counter-terms were given. However,
it was not clear if and how could one continue with the
construction of higher order counter-terms. A framework for addressing
marginal deformations with a singular OPE was presented
in~\cite{Fuchs:2007yy,Kiermaier:2007vu}\footnote{\ATFV a novel approach
towards regular marginal deformation appeared, which could presumably
also be generalized to the singular case. In this construction the main
building blocks are
``boundary condition changing operators''~\cite{Kiermaier:2010cf}.}.
We turn now to describe regular marginal deformations
in~\ref{sec:RegMarginal}.
The case of marginal deformations with a singular
OPE is presented next, in~\ref{sec:SingMarginal}.

\subsection{Regular marginal deformations}
\label{sec:RegMarginal}

The solutions of~\cite{Kiermaier:2007ba,Schnabl:2007az} obtain
a simple form when represented as formal pure-gauge solutions.
The key element for a pure-gauge representation of these solutions
lies in the introduction of the formal
string field $J$~\cite{Erler:2007rh,Okawa:2007ri},
which inverts the BRST
operator $Q$,
\begin{equation}
\label{QJ1}
Q J=\ket{1}.
\end{equation}
This looks very similar to~(\ref{QA1}). The difference is that here $Q$
is the usual BRST operator, whose cohomology is not empty. Thus,
this state should not exist.
We can nevertheless write down a formal solution to~(\ref{QJ1}),
\begin{equation}
\label{JBW}
J=-\frac{2}{\pi}B \int_0^{-1} W_t\, dt\,.
\end{equation}

The string field $J$ in~(\ref{JBW}) looks very similar to $A$
of~(\ref{A}). The only difference is the range of integration, which is
$(0,-1)$ for $J$ and $(0,1)$ for $A$.
This range of integration is what causes $J$ to be
a non-legitimate string field,
since the local coordinate patch is not included in this range.
Recall that the local coordinate
patch is reserved for the test state, with which the surface state should
be contracted and so should be kept intact. A wedge state can be represented
by a Gaussian wave function in the infinite dimensional space of string
modes.
In the limit in which the identity string field $W_0$ is approached, i.e.,
the limit in which the surface reduces to solely the local coordinate patch,
this Gaussian wave function approaches a product of delta functions in
some directions and constants in the others. That is to say, the standard
deviation $\sigma$ in $\exp({-\frac{X^2}{2\sigma}})$ approaches zero or infinity
in all directions. When one tries to go beyond the identity string field to
wedge states with $t<0$, $\sigma$ changes sign and the wave function
diverges badly.
Despite the above, one can make sense of $J$ when it multiplies states that
have enough strip length that can be removed from them.
If the states multiplying $J$ from both sides are of the form $\cO(0)\ket{0}$
and the OPE of $\cO$ with itself has a zero, then~(\ref{QJ1}) indeed holds.

Using $J$ the form of the solution is very simple. The gauge string field equals
\begin{equation}
\Lambda=\lambda J \Psi_1\,,
\end{equation}
where $\Psi_1$ is given by~(\ref{marg1stOrder}).
Working in the left-scheme this leads to
\begin{equation}
\label{LeftScheme}
\Psi=Q\Lambda\frac{1}{1-\Lambda}=\Psi_1\frac{\lambda}{1-\lambda J \Psi_1}\,,
\end{equation}
where we used~(\ref{QPsi1}) and (\ref{QJ1}).
This is a solution by construction (being given in a pure-gauge form),
provided it is a legitimate string field despite the usage of $J$ in
its definition. We can expand the solution and write
\begin{equation}
\label{RegMargJdep}
\Psi_n=\Psi_1 J \Psi_1 J \cdots J \Psi_1\,,
\end{equation}
with $n$ appearances of $\Psi_1$.
Since $J$ removes a varying amount of strip we can write the above
(in split-string notations) as
\begin{equation}
\label{RegMarDefPsiN}
\Psi_n=W_{\frac{1}{2}} (cV)\int_0^1 dt_1\,W_{t_1}
 (cV)\int_0^1 dt_2\,W_{t_2} (cV)
       \cdots \int_0^1 dt_{n-1}\,W_{t_{n-1}} (cV) W_{\frac{1}{2}}\,.
\end{equation}
We see that in the lower limit of the integrations one gets a collision of
two $cV$ vertex operators. If there are singularities in the OPE it may lead
to an expression that is not well behaved and to a breakdown of the equation
of motion. Nevertheless, this is a perfectly well behaved solution
for the regular marginal deformations\footnote{Singularities could in
principle emerge from collisions of more than two operators. Deformations
that behave in this way should be dealt with using the methods
of~\ref{sec:SingMarginal}. In particular cases such a behaviour might imply
that the deformation is not exactly marginal. It is not expected that a
string field theory solution would exist for such deformations.}.

The above method for constructing (regular) marginal deformations
can be used for getting the marginal deformation associated with the
rolling tachyon, which was found in~\cite{Kiermaier:2007ba,Schnabl:2007az}.
The coefficient of the tachyon field was evaluated as a series in
$e^{X_0}$ and it seemed that the series converges uniformly for all $X_0$
and represents an oscillatory tachyon with an ever growing amplitude.
This is in accord with similar behaviour observed (in the Siegel gauge)
in level truncation analysis~\cite{Moeller:2002vx}. This behaviour is
counterintuitive and it seems to be in conflict with the exponential
grow result found using BSFT\footnote{A way to avoid this behaviour was
recently found by Hellerman and Schnabl~\cite{Hellerman:2008wp}.
They studied tachyon condensation with a tachyon
perturbation evolving along a light-like, rather than a time-like direction,
including a dilaton background that introduces an effective friction term.
Their solution does not suffer from oscillations.}.
In~\cite{Moeller:2002vx} it was claimed that this
behaviour is acceptable in principle for a system with infinitely many
time derivatives. Later, it was shown that the apparent contradiction between
the smooth rolling of the tachyon in BSFT and the divergent oscillatory one
of cubic string field theory can be attributed to the (non-local)
field redefinition between the two
theories~\cite{Coletti:2005zj}\footnote{An interesting approach for obtaining
(level-truncated) solutions of tachyon condensation
is to use the diffusion equation in order to
relate solutions of the non-local system to analogous solutions of a local
system~\cite{Calcagni:2007wy}. Note, however, that due to an improper
limiting procedure, one of the solutions obtained there
has a cusp singularity at $t=0$ and is actually
a gluing at $t=0$ of two genuine solutions~\cite{Coletti:2005zj}.
The authors of~\cite{Calcagni:2007wy} end up discarding this solution,
albeit for other reasons.}.
Indeed, a direct evaluation of the partition function of the rolling
tachyon solution of~\cite{Kiermaier:2007ba,Schnabl:2007az},
produced a result very similar to the one obtained in
BSFT~\cite{Jokela:2007dq}.

Moreover, the numerical result of~\cite{Coletti:2005zj} was shown
to hold analytically by Ellwood~\cite{Ellwood:2007xr}. Ellwood constructed a
gauge equivalent family of solutions for the rolling tachyon marginal
deformation. In one limit, the solution reduces to a solution based on the
identity string field. Ellwood interpreted that as an IR
limit\footnote{Related work appeared in~\cite{Kishimoto:2007bb}.}. An
insertion over the identity string fields looks like a local vertex
operator. Hence, this description resembles in some sense the BCFT one.
Heuristically it could be thought of as ``looking at the string field'' from
far away. Then, the internal structure is lost and the string field reduces
to a local insertion over the identity. This naive IR limit might be
singular, as is the case in other theories, e.g., Fermi's interaction. This
description sheds some light on the relation between the identity based
solutions and the more regular ones.

Still, it may seem strange that in a fixed gauge, namely Schnabl's gauge,
the rolling tachyon solution oscillates widely and does not seem to
converge to the tachyon vacuum, i.e., Schnabl's solution, found in the
same gauge\footnote{Strictly speaking the tachyon marginal deformation
cannot converge to the tachyon vacuum. The fact that it is a marginal
deformation implies, e.g., that it has the same energy as the perturbative
vacuum, which is different from that of the tachyon vacuum. However, the
marginal tachyon deformation is related to the actual physical process of
D-brane decay. Hence, this energy should be concentrated in the tachyon
matter~\cite{Sen:2002in}, which should flow to spatial infinity at large
time/marginality parameter and the tachyon vacuum solution should be
left locally. A related issue is the type of limit used, e.g., for any
radiation process (be it the radiation of a D-brane or that of an electron),
in a non-compact space,
a point-wise limit of all space points as $t\rightarrow \infty$ leads
to a state with no radiation. This type of limit does not respect energy
conservation. It is in this
sense that one should expect to obtain the tachyon vacuum as a limit.
It might make more sense physically to consider these
questions with a light-like expansion of a tachyon bubble, as
in~\cite{Hellerman:2008wp}.}.
One possible resolution of the problem can come from the fact
that the Schnabl gauge is not a complete gauge choice. An exact string field
exists in this gauge, namely $Q\La$ on which Schnabl's solution is
based~(\ref{LambdaSolution}). Thus, it might be the case that the limit
of the oscillatory solution is a solution that is gauge equivalent to
Schnabl's one despite the fact that both solutions respect the Schnabl
gauge. This explanation is not very satisfactory for
a couple of reasons.
First, Schnabl's gauge gives almost a complete gauge
fixing. Relating the apparent different behaviour to the small residual
gauge symmetry seems unnatural.
Second, it does not explain the relation between the two solutions.
Hence, it is not clear how and in which sense does the rolling tachyon solution
approach the tachyon vacuum at late time\footnote{Alternatively, one can
try to study the rolling tachyon solution directly in the vicinity of
the tachyon vacuum. A preliminary study of this question was presented
in~\cite{Kwon:2008ap}.
However, it is not clear how to relate the solution there neither to
the tachyon vacuum of~\cite{Schnabl:2005gv} nor to the rolling tachyon
solution of~\cite{Kiermaier:2007ba,Schnabl:2007az}.}.

Another resolution to the problem was studied by Ellwood
in~\cite{Ellwood:2007xr}, where it was suggested that in spite of the
apparent different behaviour, the rolling tachyon solution does
approach the tachyon vacuum at late time. The rolling tachyon solution
was expanded as an integral over strip length,
\begin{equation}
\Psi=\int_0^\infty \Psi_r\,.
\end{equation}
The ghost part of $\Psi_r$ is simple, but its matter dependence is
complicated, since it contains contributions from all $\Psi_n$, as
can be seen from the explicit expression~(\ref{RegMarDefPsiN}).
This is due to the appearance of $J$ in~(\ref{RegMargJdep}).
The dependence of $\Psi_r$ on $X_0$ can be described by a function
$F_r(X_0)$ as,
\begin{equation}
\Psi_r^{mat}=F_r(X_0) W_r\,.
\end{equation}
Next, it was assumed that as $x_0\rightarrow \infty$ a limit exists
for $F_r(X_0)$ and it was further assumed that this limiting function
can be written as
\begin{equation}
\lim_{x_0\rightarrow \infty}F_r(X_0)=f_r(x_0)\,.
\end{equation}
With this assumption it was found that at late time the rolling tachyon
solution indeed reduces to the tachyon vacuum solution.
It is not quite clear how to reconcile this result with the numerical study
of the tachyon coefficient described above. At any rate, this result
demonstrates that the two solutions are somehow related.
Further study of this point is certainly worthwhile.

\subsection{Singular marginal deformations}
\label{sec:SingMarginal}

The method of~\cite{Kiermaier:2007ba,Schnabl:2007az} is not easily
generalized to the case of a singular OPE\footnote{Some ideas
regarding the origin of the singularities and the way they should be
dealt with appeared in~\cite{Lee:2007ns}.}.
A different method for constructing a formal pure-gauge solution for this
case was developed in~\cite{Fuchs:2007yy}. There, a specific marginal
deformation, namely
\begin{equation}
\label{V=dX}
V=i \partial X^\mu\,,
\end{equation}
for some $\mu$, which we leave implicit henceforth, was addressed.
The factor of $i$ is needed to ensure the reality of the solution.
This deformation describes a Wilson loop, for the case of a
compact $X$. If $X$ is not compact, this deformation describes a
gauge transformation.
While the
method used is especially simple for the $\partial X$ case,
it can also be generalized. One
root towards this generalization was sketched in~\cite{Fuchs:2007yy},
while another one was developed in~\cite{Kiermaier:2007vu}.
We believe that these two methods are essentially equivalent.
For now let us continue with the $\partial X$ deformation.
We return to comment on possible generalizations later on.

The observation of~\cite{Fuchs:2007yy} was that in the case~(\ref{V=dX})
one can write (we use $\al'=2$ conventions),
\begin{equation}
\label{Qx0}
cV(0)\ket{0}=Q  \frac{i x_0}{2}\ket{0}.
\end{equation}
This suggests that the deformation is a pure-gauge one. This is indeed
the case for a non-compact $X$. However, for a compact $X$ direction,
the zero mode $x_0$ does not exist and the expression~(\ref{Qx0}) is
formal\footnote{The zero mode $x_0$ was similarly used in the context
of closed string field theory in~\cite{Belopolsky:1995vi}.}.

The zero more $x_0$ would have been a mode of the field $X$. It is
very convenient to work in terms of conformal fields.
Therefore, we pretend that the field $X$ has in its expansion
the zero mode $x_0$ and use $X$ as our building block.
The use of $X$ introduces some subtleties, which we would like to address.
Our marginal deformation is a boundary deformation.
Thus, it seems that we have to work with boundary normal ordering in order
to obtain well-defined expressions.
However, we find it more convenient to use the doubling trick, which we
already used implicitly by writing the holomorphic derivative $\partial X$
instead of the boundary derivative $\partial_x X$.
Thus, we decompose $X$ into a holomorphic and anti-holomorphic parts,
\begin{equation}
\label{XXX}
X(z,\bar z)=X(z)+\bar X(\bar z)\,.
\end{equation}
The holomorphic part is the primitive of $\partial X$ and is, therefore,
the object that we are interested in. This is what we refer to by $X$
from now on. The mode expansion of $X$ is given by,
\begin{equation}
\label{Xexpansion}
X(z)=\frac{x_0}{2}-2i p\log z+i \sum_{m\neq 0} \frac{\al_m}{m z^m}\,.
\end{equation}
The zero mode $x_0$ cannot be obtained from the integration and the factor
of $\frac{1}{2}$ is obtained from equally dividing the zero mode between
$X$ and $\bar X$. This factor of $\frac{1}{2}$ effectively
substitutes the use of boundary normal ordering.
The second term in the expansion seems problematic due to the
presence of the (non-holomorphic) logarithm. Indeed, $X(z)$ cannot be
considered as a genuine holomorphic field. Nonetheless, we are free
to continue with our construction, since $Q$ annihilates this term
and since the only instance in which it would contribute is when
considering several $X$ insertions, in which case, we would be able to
evaluate the normal ordering in a sensible way. At any rate,
the logarithm should not worry us, since only the formal gauge
string fields would essentially depend on $X$. The solution itself
could be brought to a form, in which it depends only on $\partial X$.

We can now write,
\begin{equation}
Q i X(0)\ket{0}=cV(0)\ket{0}.	
\end{equation}
We use this observation in order to write our formal gauge field as,
\begin{equation}
\label{Lambda1}
\Lambda_1= i X(0)\ket{0},
\end{equation}
which leads to the desired first order result
\begin{equation}
\label{Psi1cdX}
\Psi_1=i c\partial X(0)\ket{0}.
\end{equation}
It may seem that this is the end of the story. However, it is easy to
see that $\Psi_2$ turns out to be linear in $x_0$, with higher order
polynomial dependence on $x_0$ of $\Psi_n$ for $n>2$. This is
not a legitimate result, since $x_0$ does not really exist in this
case\footnote{One can consider also $x_0$-dependent solutions,
provided that the dependence is compatible with the compactification,
i.e., periodic solutions. We expect to have such form of solutions
for the periodical lump system. \ATFV the first analytical solutions
describing lump solution were constructed in~\cite{Bonora:2010hi}.}.
The resolution of this problem comes from adding
higher order terms to the gauge string field,
\begin{equation}
\La=\sum_{k=1}^\infty \la^n \La_n\,,
\end{equation}
while identifying~(\ref{Lambda1}) as the leading order term, as is
implied by the notations.
The higher order terms should be chosen such that there is
no $x_0$-dependence of the solution, while not changing its physical
content.

There are many ways to complete $\La$ into an $x_0$-independent
solution. One particularly simple choice is
\begin{equation}
\label{La_n}
\La_n=-\frac{(-i)^n}{n!}(X^n,\underbrace{1,\ldots,1}_{n-1})\,.
\end{equation}
Here we introduced a notation, similar to the split-string notation.
A vector of length $n$ represents the star product of $n$ vacuum
states with insertions. A
vacuum state with no insertion is represented by an insertion of an
identity. The fact that we are using insertions of the form
$X^n$ may seem problematic. One obvious objection is the singularities
from the OPE of the $X$ fields. This is resolved by defining the products
to be normal ordered at each point.
Eventually, we would have to turn the expression into a fully normal
ordered one, but first we want to verify that this definition implies
that the solution described by~(\ref{La_n}) is indeed $x_0$-independent.

One can define the normal ordered $X^n$ operators using point-splitting.
In practice, however, it seems that the simplest way to define it is using,
\begin{equation}
X^n \equiv \partial_k^n e^{k X}\!\big|_{k=0}\,.
\end{equation}
Here, $e^{k X}$ is implicitly normal ordered. It is a primary conformal
field that in our conventions carry conformal weight $-\frac{k^2}{2}$.
From this we infer the relation,
\begin{equation}
\label{Qexp}
[Q,e^{k X}(z)]=\Big(\big(k c\partial X -\frac{k^2}{2}\partial c \big)
   e^{k X}\Big)(z)\,.
\end{equation}
From this relation it follows that
\begin{align}
\label{QX}
[Q,X^n] = (\partial_k)^n[Q,e^{k X}]\Big|_{k=0}
     = n c \partial X X^{n-1} - \frac{n(n-1)}{2}\partial c X^{n-2}\,.
\end{align}
Let us restrict ourselves to ans\"atze that generalize~(\ref{La_n}),
i.e., with $\La_n$ being the star product of $n$ states of the form
$X^{k_i}\ket{0}$ with $\sum_{i=1}^n k_i=n$. For this class of gauge string
fields the $x_0$-independence condition is satisfied provided the following
recursion relation holds,
\begin{equation}
\partial_{x_0}\La_n = -i \La_{n-1}W_1\,.
\end{equation}
The recursion relation and the initial condition~(\ref{Lambda1})
can be compactly written as
\begin{equation}
\label{LeftDE}
\partial_{x_0}\La = i\lambda(1-\La)W_1\,.
\end{equation}
It can be seen that the expression~(\ref{La_n}) indeed obeys this equation
and is therefore a solution.

Many more solutions exist with the form of the suggested ansatz.
All of them are supposedly gauge equivalent, so one may ignore this
degeneracy and choose to work with the simple expression~(\ref{La_n}).
This will not be good enough, though, when one wants to impose the reality
condition on string fields~(\ref{RealPsi}).
The reality condition states that the string field $\Psi$ is real
with respect to the involution defined by composing hermitian and BPZ
conjugations. This will be the case if $\La$ is pure imaginary.
For our ansatz the conjugation is equivalent to inverting the orientation
of the strip (while retaining the formal Grassmann ordering) and
complex conjugating coefficients. The inversion of orientation implies also
a minus sign for every derivative present in the expression.
The first order term~(\ref{Lambda1}) is imaginary by construction.
The higher order terms, on the other hand,
carry no definite symmetry property and so fail to be real or
imaginary\footnote{The condition of being imaginary should be applied
to the $\La$ of the ``canonical scheme''~(\ref{PureGauge}), while here
we use the left scheme.
Transforming to the correct scheme does not introduce any concrete
symmetry properties to $\La$.}.

We can trace the origin of the lack of symmetry to the fact that we chose
a solution that takes a simple form in
the left-scheme~(\ref{LeftScheme}). This scheme is very easy to use. In
particular it was easy to write a closed form expression for the
$x_0$-independent condition~(\ref{LeftDE}).
Another simple scheme is the right scheme, which is the ``mirror image''
of the left scheme. In this scheme the solution is represented as
\begin{equation}
\Psi=\frac{1}{1+\La}Q\La\,.
\end{equation}
The simplest gauge potential generating an $x_0$-independent solution is
\begin{equation}
\label{La_nRight}
\La_n=\frac{i^n}{n!}(\underbrace{1,\ldots,1}_{n-1},X^n)\,.
\end{equation}
The left and right schemes are easy to use since
the rational function $(1-\Phi)^{-1}$ has a very simple variation,
\begin{equation}
\delta\frac{1}{1-\Phi}=\frac{1}{1-\Phi} \delta \Phi \frac{1}{1-\Phi}\,.
\end{equation}
For comparison, the variation of an exponential function can be most
nicely written in terms of an integral\footnote{To prove this identity,
expand the l.h.s, change one summation index and use the familiar integral
representation of the Beta function. An equivalent, presumably more useful
representation can be found in section 4.2.a.2 of~\cite{Gates:1983nr}.},
\begin{equation}
\delta e^\Phi=\int_0^1 dt e^{t\Phi}\delta \Phi e^{(1-t)\Phi}\,.
\end{equation}
Trying to work directly in more complicated schemes results in
complicated reality conditions. It is nevertheless possible to define
a real solution by an iteration prescription~\cite{Fuchs:2007gw}. 
While the iteration formula is given in a closed form, the resulting
expressions are quite complicated and no closed form expression
(other than the iteration formula) was found.

Another way of defining real solutions was
given in~\cite{Kiermaier:2007vu}.
This relies on the gauge transformation relating the left and right
schemes~\cite{Fuchs:2007yy},
\begin{equation}
\label{LRgaugeTrands}
e^\La=\sum_{n=0}^\infty \frac{(i \la)^n}{n!}\underbrace{(X_n-X_1)^n}_n\,.
\end{equation}
In the above expression the $n^{th}$ order term is built upon the wedge
state $\ket{n}$. It is clear from this representation that $e^\La$ is
$x_0$-independent, which implies that this is a genuine gauge
transformation. Writing the transformation explicitly one has,
\begin{equation}
\Psi_r=e^{-\La}(\Psi_l+Q)e^\La\,,\qquad
\Psi_l=e^\La(\Psi_r+Q)e^{-\La}\,,
\end{equation}
where the subscripts $l$ and $r$ stand for ``left'' and ``right''
respectively.
The real solution of~\cite{Kiermaier:2007vu} is based on ``going
half the way'' between the left and right solutions,
\begin{equation}
\Psi=e^{-\frac{\La}{2}}(\Psi_l+Q)e^{\frac{\La}{2}}=
    e^{\frac{\La}{2}}(\Psi_r+Q)e^{-\frac{\La}{2}}\,.
\end{equation}
Using the above defining expressions and the relation
\begin{equation}
\label{KOReal}
\Psi_r=-\Psi_l^*\,,
\end{equation}
immediately implies the reality of $\Psi$.

While the solution~(\ref{KOReal}) is given in a nice closed form,
it relies on the gauge transformation~(\ref{LRgaugeTrands}), whose
form is misleadingly simple. This gauge transformation is not
some sort of an exponent, since the size of the wedge states varies.
Explicitly one may write,
\begin{equation}
e^\La=1+\frac{(i\la)^2}{2}\big((X^2,1)+(1,X^2)-2(X,X)\big)+\ldots\,.
\end{equation}
Note, in particular, that the first order term drops out. The power of
$\la$, the wedge size and the total $X$ power are correlated. This will
remain so in calculating any function of $e^\La$ such as its inverse
$e^{-\La}$. The exact form of $e^{-\La}$ does not correspond to
$\la\rightarrow -\la$. Explicit calculation gives
\begin{equation}
\begin{aligned}
e^{-\La}=1&-\frac{(i\la)^2}{2}(X_2-X_1)^2-\frac{(i\la)^3}{3!}(X_3-X_1)^3\\
&+\frac{(i\la)^4}{4!}\big(6(X_1-X_2)^2(X_3-X_4)^2-(X_1-X_4)^4\big)+\ldots\,.
\end{aligned}
\end{equation}
Here we left the wedge size implicit.
Note that already the second order term has ``the wrong sign'' relative
to a $-\la$ assignment, while the fourth order term is completely different
from the corresponding one in $e^\La$. Calculating the functions
$e^{\pm\frac{\La}{2}}$ needed for~(\ref{KOReal}) results in quite
complicated expressions and the practical evaluation of a real solution
at a given level may turn out to be more complicated than evaluating the
recursion relation. Nevertheless, one may think of several cases
where the closed form of~(\ref{KOReal}) makes it more adequate for
calculations than its counterpart.

We have seen that the explicit form of real solutions is more
complicated than that of the non-real solutions in the left and right
schemes. These solutions are gauge equivalent in the theory without
the reality condition. One may decide then, to enlarge the gauge orbits
of the real solutions, in order to include these ones as well. The theory
would then include all the gauge orbits that
contain a real representative. This procedure does not add any degrees
of freedom to the theory with the reality condition,
since the added solutions are all gauge equivalent to existing
real ones. In particular, we do not complexify any component fields,
rather we ``complexify'' gauge orbits.
Thus, it seems that one can safely forget about imposing the
reality condition and work with simpler non-real solutions, provided that
one can prove that they are gauge equivalent to real ones.

Let us now consider the $\la$-expansion of the solution $\Psi_L$ itself.
At the first order we obtain the unintegrated vertex operator of the
marginal deformation~(\ref{Psi1cdX}), which was our starting point.
The next order gives,
\begin{equation}
\Psi_2=Q\La_2+Q\La_1 \La_1=\frac{1}{2}Q(X^2,1)-(QX,X)=
  (c\partial X X-\frac{\partial c}{2},1)-(c\partial X,X)\,.
\end{equation}
We see that the zero mode indeed cancels out. A function that
depends on $X$, but not on its zero mode, can be regarded as depending on
$\partial X$, which holds exactly this information. Specifically, it might
seem that one can write the $X$-dependent terms as,
\begin{equation}
\label{MDPsi2}
(c\partial X X,1)-(c\partial X,X)=-(c\partial X)(-\frac{\pi}{4})
  \int_{-\frac{\pi}{4}}^{\frac{\pi}{4}}\partial X(z)dz\,.
\end{equation}
This is almost correct. Indeed, if it was not for the singularity of the OPE
$\partial X X$, that would have been correct. In particular, if one chooses
a light-like direction for $X$, this expression does hold. Note, that the
expressions is constructed from a single unintegrated vertex operator at
the leftmost position, followed by an integral of the integrated vertex
operator. However, unlike in the solutions of this problem presented in
the previous subsection, the strip length is fixed and one only integrates
the integrated vertex insertion over the fixed strip.
Similarly evaluating $\Psi_{n>2}$ results in a strip of fixed length with
a single unintegrated vertex operator to the left, followed by $n-1$
integrated vertex operators.

Let us now consider the consequence of the OPE singularity. To that end, let
us restore the normal ordering. The matter part of the l.h.s of~(\ref{MDPsi2})
reads,
\begin{equation}
\label{Psi2NO}
\Psi_2^m=\,:\partial X X(-\frac{\pi}{4}):
      -\partial X(-\frac{\pi}{4}) X(\frac{\pi}{4})\,.
\end{equation}
Note, that the first term is normal ordered, but the second one is not.
Hence, we cannot simply write the r.h.s of~(\ref{MDPsi2}) with implicit normal
ordering. The resolution is, however, clear. We should first normal order
the second, regular, term,
\begin{equation}
\label{NorOrdX}
\partial X(-\frac{\pi}{4}) X(\frac{\pi}{4})=\,
  :\partial X(-\frac{\pi}{4}) X(\frac{\pi}{4}):-k\,,
\end{equation}
with $k$ being a constant. Note, that one has to use the expression for
the normal ordering in the cylinder coordinate for the evaluation of this
constant.
The result of the normal ordering is then nothing but the addition to the
solution of the term,
\begin{equation}
\delta \Psi_2=k c (-\frac{\pi}{4})\,.
\end{equation}
At higher orders similar terms multiply also $X$ insertions. This
cannot change the fact that the solution is $x_0$-independent and all the
$X$-dependence can be recast in terms of integrals of normal ordered
$\partial X$ insertions.

Next, we would like to generalize the construction to an
arbitrary exactly marginal deformation.
The first step involves defining a primitive for the marginal operator.
This is not a problem as far as the cohomology is concerned.
One can always enlarge the space in order to formally trivialize
it. Morally speaking this is not different from using complex numbers for
solving problems with the reals. On the other hand, in order to be
able to make sense out of the CFT, one should also specify
the OPE's of this primitive with the other conformal fields. This is
quite non-trivial in general.

Let us consider a specific example where we know how to define the
primitive, namely the deformation induced by the operator
$:\!\!\cos X\!\!:\,$. This
operator is a superposition of two exactly marginal
operators, $e^{\pm i X}$. While each of these operators has
a regular OPE with itself, their mutual OPE contains a double pole.
Hence, issues of normal ordering will emerge for this operator. For defining
the primitive, we use the fact that at the self-dual radius this operator
is dual to a scalar~\cite{Callan:1994ub}, i.e., it can be written as,
\begin{equation}
:\!\cos X\!:\,=\partial Y\,.
\end{equation}
Now, the solution takes exactly the same form as the $\partial X$ solution.
The only change is that $X$ should be replaces by $Y$ in~(\ref{La_n}).
This is not the end of the story, since one would like to
obtain an expression from which it is possible to evaluate, for example, the
expectation values of some given coefficient fields. Consider $\Psi_2$.
The expression that is analogous to~(\ref{Psi2NO}) takes the form,
\begin{equation}
\label{Psi2mCos}
\Psi_2^m=\dcirc\!\partial Y Y(-\frac{\pi}{4})\!\dcirc
      -\partial Y(-\frac{\pi}{4}) Y(\frac{\pi}{4})\,.
\end{equation}
Note, that we introduced a new normal ordering, which is the one associated
with $Y$, which is different from the one associated with $X$.
We can now proceed as in~(\ref{NorOrdX}) and write,
\begin{equation}
\label{NorOrdY}
\partial Y(-\frac{\pi}{4}) Y(\frac{\pi}{4})=
 \dcirc\!\partial Y(-\frac{\pi}{4}) Y(\frac{\pi}{4})\!\dcirc-k\,.
\end{equation}
Substituting the first term in the r.h.s back into~(\ref{Psi2mCos}) gives,
\begin{equation}
\begin{aligned}
\label{beforeLastCosMarg}
\dcirc\!\partial Y Y (-\frac{\pi}{4})\!\dcirc
 &-\dcirc\!\partial Y(-\frac{\pi}{4}) Y(\frac{\pi}{4})\!\dcirc
 =-\dcirc\!\partial Y(-\frac{\pi}{4})
  \int_{-\frac{\pi}{4}}^{\frac{\pi}{4}}\partial Y(z)\!\dcirc dz\\
  &=-\dcirc\!:\!\cos X(-\frac{\pi}{4})\!:
  \int_{-\frac{\pi}{4}}^{\frac{\pi}{4}}:\!\cos X(z)\!:\!\dcirc dz\,.
\end{aligned}
\end{equation}
Note that now the expression involves two different normal orderings,
the one associated with $X$, which is what we want to retain for the
evaluation of the coefficient fields as well as the $Y$ normal ordering.
What we have to do now is to ``undo'' the $Y$ normal ordering,
\begin{equation}
\begin{aligned}
\label{lastCosMarg}
-\dcirc\!\partial Y(-\frac{\pi}{4})
  \int_{-\frac{\pi}{4}}^{\frac{\pi}{4}} & \partial Y(z)\!\dcirc dz
  =-\int_{-\frac{\pi}{4}}^{\frac{\pi}{4}}\partial Y(-\frac{\pi}{4})
      \Big(\partial Y(z)+k(z)\Big) dz
  \\ &=
  -\int_{-\frac{\pi}{4}}^{\frac{\pi}{4}}
  :\!\cos X(-\frac{\pi}{4})\!: \Big(:\!\cos X(z)\!:
     +k(z)\Big) dz\,,
\end{aligned}
\end{equation}
with $k(z)$ a known function with a double pole at $z=0$.
The $X$-related normal ordering of the two $\!\cos X\!$
factors eliminates the double pole and results in a well-defined
expression\footnote{Note, that~(\ref{beforeLastCosMarg}) is manifestly
regular. However, the simultaneous use of several normal
orderings makes it less useful than the final, explicit
expression~(\ref{lastCosMarg}).}.
The procedure described here can be carried further to higher orders in $\la$.

Two key features of the construction described here were identified
by Kiermaier and Okawa in~\cite{Kiermaier:2007vu}, namely, the fact that the
main building block of the solution is the integrated vertex operator and
the importance of defining a proper regularization scheme. In their
construction one starts with the integrated vertex operator and defines
its renormalized version. The solution is then constructed directly in
terms of this renormalized integrated vertex operator. However, a complete
description of this renormalization procedure for the most general exactly
marginal boundary deformation was not gives. This stems from the lack of
a complete classification of exactly marginal boundary deformations.
We already mentioned that a necessary condition for the exactness of a
marginal boundary deformation is the absence of a simple pole in the $VV$
OPE. A sufficient condition is that $V$ is a ``self-local'' operator,
which means that the $VV$ OPE includes only even powers of the
separation~\cite{Recknagel:1998ih}. Nevertheless, a condition which is
both necessary and sufficient have not been found yet. Kiermaier and Okawa
found a sufficient condition for the construction of their renormalized
vertex. This condition is different from
self-locality. One should expect that this condition is also a sufficient
condition for exactness of the marginal deformation, since one should not
expect that a solution of string field theory would exist for a marginal
deformation which is not exactly marginal. It would be interesting to
understand the relation, if any, of their condition and the self-locality
condition.

As far as the solutions are concerned, it seems that the construction
of~\cite{Kiermaier:2007vu} is equivalent to that of~\cite{Fuchs:2007yy},
described so far. For the $\partial X$ marginal deformation that has been
actually proven in~\cite{Kiermaier:2007vu} and it might well be the case
that one can generalize the construction of the primitive presented for
the $:\!\cos X\!:$ deformation to those deformations that can be
renormalized according to the criterion of~\cite{Kiermaier:2007vu}.
While the form of the solution that uses the primitive is compact and
simple, the one of~\cite{Kiermaier:2007vu} seem to be more systematic and
the lack of an extension of the space by a primitive might better suit a
background independent approach. Nonetheless, regardless of the specific
approach used, a case by case analysis is necessary, either for defining the
primitive and its OPE's or for defining the renormalization of the vertex
operator. It is yet to be shown the all exactly marginal deformations can be
dealt with using any of these methods. A better understanding of this issue
from both the BCFT and the string field theoretical side is still needed.

\section{Oscillator representation}
\label{sec:oscillator}

In this section we discuss the recent developments in string field
theory in the oscillator basis formalism. As already stated, this
formalism is related to a given background and as such was not
relevant to most of the recent advance in the field. Nonetheless, it
is plausible that this formalism will play a more dominant role
in constructing analytical solutions describing lumps.

We start in~\ref{sec:oscillatorIntro} by introducing
the tools needed, i.e., the form of the string vertices in the
oscillator representation and the continuous (kappa) basis,
as well as introducing the problem of ``normalization anomaly'',
which used to be a stumbling block for the derivation of
analytical results in the oscillator representation. 
Next, in~\ref{sec:oscillatorRecent} we describe the form of Schnabl's
operators in the oscillator representation and explain how to
solve the problem of normalization anomalies.

\subsection{Representation of string vertices and the continuous basis}
\label{sec:oscillatorIntro}

For the construction of the oscillator representation recall the
form of the string field in flat background~(\ref{LinEOM}). All
possible values for the coefficient fields span the (classical)
string field Hilbert space\footnote{\label{foot:hilbert}To
have strictly a Hilbert space
one should also give a positive definite inner product that will restrict
the possible values of the coefficient fields. However, there is no
natural candidate for such an inner product. One reason for that is
that the ghost fields and the time direction induce a negative norm
in the inner product of the first quantized string. On the other hand,
when not restricted, star multiplication of string fields
leads to ``associativity anomalies''~\cite{Horowitz:1987yz,Bars:2002bt}.
It is common in the literature to refer as a Hilbert space (or Fock space)
to the subspace of string fields with a finite number of non-zero
coefficients multiplying a finite number of oscillators acting on the vacuum.
This is of course not a Hilbert space. It is also not a closed space with
respect to the star product. However, as we have nothing
new to say about it, we shall follow the common lore
by simply ignoring this issue.}. Our first goal is to define the string
vertices in this representation.

For the description of interaction the string vertices should be defined
in terms of the oscillators. The $n$-string vertex is a state in the
$n^{th}$ tensor power of the single string Hilbert space. In this space the
string vertex is (up to zero-modes) a squeezed state in the matter as well
as in the ghost sector. One can write
\begin{align}
&\int \Psi=\braket{V_1}{\Psi},\\
\label{V2}
 & \int \Psi_1 \star \Psi_2=\phbra{12}{V_2}\ket{\Psi_1}_1\ket{\Psi_2}_2,\\
 & \int \Psi_1 \star \Psi_2 \star \Psi_3=\phbra{123}{V_3}
      \ket{\Psi_1}_1\ket{\Psi_2}_2\ket{\Psi_3}_3,
\end{align}
and so on for higher vertices if needed. Here, the subscripts next to the
bras and kets refer to the Hilbert space index, since as mentioned, the $V_n$
for $n>1$ live in multi-string spaces, so in~(\ref{V2}), for example,
the vertex lives in the spaces one and two. The string field $\Psi_1$
lives in space number one and the string field $\Psi_2$
lives in space number two.
Next, we define $\ket{V_2}$ as the inverse of $\bra{V_2}$ by the relation
\begin{equation}
\phbra{12}{V_2}\ket{V_2}_{23}=\One_{13}\,,
\end{equation}
where
\begin{equation}
\One_{13}\ket{\Psi}_1=\ket{\Psi}_3.
\end{equation}
The vertex $\ket{V_2}$ implements (inverse) BPZ conjugation,
up to a sign that may originate from the ghost zero modes.

The concrete realization of the vertices $V_1$ and $V_2$ is very simple.
It can be derived from their geometrical interpretation as inducing
a gluing of surfaces.
In the matter sector the gluing of a string about its middle
can be represented by setting to zero the coefficients of the odd
modes as well as of the momenta of the even modes. In terms of oscillators
this infinite product of delta functions is represented by
\begin{equation}
\begin{aligned}
\label{V1Def}
\bra{V_1^m}=&\int d^{26} k \bra{k}\delta(k)
 \exp\Big(-\frac{1}{2}\sum_{n,m=1}^\infty a_n^\mu C_{n,m} a_m^\nu
    \eta_{\mu\nu}\Big)
  \\=&\bra{0}\exp\Big(-\frac{1}{2}\sum_{n=1}^\infty (-1)^n a_n^\mu a_n^\nu
    \eta_{\mu\nu}\Big)\,,
\end{aligned}	
\end{equation}
where the implicitly defined matrix $C$ is the twist matrix,
\begin{equation}
C=(-1)^n \delta_{n,m}\,,
\end{equation}
and $\eta_{\mu\nu}$ is the flat 26-dimensional metric. The summation on
the indices $\mu,\nu$ is implicit. Henceforth, we shall refrain from writing
the metric and the spatial indices explicitly unless needed, as they do not
play an important role in most of the following discussion. One should keep
in mind though, that there are 26 coordinate dimensions.
The gluing defined by the two-vertex is imposed by identifying the even
modes of the two strings and the odd modes of their momentum.
This results in
\begin{equation}
\begin{aligned}
\phbra{12}{V_2^m}=&\int d^{26} k_1 d^{26} k_2 \, \phbra{12}{k_1,k_2}
 \delta(k_1+k_2)\exp\Big(-\frac{1}{2}\sum_{n,m=1}^\infty a_n^1 C_{n,m} a_m^2\Big)
  \\=&\int d^{26} k\, \phbra{12}{k,-k}
    \exp\Big(-\sum_{n=1}^\infty (-1)^n a_n^1 a_n^2\Big)\,,
\end{aligned}	
\end{equation}
where the superscripts on the oscillator modes represent the space-number
on which they act.

The application of these simple geometric ideas to the three-vertex
$V_3$ is technically more complicated. The easiest way to find it is
first to assume the following squeezed state form
\begin{equation}
\begin{aligned}
\phbra{123}{V_3^m}=&\int d^{26} k_1 d^{26} k_2 d^{26} k_3
 \, \phbra{123}{k_1,k_2,k_3}\delta(k_1+k_2+k_3)\cdot\\
 & \cdot\exp\Big(-\sum_{s,r=1}^3 \big( \frac{1}{2}\brao{a^r} V^{rs} \keto{a^s}+
 \braketo{a^r}{V^{rs}_0} k_s + k_r V^{rs}_{00} k_s \big)\Big).
\end{aligned}	
\end{equation}
Here we introduced the notation of round brackets to represent the infinite
vectors of different string modes. Now, the zero modes enter into the exponent
and, when singled out as above, the data defining the three-vertex reduces to
the matrices (in mode space) $V^{rs}$ the vectors $\keto{V^{rs}_0}$ and the
scalars $V^{rs}_{00}$ (in all this cases $r,s=1\ldots 3$).
To find the coefficients $V^{rs}_{nm}$ we note that (up to the
infinite volume factor $\delta(0)$) one can write
\begin{equation}
V^{rs}_{nm}=-\phbra{123}{V_3^m} a^{r\,\dag}_n a^{s\,\dag}_m\ket{0}_{123}\,.
\end{equation}
Writing the creation operators using a contour integral of $\partial X$
one can use the CFT expression~(\ref{CFTvertex}) and the explicit form
of $f^{(3)}_r$~(\ref{fDef})\footnote{We henceforth denote $f^{(3)}_r$ by
$f_r$ for simplicity.} to get to the following
expression~\cite{LeClair:1989sp,LeClair:1989sj}
\begin{equation}
V^{rs}_{nm}=-\frac{1}{\sqrt{nm}}
  \oint \frac{dz dw}{(2\pi i)^2 z^n w^m}
      \frac{f'_r(z)f'_s(w)}{(f_r(z)-f_s(w))^2}\,.
\end{equation}
It is easy to see in this representation that the matrices obey the symmetry
properties one would expect them to have, namely $V^{rs}$ depends only on
$r-s$ modulo 3.
Analogous expressions hold also for the vectors and the scalars.

A consequence of the fact that all the string vertices are given by squeezed
states is that squeezed states form a subalgebra of the star product and
the expressions for multiplying squeezed states\footnote{By squeezed states
we also mean states with linear terms. In particular coherent states that do
not contain the quadratic terms are also part of this subalgebra.} are
formally very simple. However, these formal expressions involve multiplying
and inverting the infinite dimensional matrices defining the states and
the vertices.

Working with infinite dimensional matrices may be complicated and one may be
forced to advance using numerical tools in the spirit of level truncation.
A natural route to simplify the work with matrices is to diagonalize them.
This cannot be done in this case, since a direct evaluation shows that
these matrices do not commute. Hence, they cannot be diagonalized simultaneously.
There is, however, a simple way out. The matrices
\begin{equation}
M^{rs}\equiv C V^{rs}
\end{equation}
are commutative and so can be simultaneously diagonalized.
This task was performed by Rastelli, Sen and Zwiebach~\cite{Rastelli:2001hh}.
They used the fact that the three matrices $M^{rr},M^{r(r\pm 1)}$ commute
with the matrix $\K_1$ related to the Virasoro operator $\LL_{-1}$~(\ref{LLm1}).
The matrix $\K_1$ has all real numbers as eigenvalues, without degeneration,
\begin{equation}
\K_1\keto{\ka}=\ka\keto{\ka}\,.
\end{equation}
The eigenvalues of the $M$ matrices are found to be,
\begin{align}
\label{muDef}
M^{rr}\keto{\ka}&=-\frac{1}{1+2\cosh(\frac{\ka \pi}{2})}	\keto{\ka}
 \equiv \mu(\ka)\keto{\ka},\\
M^{r(r\pm 1)}\keto{\ka}&=
  \frac{1+\exp(\pm\frac{\ka \pi}{2})}
     {1+2\cosh(\frac{\ka \pi}{2})}\keto{\ka}.
\end{align}
The eigenvectors $\keto{\ka}$ were normalized in~\cite{Okuyama:2002yr}.
Given a squeezed state defined by a matrix $S$, one can consider the 
matrix $T=CS$. Squeezed states whose matrix $T$ is
diagonal in the continuous basis form a subalgebra that can be
easily manipulated.

The twist matrix $C$ also has a simple form in the continuous basis, as it
only interchanges the eigenvectors $\keto{\pm \ka}$. With the standard
definition of the eigenvectors it is given in this subspace by the matrix
\begin{equation}
C_{-\ka,\ka}=\mat{\phantom{-}0 & -1\\-1 & \phantom{-}0}.
\end{equation}
This means that while the matrices defining the string vertex are not mutually
diagonalizable and in particular are not diagonal in the continuous basis,
they are block diagonal in this basis with two by two blocks.
Again, block diagonal squeezed states form a subalgebra~\cite{Fuchs:2002zz}.
Within this subalgebra
the operations involving infinite matrices reduce to manipulating
two by two matrices and integrating over $\ka$.

Let $\keto{n}$ be the natural discrete basis states, i.e.,
\begin{equation}
\keto{n}=(\underbrace{0,..,0,1}_{n},0,..)\,.
\end{equation}
The transformation between the discrete $\keto{n}$ basis and the
continuous $\keto{\ka}$ basis is given by
\begin{equation}
\keto{\ka}=\sum_{n=1}^{\infty} v_n^\ka \keto{n},\qquad
\keto{n}=\int_{-\infty}^{\infty}d\ka\, v_n^\ka \keto{\ka}.
\end{equation}
The transformation coefficients $v_n^\ka$ obey the usual
unitarity properties\footnote{Some other properties of these
coefficients are collected in the appendix of~\cite{Erler:2003eq}.}
\begin{align}
\label{vnvnDelta}
\sum_{n=1}^{\infty} v_n^\ka v_n^{\ka'} &= \delta(\ka-\ka')\,,\\
\int_{-\infty}^{\infty} d\ka \,v_n^\ka v_m^\ka &= \delta_{nm}\,,
\end{align}
and are defined by the generating function,
\begin{equation}
f_\ka(z)=\sum_{n=1}^\infty\frac{v_n^\ka}{\sqrt{n}}z^n
        =\frac{1}{\sqrt{\N(\ka)}}\frac{1-
e^{-\ka \tan^{-1}z}}{\ka}\,,\qquad
\N(\ka)=\frac{2}{\ka}\sinh\big(\frac{\ka\pi}{2}\big)\,.
\end{equation}
Transforming the creation and annihilation operators to the new
basis results in the commutation relation
\begin{equation}
\label{aKaaDagKa}
[a_\ka,a_{\ka'}^\dag]=\delta(\ka-\ka')\,.
\end{equation}

The continuous basis proved to be useful for dealing with many
questions. In particular, it was used in~\cite{Douglas:2002jm} to formulate
the continuous Moyal representation of the star algebra. Analytical
calculation of the tension ratio was performed in~\cite{Okuyama:2002tw} while
in~\cite{Okuyama:2002yr} the equivalence of two definitions of the
kinetic operator of vacuum string field theory was proven\footnote{A-priori
there could have been many ways for defining the kinetic operator of vacuum
string field theory. Two possible ``canonical'' definitions were proposed
in~\cite{Hata:2001sq,Gaiotto:2001ji}. While the former paper used
implicit expressions \`a la Kostelecky-Potting~\cite{Kostelecky:2000hz},
the latter used explicit simple expressions. Both expressions take simple
(and identical) form in the continuous basis.}.
Furthermore, the continuous basis is strongly related to wedge states.
Of all surface states, wedge states are the only ones whose matrix $T$ is
diagonal and the so called (hybrid) butterfly
states~\cite{Gaiotto:2001ji,Schnabl:2002ff,Gaiotto:2002kf,Fuchs:2002zz}
are the only (other) ones with a
block diagonal matrix~\cite{Fuchs:2004xj,Uhlmann:2004mv}.

There are also some difficulties related to the continuous basis.
The Virasoro generators take a very singular form when
expressed in this basis. They are more singular than delta
functions~\cite{Douglas:2002jm,Erler:2002nr}. In~\cite{Fuchs:2002wk}
it was demonstrated that they can nevertheless be expressed by
delta functions whose arguments are complex. It was explained in
this paper how to deal with such expressions\footnote{Another
expression for some of the Virasoro generators was given
in~\cite{Belov:2002te}.}.

Another problem one may encounter with the continuous basis is
related to the calculation of normalization factors. It is quite
common that a state has a vanishing normalization in the matter sector
and an infinite one in the ghost sector or vice versa. Thus, in this
type of calculations one has to introduce the ghost sector and properly
regularize both sectors, in order to get a sensible
result\footnote{This state of affairs does not imply by itself
that the string field exists in the limit where the regularization
cutoff is removed~\cite{Moore:2001fg}. However, when dealing
only with states that are well defined in the CFT no problem
of principle should arise.}.
The ghost sector can be incorporated either using the $bc$ system
or its bosonized form. In the first case the string vertices
are given by a product of ghost zero modes, which multiplies a
zero-ghost-number squeezed state. A ghost squeezed state of zero
ghost number is necessarily of the form $\exp(b S c)\ket{0}$.
It is possible to define the continuous modes $b_\ka,c_\ka$, in a
manner similar to~(\ref{aKaaDagKa}). It turns out that the one-vertex, i.e.,
the identity string field, takes a very singular form in this
representation. It was not clear how to overcome this problem.
Therefore, the bosonized ghost approach was usually used within
the continuous basis.

In the bosonized ghost formalism the field $\phi$ is added to the 26
spatial bosons. This field is a linear dilaton whose total central
charge equals $-26$. The string vertices related
to this field have the same quadratic terms defining them, as the
matter sector vertices, but somewhat different linear terms.
When trying to evaluate the inner product of two wedge states
in the oscillator representation with bosonized ghosts, erroneous
results were encountered~\cite{Belov:2002pd}.
Namely, it was found that the result is not unity.

From CFT considerations it is clear that the inner product of two
arbitrary surface states (wedge states in particular) should equal
unity~\cite{LeClair:1989sp}\footnote{Of course, in order to get a
non-zero result the ghost zero modes should be saturated. Also,
in the inner product an infinite factor comes from the momentum
conservation delta function, which as usual represents the infinite
volume of space-time. In our current discussion it is assumed
that all zero modes (bosonic and fermionic) were taken care of.
In the matter sector this simply amounts to disregarding the infinite
volume factor. In the ghost sector, this can be achieved by
inserting the zero modes $c_{-1}c_0 c_1$
on either one of the vacua, or between the two states. The exact
choice does not change the result. This can be seen from the identity
$[L_0,c_{-1}c_0 c_1]=0$, the requirement that the combination
$c_{-1}c_0 c_1$ must remain in order to obtain a non-trivial result,
together with the fact that no constants appear
while commutating the Virasoro operators, since $c=0$.
While this shows that~(\ref{SSinnerProd}) indeed holds, the
prescription for the insertion is somewhat arbitrary. Everything
appears much more systematic when we consider string vertices, as we
describe next, instead of the inner product of states.
The string vertices explicitly depend on the ghost zero modes
and their descent relations lead to the same identities.},
\begin{equation}
\label{SSinnerProd}
\braket{S}{V}=\bra{0}e^{\sum_{n=2}^\infty s_n L_n}
   e^{\sum_{m=2}^\infty v_m L_{-m}}\ket{0}=1\,.
\end{equation}
Here $s_n,v_m$ are the coefficients defining the two surface
states $S,V$ according to~(\ref{surfS})
and the final result comes from using the Virasoro algebra
with central charge zero in order to move all positive operators
to the right and all negative ones to the left.
Moreover, since (the integer) wedge states are the surfaces on
which the string vertices are defined,
the above expression is related to the descent relations
of the string vertices,
\begin{equation}
\braket{V_{n+1}}{V_1}=\bra{V_n}\,.
\end{equation}
These relations should hold, without additional normalization
factors, to ensure the consistency of string field theory.

Let us concentrate on evaluating the inner product of the
wedge states $\ket{1}$ and $\ket{3}$. This gives the
fundamental descent relation,
\begin{equation}
\braket{V_1}{V_3} = \gamma_{13} \ket{V_2},
\label{DescRel}
\end{equation}
where we introduced the normalization factor that should equal
unity from CFT considerations.
Squeezed state algebra reveals~\cite{Kostelecky:2000hz},
\begin{align}
\nonumber
\gamma_{13}=\det(\One-& C V_3^{11})^{-\frac{d+1}{2}}\cdot\\ \cdot
\nonumber
  \exp\bigg(\frac{9}{4}\Big(&
    -\frac{1}{2}\brao{V_{30}^{g11}}(\One-C V_3^{11})^{-1} C
       \keto{V_{30}^{g11}}
   +\brao{V_{10}^{g\vphantom{1}}}(\One-V_3^{11} C)^{-1}
      \keto{V_{30}^{g11}}\\
\label{ga13}
&-\frac{1}{2}\brao{V_{10}^{g\vphantom{1}}}(\One-V_3^{11} C)^{-1}
   V_3^{11} \keto{V_{10}^{g\vphantom{1}}}
     -\frac{1}{2}V_{300}^{g11}\Big)\bigg).
\end{align}
Here, the twist matrix comes from its part in defining the
one-vertex~(\ref{V1Def}), the factor of $d+1=27$ in the determinant
comes from the quadratic terms of the $d$ space-time dimensions
and from the bosonized ghost direction and the terms in the exponent
all come from linear terms of the bosonized ghost, which are absent
for the space-time directions. We marked here all ghost terms with a
superscript $g$, which we omit later in order to improve readability.
The first term in the exponent comes from terms in the three-vertex
that are
linear in momentum. This term is absent in the matter sector
since the momentum there is zero. Saturation of the ghost zero mode
requires $p^g=\pm \frac{3}{2}$, which results in the factor of
$\frac{9}{4}$ common to all the terms in the exponent.
The factor $\keto{V_{10}^{g}}$ is the linear term of the one-vertex.
Such a factor could not exist in the matter sector,
since the momentum there is zero. Both the determinant and the
exponent formally diverge.

In order to be able to compare the divergence of the exponent
and determinant we use the usual trick of expressing the determinant
in terms of a trace of the logarithm of the matrix,
\begin{align}
\nonumber
&\det(\One-C V_3^{11})=
  \exp\!\Big(\!\tr\big(\log(\One-C V_3^{11})\big)\!\Big)\!=
  \exp\!\Big(\sum_{n=1}^\infty\brao{n}
       \log(\One-C V_3^{11})\keto{n}\!\Big)\\&=
\label{TooSimpleReg}
  \exp\!\Big(\sum_{n=1}^\infty\int d\ka\,d\ka'\,
    \braketo{n}{\ka}\brao{\ka}
       \log(\One-C V_3^{11})\keto{\ka'}\braketo{\ka'}{n}\!\Big)\\&=
\nonumber
  \exp\!\Big(\sum_{n=1}^\infty\int d\ka\,
   \braketo{n}{\ka}\log(1-\mu(\ka))\braketo{\ka}{n}\!\Big)\!=
  \exp\big(\delta(0)\int d\ka\,\log(1-\mu(\ka))\big).
\end{align}
Here, we wrote the trace explicitly in the second equality. In the
third one a unity was inserted twice in order to pass to the
continuous basis. Next, we used~(\ref{muDef}) to get $\delta(\ka-\ka')$.
Then, a sum over $n$ left us with another delta function $\delta(0)$,
which is the manifestation of the divergence. Similar expressions
are obtained also for the exponent part of~(\ref{ga13}).

In order to regularized the $\delta(0)$, a cutoff on $n$
was imposed on~(\ref{vnvnDelta}) for $\ka=\ka'$. This cutoff $l$,
called the level\footnote{Note that while resembling level truncation,
this is not the same. In level truncation one keeps all
fields up to a given level, here we effectively keep all fields, which are
built from oscillators up to a given level, i.e., we keep an infinite
number of fields. This approximation is occasionally referred to as
``oscillator level truncation'' in order to distinguish it from
the usual level truncation.} gives the
prescription~\cite{Rastelli:2001hh,Belov:2002pd,Fuchs:2002wk,Belov:2002sq},
\begin{equation}
\label{BadContReg}
\delta(0)\rightarrow
   \frac{2\log l-\psi(\frac{i \ka}{2})-\psi(-\frac{i \ka}{2})}{4\pi}\,,
\end{equation}
where $\psi$ is the polygamma function. It turns out that while the
infinite ($\log l$) part cancels out exactly for $d=26$, the finite part
does not cancel out. Rather, it gives~\cite{Fuchs:2003wu},
\begin{equation}
\gamma_{13}= 3^{3/8} e^{\frac{3}{2} \gamma+36\zeta'(-1)}
\Big(\frac{\Gamma \big(\frac{1}{3}\big)}{\sqrt{\pi}}\Big)^9
 \approx 0.382948\,,
\end{equation}
where $\Gamma$ is the gamma function, $\gamma$ is Euler's constant
and $\zeta'$ is the derivative of the zeta function.

A possible resolution of this problem was suggested
in~\cite{Belov:2003df,Belov:2003qt}. According to this proposal, the
normalization constant $\gamma_{13}$ is not an artifact but a correct result.
All string vertices should have normalization factors, i.e,
$\ket{V_n}\rightarrow Z_n\ket{V_n}$. The normalization factors
supposedly originate from the partition function on a surface
with a conical singularity of the sort used for the definition
of the vertex. The two-vertex is defined over a surface without a conical
singularity and so carries a trivial normalization, $Z_2=1$.
Hence, the descent relation holds, provided that
$\gamma_{13} Z_1 Z_3=1$.
Similar relations should also hold for the other descent relations.
This proposal have some problems. First, these partition functions
were nowhere directly calculated and no other test was performed
in order to show the consistency and universality of the proposal.
More importantly, the proposal
is inconsistent with~(\ref{SSinnerProd}).

One faces several options for an alternative clarification of the source
of the anomaly. It can be related to the use of the continuous basis,
to the use of the oscillator level truncation needed for defining the
normalization or to the use of the bosonized form of the ghosts.
In~\cite{Fuchs:2005ej}\footnote{See also~\cite{Aref'eva:2006pu}
and the recent papers~\cite{Bonora:2007tm,Bonora:2008rt}. \ATFV
the study of the last two papers was completed
in~\cite{Bonora:2009he,Bonora:2009hf}, leaving us with a reliable fermionic
oscillator formalism,
both in the continuous representation and in the discrete one.
In cubic superstring field theory this problem was addressed
in~\cite{Aref'eva:2007ka}.} these possibilities were studied.
It was found that when carefully treated, the discrete basis
gives the same normalization constant as the continuous one, suggesting
that the use of the continuous basis is not the source of the problem.
Also, it was found that while the use of the ghosts in their $bc$
form gives a different normalization factor, this factor is also
not equal to unity. All that seems to suggest that oscillator
level truncation is a bad
regularization. Without a proper resolution of this problem
it seems that one cannot really use the oscillator basis as a
complete framework for the study of string field theory.

\subsection{Schnabl's operators and the resolution of normalization anomalies}
\label{sec:oscillatorRecent}

Schnabl's construction uses the CFT language. Despite
that and regardless of the problems with the oscillator formalism,
one would like to formulate his solution in the oscillator language as well.
An obvious reason would be the fact that much advance in string field theory
was achieved in the past in this formalism. Another reason is that
Schnabl's solution is based on wedge states, which have a very simple
representation in the oscillator formalism. In particular, they are diagonal
in the continuous basis. Thus, this representation may turn out to be
as simple as the CFT one. Alternatively, it may happen that it
may teach us something about the oscillator formalism itself.
As the primary building blocks of the solution are
the operators $\LL_0,\LL_0^\dag$, the first step towards an oscillator
representation is to cast these operators into an oscillator form,
preferably in the continuous basis~\cite{Fuchs:2006an}.

An obvious obstacle for this program comes from the fact that Schnabl's
operators are combinations of Virasoro operators, which take a very singular
form in the continuous basis. Remarkably, it turns out that exactly for the case of
Schnabl's operators the highly singular (complex delta function) terms conspire
to vanish. The other terms add nicely to give a mild singularity, which
corresponds to an almost (block) diagonal form of the operators.
The defining matrices in the continuous basis are proportional to the delta
function and to its derivative,
\begin{align}
\label{Lcont}
\LL_0 = \int_{-\infty}^\infty d\ka d\ka'
  \bigg(\Big(&
    \frac{\ka\pi}{4}\coth\big(\frac{\ka\pi}{2}\big)\delta(\ka-\ka')+
    \frac{\ka+\ka'}{2}\delta'(\ka-\ka')\Big) a_\ka^\dagger a_{\ka'} \\+&
  \frac{\pi\delta(\ka+\ka')}{2\N(\ka)} a_\ka a_{\ka'}
  \bigg).
\end{align}
Similar results were also found for the bosonized ghost sector.
However, when trying to verify the basic algebra~(\ref{SL2Algebra}) one seem
to gets an anomaly\footnote{This is of course the result at the critical
dimension. For $d\neq 26$ the central term diverges logarithmically as a
function of the level $l$.},
\begin{align}
[\LL_0,\LL_0^\dagger]\stackrel{?}{=}\LL_0+\LL_0^\dagger+\frac{15}{4}\,.
\end{align}
This anomaly is similar to the anomaly encountered in the evaluation
of the descent relations, but simpler, since it involves only linear
expressions. This simple form of the normalization anomaly was used
in~\cite{Fuchs:2006gs} in order to reconsider the issue of regularization in
the oscillator formalism. 

There are some features common to good regularization schemes. First, they
tend to be analytic in the regularizing parameter. Second, they tend to
leave intact (or to simply deform) the main symmetries of the theory considered.
In the case at hand it is clear that oscillator level truncation does not
enjoy these qualities. The cutoff is performed using a step function, which
is obviously not an analytic function and there is no treatment of the
relevant symmetries.
One may wonder, at this stage, how is that different from the usual level
truncation. One difference is that for level truncation the cutoff commutes
with the Virasoro algebra.
Infinities may arise in relations such as~(\ref{SSinnerProd}) when the
central charge is non-zero. Thus, the Virasoro symmetry is the relevant one
in our case. While it seems that there is no regularization that keeps the
Virasoro algebra intact, it is nevertheless possible to deform it in a way
that does not alter the role of the central charge. Moreover, this deformation
depends analytically on a parameter. The deformation is performed by defining
(for $s\leq 1$)
\begin{equation}
L_n^s = s^{|n|}L_n\,,
\end{equation}
which for $s\rightarrow 1$ reduces to the non-regularized case.
The deformed operators satisfy the regularized Virasoro algebra
\begin{equation}
\label{VirasoroAlgebra}
[L_n^s, L_m^s] = s^{|n|+|m|-|n+m|}
    \left( (n-m)L_{n+m}^s + \frac{c}{12}(n^3-n)\delta_{n+m} \right).
\end{equation}
For the oscillators the regularization is similarly applied,
\begin{equation}
a_n^s = s^n a_n\,,\qquad a_n^{\dag\,s} = s^n a_n^\dag\,.
\end{equation}

\begin{figure}
\center{\includegraphics[width=13cm]{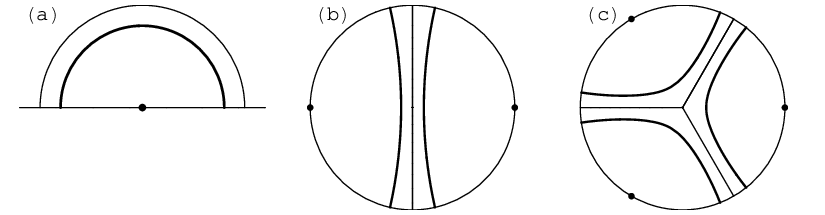}}
\caption{The geometric effect of the regularization on:
(a)~The definition of the state. (b)~The two-vertex. (c)~The three-vertex.}
\label{fig:RegVertices}
\end{figure}

A very simple geometric interpretation can be given to the regularization
if one thinks of the positive Virasoro modes as related to bra states and
of the negative ones as related to ket states. In such a case the power of
the cutoff is simply the conformal weight of the state created
and so the regularization simply amounts to squeezing the surface by the
factor $s$, i.e., acting on the state by the scale
transformation\footnote{Recall that $\xi$ are the coordinates on
the upper half plane.}
\begin{equation}
f_s(\xi)=s \xi\,.
\end{equation}
This illustrates how the singularity is resolved in yet another context.
The string vertices are singular surfaces, since they induce gluings,
i.e., delta function interactions. This translates also to the
singularity of the relevant surfaces in the oscillator formalism.
The regularization that we are using resolves exactly this delta function.
We present the effect of this regularization for the definition of a single
state and for the interaction using the two- and the three-vertex
in figure~\ref{fig:RegVertices}.

Having established the good formal properties of the regularization, the
next task is to check whether it indeed works. The simplest possibility
is to check numerically the descent relations. Indeed, a numerical
evaluation of $\gamma_{13}$ strongly suggests that it equals unity
when using the above regularization. Next, we would like to have an
expression describing the regularization in the continuous basis, where
analytical results are usually easier to obtain.
The problem with the level truncation regularization in the continuous
basis can be traced to the manipulations performed in~(\ref{TooSimpleReg}).
There, we ended with a $\delta(0)$ that we had to regularize. However, some
of the manipulations used to get to the final expression also used
delta functions. It seems inconsistent to regularize one delta function
and not the other. Indeed, the $s$-regularization is effectively
performed by regularizing the identity operator inserted between
various terms that contain oscillators,
\begin{equation}
\One_s \equiv \sum_{n=1}^\infty \keto{n}s^{2n} \brao{n}\,.
\end{equation}
Let us switch to the variable
\begin{equation}
\ep=1-s\,.
\end{equation}
Expressing $\One_{1-\ep}$ in the continuous basis gives
\begin{equation}
\One_{1-\ep}=\sum_{n=1}^\infty \int d\ka\, d\ka'\,\keto{\ka}\braketo{\ka}{n}
  (1-\ep)^{2n} \braketo{n}{\ka'}\brao{\ka'}=\int d\ka\, d\ka'\,\keto{\ka}
    \rho_s(\ka,\ka')\brao{\ka'},
\end{equation}
where we defined the regularized kernel
\begin{equation}
\rho_\ep(\ka,\ka')=\sum_{n=1}^\infty v_n^\ka v_n^{\ka'} (1-\ep)^{2n}\,.
\end{equation}
Evaluating this kernel is neither simple nor illuminating. Thus, we simply
state the final result,
\begin{align}
\label{GoodContReg}
\rho_\ep(\ka,\ka')=\frac{1}{\ka \ka'\sqrt{\N(\ka)\N(\ka')}}
     \left( \frac{e^{i \frac{\ka-\ka'}{2} \log \ep}}
       {B(\frac{i \ka'}{2},-\frac{i \ka}{2})}+
       \frac{e^{-i \frac{\ka-\ka'}{2} \log \ep}}
       {B(\frac{i \ka}{2},-\frac{i \ka'}{2})}\right),
\end{align}
where $B$ is the Beta function.
This expression is correct up to $\cO(\ep)$ corrections. However, since the
singularity is of the order of $\log (\ep)$, these corrections can be
safely neglected.
Using some residue calculus while consistently neglecting terms of the
form $\ep^n \log(\ep)$, one can see that these kernels have some nice
properties, such as
\begin{equation}
\int d\ka'\, \rho_\ep(\ka,\ka')\rho_{\ep'}(\ka',\ka'')=
   \rho_{\ep+\ep'}(\ka,\ka'')\,.
\end{equation}
This may seem like a surprising identity. However, it is simply the
(approximate) continuous basis form of the trivial identity
\begin{equation}
\One_s \One_{s'}=\One_{s s'}\,,
\end{equation}
which is exact for arbitrary values of $s,s'$.
It is important to note that~(\ref{GoodContReg}) reduces
to~(\ref{BadContReg}) in the limit $\ka\rightarrow \ka'$, with
$\ep \sim l^{-1}$. Thus,~(\ref{BadContReg}) can be considered as
the kernel~(\ref{GoodContReg}) restricted to its diagonal and
multiplied by a $\delta(\ka-\ka')$ factor. It is exactly this
extra delta function factor that this regularization helps to resolve.

Using the result for $\rho_\ep(\ka,\ka')$ it was analytically proved
in~\cite{Fuchs:2006an} that Schnabl's algebra holds in the continuous basis,
without a central term, as it should. The nonlinear expressions for the
inner product $\braket{n}{m}$\footnote{This is a generalisation to
arbitrary wedge states of our previous discussion. Recall that the case
$n=1$ and $m$ an integer is equivalent to the descent relations.}
were evaluated analytically as a power series in
$n,m$ and it was shown analytically that the lowest orders of the series
vanish. More terms were shown to vanish using a numerical evaluation.
Finally, the most potentially singular inner product, namely $\braket{1}{1}$
was proven to equal unity.
All that gives high credibility to the advocated regularization, thus
allowing for the use of the oscillator basis as a practical and reliable
formalism for string field theoretical calculations.

\section{Superstring field theory}
\label{sec:SUSY}

There are several versions of open covariant superstring field theory,
most of which are defined in the RNS approach to superstring field theory.
The only exception being the pure-spinor open string field theory
of~\cite{Berkovits:2005bt}, which we do not discuss here.
The RNS theories are most naturally defined using the fermionized version
of the superghosts~\cite{Friedan:1985ge}.
We start this section by introducing the fermionization and the related
concept of pictures in~\ref{sec:SUSYintro}. Next, in~\ref{sec:SUSYtheories}
we describe the various superstring field theories. The recent results are
described in~\ref{sec:Map}, where a mapping of solutions
between string field theories is studied and in~\ref{sec:SUSYsolutions}
where analytical solutions of superstring field theories are described.

\subsection{Superstring variables, pictures and conventions}
\label{sec:SUSYintro}

In the RNS\footnote{We focus exclusively on the NS sector of the
open superstring.
The inclusion of the Ramond sector may also be possible, but is
irrelevant to the results that we want to present here.}
formulation of the superstring, in addition to the
world-sheet-scalar space-time-vector $X^\mu$ there is also a
world-sheet-fermion space-time-vector $\psi^\mu$.
These fields are referred to as matter fields.
Supersymmetry is
manifest on the world-sheet. This results in the elevation of
the conformal symmetry to a superconformal one.
This symmetry is related to the (matter) energy-momentum tensor
$T_m$ (we work here in the $\al'=2$ convention)\footnote{Operators
inserted at the same point carry implicit normal ordering.},
\begin{equation}
T_m(z)=-\frac{1}{2} \partial X^\mu \partial X_\mu(z)-
\frac{1}{2}\psi^\mu \partial \psi_\mu(z)\,,
\end{equation}
as well as to the
superconformal (matter) generator $G_m$,
\begin{equation}
G_m(z)=i\psi^\mu \partial X_\mu(z)\,.
\end{equation}

Fixing the superconformal symmetry introduces in addition to the fermionic $bc$
ghosts with central charge $c_{bc}=-26$ also the bosonic $\beta \gamma$
superghosts with central charge $c_{\beta\gamma}=11$. It is convenient to
fermionize the superghosts. The first step in this direction consists of
introducing a field $\phi$ in order to fermionize the superghost-number
current as
\begin{equation}
\partial \phi(z)=\beta \gamma (z)\,,
\end{equation}
with the field $\phi$ satisfying
\begin{equation}
\phi(z)\phi(w)\sim -\log(z-w)\,.
\end{equation}
One can see that the superghost number is shifted by $q$ units under the
action of the operator
$e^{q \phi}$. Thus, the $\beta, \gamma$ operators should be related to the
operators $e^{\pm \phi}$.
We define the operator $e^{q \phi}$ to carry picture number $q$. From the
discussion above it may seem that this is just the superghost number.
However, there is some freedom in the definition that we are about to
exploit. As a result, picture number turns out to be an independent
quantum number.

One can use the Sugawara construction~\cite{Sugawara:1967rw} and define
the energy momentum tensor of the superghost current,
\begin{equation}
\label{Tphi}
T_\phi(z)=-\frac{1}{2}\partial\phi\partial\phi - \partial^2\phi\,.
\end{equation}
This definition gives the correct OPE for the energy momentum tensor and the
current.
Yet, it defines a system with central charge $c_\phi=13$, so this
cannot be the whole story. Also, one can check that the
operators $e^{\pm \phi}$ anticommute, while the $\beta,\gamma$ system
is bosonic. The OPE's of $e^{\pm \phi}$ are also not quite what is needed.
To compensate for the lost central charge and the wrong statistics
one appends to $\phi$ the conjugate fermions $\eta,\xi$
with conformal dimensions $(1,0)$ respectively.
The energy momentum tensor of this system is given by
\begin{equation}
\label{Txieta}
T_{\eta\xi} = -\eta\partial\xi\,.
\end{equation}
The complete fermionization formulas then read\footnote{A co-cycle is
implicitly assumed when writing $e^{\pm \phi}$, in order for
these operators to anticommute also with the other anti-commuting
variables of the theory.},
\begin{equation}
\label{fermiExp}
\beta=e^{-\phi}\partial \xi\,,\qquad \gamma=\eta e^\phi\,.
\end{equation}
Now, we have to assign picture number to the new variables. We want
all fields not in the fermionized ghost sector,
including the original $\beta,\gamma$, to have zero picture
number.
This would imply the neutrality of the BRST charge $Q$.
Thus, we assign the picture numbers as $n_p(\xi)=1$,
$n_p(\eta)=-1$. This is indeed a new quantum number.
We write down for convenience the picture number, the ghost
number, the conformal weight and the parity of some
fields including the BRST current $J_B$, its ``inverse'' $P$
and the picture changing operators $\X,Y$
(introduced below) in table~\ref{tab:ghPic}.
\begin{table}
\begin{center}
\begin{tabular}{|c||c|r|r|c|}
\hline
operator & $h$ & $n_g$ & $n_p$ & parity\\
\hline
$\partial X^\mu$ & 1 & 0 & 0 & 0\\
$\psi^\mu$ & $\frac{1}{2}$ & 0 & 0 & 1\\
$b$ & 2 & -1 & 0 & 1\\
$c$ & -1 & 1 & 0 & 1\\
$\eta$ & 1 & 1 & -1 & 1\\
$\xi$ & 0 & -1 & 1 & 1\\
$e^{q\phi}$ & $-\frac{q(q+2)}{2}$ & 0 & $q$ & $q \mod 2$\\
$\beta$ & $\frac{3}{2}$ & -1 & 0 & 0\\
$\gamma$ & $-\frac{1}{2}$ & 1 & 0 & 0\\
$J_B$ & 1 & 1 & 0 & 1\\
$P$ & 0 & -1 & 0 & 1\\
$\X$ & 0 & 0 & 1 & 0\\
$Y$ & 0 & 0 & -1 & 0\\
\hline
\end{tabular}
\end{center}
\caption{The conformal weight $h$, ghost number $n_g$, picture number
$n_p$ and parity of the fields relevant to us.}
\label{tab:ghPic}
\end{table}
The normal ordering of some of the fields is given explicitly
by\footnote{Note that the order of terms in~(\ref{fermiExp}) is important for the
consistency of~(\ref{OPEetaxi}) and~(\ref{OPEphi})
with~(\ref{OPEbetagamma})
(recall that $e^{\pm \phi}$ are anti-commuting).},
\begin{subequations}
\label{OPEeqs}
\begin{align}
\partial X^\mu(z)\partial X^\nu(w)=& :\partial X^\mu(z)\partial X^\nu(w):-
  \frac{\eta^{\mu \nu}}{(z-w)^2}\,, \\
\psi^\mu(z) \psi^\nu(w)=& :\psi^\mu(z) \psi^\nu(w) :
  +\frac{\eta^{\mu \nu}}{z-w}\,,\\
b(z)c(w)=& :b(z)c(w):+\frac{1}{z-w}\,,\\
\label{OPEbetagamma}
\beta (z)\gamma(w)=& :\beta (z)\gamma(w):-\frac{1}{z-w}\,,\\
\label{OPEetaxi}
\eta(z)\xi(w)=& :\eta(z)\xi(w):+\frac{1}{z-w}\,,\\
e^{a \phi}(z) e^{b \phi}(w)=&\, \frac{:e^{a \phi}(z) e^{b \phi}(w):}{(z-w)^{a b}}\,.
\label{OPEphi}
\end{align}
\end{subequations}

In the fermionization equations~(\ref{fermiExp})
the field $\xi$ appears only through its
derivative. That is, the zero mode $\xi_0$ is not needed in order to describe
the $\beta,\gamma$ system. Its inclusion would double the Hilbert space,
since it is a two-level fermionic variable. It is common to refer to the Hilbert
space without the zero mode as the ``small Hilbert space'' and to the doubled
space with the $\xi_0$ included as the ``large Hilbert space''.
In the large Hilbert space CFT correlators are normalized as
\begin{equation}
\label{LargeSpaceCFT}
\left<\xi \partial^2 c\partial c c e^{-2\phi} \right >_\text{Large}=2\,.
\end{equation}
The correlator is non-zero only for ghost number $n_g=2$ and picture number
$n_p=-1$.
Since the small Hilbert space is defined by the absence of the zero mode
$\xi_0$, the factor of $\xi$ is omitted in defining correlators,
\begin{equation}
\label{SmallSpaceCFT}
\left<\partial^2 c\partial c c e^{-2\phi} \right>_\text{Small}=2\,.
\end{equation}
Now, the total picture number $n_p=-2$ is needed for a non-trivial result
and the ghost number should be $n_g=3$, as in the bosonic
case\footnote{One way to think of it is to say that the left vacuum
in~(\ref{SmallSpaceCFT}) equals the left vacuum of~(\ref{LargeSpaceCFT}),
with $\xi_0$ acting on it.}.

The existence of the superconformal algebra implies that a BRST charge $Q$
should be defined. The BRST charge of a constrained
system that forms a (super-)Lie algebra contains two types of terms.
First, the ghost multiplied by the constraints are added. To them one
has to add couplings of two ghosts and one antighost multiplied by the
algebra structure constants.
All that can be nicely summed up, writing the BRST charge as an integral
of the BRST current $J_B$,
\begin{equation}
Q = \frac{1}{2\pi i}\oint dz J_B(z)\,,
\end{equation}
which is defined up to a total derivative.
In the case at hand, the set of constraints is encoded
in the (matter) energy momentum tensor $T_m$, whose ghost and anti-ghost are
the $c$ and $b$ respectively and by the superconformal (matter)
generator $G_m$, with $\gamma$ and $\beta$ as the ghost and antighost.
Thus, the first part of the BRST current is
\begin{equation}
J_B^{(0)}(z) = c T_m + \gamma G_m\,.
\end{equation}
Adding the terms that encode the super-Virasoro algebra one gets
\begin{equation}
\begin{aligned}
\label{JbSUSY}
J_B(z) &= c T_m + \gamma G_m + c\partial c b 
   -\frac{c}{2}(3\beta \partial \gamma+\partial \beta \gamma)- b\gamma^2\\
    &= c (T_m + T_{\eta\xi}+T_\phi + \partial c b) + \eta e^{\phi}G_m
      -b\eta\partial\eta e^{2\phi}\,.
\end{aligned}
\end{equation}
In the second line we converted the expression to the more convenient
variables $\phi,\eta,\xi$ (recall~(\ref{fermiExp})) and the
energy-momentum tensors of the fermionized ghost systems are
given by~(\ref{Tphi}) and (\ref{Txieta})\footnote{Note that
normal ordering is implicit in both lines of~(\ref{JbSUSY}).
In order to convert, one has to use~(\ref{OPEeqs}) in order to
first undo the $\beta \gamma$ normal ordering and then introduce
the ones of the $\phi,\eta,\xi$ systems.}.

An interesting property of the BRST charge in this formulation
is that it defines a trivial cohomology in the large Hilbert
space.
This fact was proven in~\cite{NarganesQuijano:1988gb}, by defining the
operator\footnote{
Alternatively, one can prove it using the similarity transformation,
presented in~\cite{Acosta:1999hi}, that generates $J_B$ from the last term
of~(\ref{JbSUSY}). This also serves as an elegant proof of the nilpotence
of $Q$.}
\begin{equation}
\label{P}
P(z)=-c\xi\partial\xi e^{-2\phi}(z)\,,
\end{equation}
which serves as a contracting homotopy for $Q$, i.e., it obeys,
\begin{equation}
\label{QP1}
[Q,P(z)]=1\,.
\end{equation}
This relation implies that any $Q$-closed state $\cO(0)\!\ket{0}$ is exact,
since it can be written as
\begin{equation}
\label{PQexact0}
\cO(0)\!\ket{0}=Q\big(P(0) \cO(0)\!\ket{0}\big)\,.
\end{equation}
One may worry that singularities may appear in the OPE of $P$ and
$\cO$. However, in such a case one may modify~(\ref{PQexact0}) to
\begin{equation}
\cO(0)\!\ket{0}=Q\Big(\frac{1}{2\pi i}
    \oint \frac{dz}{z} P(z)\cO(0)\!\ket{0}\Big)\,.
\end{equation}
No singularities can remain in the above equation and its proof
is as straightforward.

An important feature of superstring theory is the abundance of vertex
operators adequate for the description of physical states.
In fact, even for the bosonic case there is more than one vertex
operator describing a given state. There, there are two variants,
the integrated vertex operator\footnote{By ``integrated vertex
operator'' we actually refer to the integrand of the vertex.}
$V_{\text{bos}}$ and the
unintegrated vertex operator $\hat V_{\text{bos}}$.
These vertex operators have to obey
\begin{align}
\label{QintegV}
[Q, V_{\text{bos}}]=&\partial \hat V_{\text{bos}}\,,\\
\label{QunintegV}
[Q, \hat V_{\text{bos}}]=&0\,.
\end{align}
The second equation, which implies that the unintegrated vertex is
closed, follows of course from the first one.
Integrating the total derivative in the r.h.s of~(\ref{QintegV})
implies that the vertex is closed in this
form as well.
Moreover, it follows from~(\ref{QintegV}) that adding a
(physically irrelevant) total derivative to $V_{\text{bos}}$
corresponds to adding a (physically irrelevant) exact term to
$\hat V_{\text{bos}}$.
A direct evaluation gives in the bosonic case\footnote{This is
a familiar expression. However, the relation~(\ref{QintegV}) is
the fundamental one, as it can be used even in formulations
with no $b,c$ system, such as the pure-spinor formulation of superstring
theory~\cite{Berkovits:2000fe,Berkovits:2002zk,Berkovits:2005bt}.},
\begin{equation}
\label{hatVcV}
\hat V_{\text{bos}}=c V_{\text{bos}}\,.
\end{equation}
It is not important which form of the vertex is used, as long as
there are exactly three unintegrated vertex operators to saturate the
three (fermionic) zero modes of the $c$ ghost. These zero modes are
in fact the reason behind the redundancy. Their fermionic nature
is the reason for having exactly two forms of each vertex operator.

It can be read from~(\ref{SmallSpaceCFT}) that (on the plane) the
superghosts have two zero modes (picture number) and one may
expects that again this will lead to a redundant description of the
vertex operators. However, as the superghosts are bosonic, one
may expect an infinite redundancy instead of the two-fold redundancy
related to the $bc$ system\footnote{In the bosonic case, the most natural
description in some sense is the one where the zero mode is fixed,
namely the unintegrated vertex. This is achieved by multiplication
by $c$. In order to fix a bosonic coordinate, as we have here, one
has to insert a factor of $\delta(\gamma)$. Luckily, this delta function
has a simple expression after fermionization, namely $e^{-\phi}$.
For this reason it is customary to refer to the $-1$ picture as
being a natural one. One still needs, of course, also the
representations of the vertex operators at other pictures.}.
The redundancy of the $bc$ system
also exists of course,
but one can treat it independently of that of the
superghosts\footnote{\label{foot:DP}
This is true for the plane. For higher genera
it is no longer possible to disentangle the
bosonic and fermionic zero modes and one should study the supermoduli
carefully. The endeavor of D'Hoker and Phong concentrated around
this issue. In particular, they showed~\cite{D'Hoker:2002gw}
that the picture changing operators, introduced below, should be
modified at genus two (see~\cite{Cacciatori:2008ay,Grushevsky:2008zm,Matone:2008td}
for interesting recent progress and~\cite{Morozov:2008wz} for a recent review).
One may hope that the framework of string field theory can be used as
a simple alternative description to the complicated study of
supermoduli spaces. To that end, a reliable formulation of
superstring field theory is needed, in which the Ramond sector
is also included.}.
Thus, each state of the superstring should correspond to one
integrated vertex operator $V^p$ and one unintegrated vertex
operator $\hat V^p$ at every integer value $p$ of the picture
number\footnote{Recall that we describe the
Neveu-Schwarz sector. In the Ramond sector the allowed picture
numbers $p$ are half-integers.}. These vertices should
obey~(\ref{QintegV}) and~(\ref{QunintegV}) separately for every
values of $p$,
\begin{align}
\label{QintegVSUSY}
[Q, V^p]=&\partial \hat V^p\,,\\
\label{QunintegVSUSY}
[Q, \hat V^p]=&0\,.
\end{align}
The relation~(\ref{hatVcV}) holds now only for specific
values of picture number.

Given an unintegrated vertex $\hat V^p$ one can construct from it
the vertex $\hat V^{p+1}$ using the following definition
\begin{equation}
\label{VpicChanged}
\hat V^{p+1}=[Q,\xi \hat V^p]\,.
\end{equation}
While the new vertex operator may seem to be exact, it is only so in the
large Hilbert space. In the physical, small Hilbert space, it is only
closed. This also shows that the only relevant part
of $\xi$ in this definition is its zero mode $\xi_0$, since the rest
gives rise to a truly exact part and so decouples from all
calculations of scattering amplitudes.
Using~(\ref{QunintegV}) one can deduce that~(\ref{VpicChanged}) can
equivalently be written as
\begin{equation}
\label{VpicChangedX}
\hat V^{p+1}=\X \hat V^p\,,
\end{equation}
where the picture changing operator $\X$ is given by\footnote{The
picture changing operator is usually represented by $X$. However,
since we tend to omit the space-time index $\mu$ from the scalars
$X^\mu$, we follow the example of~\cite{Ohmori:2001am} and use $\X$
instead.},
\begin{equation}
\label{picChange}
\X=[Q,\xi]=c\partial\xi+e^\phi G_m+e^{2\phi}b\partial \eta
   +\partial(e^{2\phi}b \eta)\,.
\end{equation}
For the integrated vertex we define
\begin{equation}
\label{unVpicChanged}
V^{p+1}=[Q,\xi V^p]+\partial (\xi \hat V^p)\,.
\end{equation}
The second, total derivative term does not contribute after integration.
It should nevertheless be added both for consistency
with~(\ref{QintegVSUSY}) as well as for obtaining a vertex that resides in
the small Hilbert space. This equation can be rewritten as
\begin{equation}
\label{unVpicChangedX}
V^{p+1}=\X V^p+\partial \xi \hat V^p\,.
\end{equation}

This procedure can be repeated ad infinitum.
In the calculation of scattering amplitudes
any set of representatives of the scattered states can be used, provided
the total picture number is such that the picture and ghost numbers
in~(\ref{SmallSpaceCFT}) are saturated~\cite{Friedan:1985ge}.
We illustrate the equivalence of different picture number distributions
in figure~\ref{fig:Picture}. One can still wonder whether the repeated
use of the picture changing scheme can produce
singularities\footnote{Note, that the products
in~(\ref{VpicChanged}),~(\ref{VpicChangedX}),
(\ref{unVpicChanged}),~(\ref{unVpicChangedX}) are simple OPE's, without
any normal ordering. Had we insisted on normal ordering in these
expressions, the argument described in figure~\ref{fig:Picture} would
have been invalidated.}, as it is easy to check that the OPE of $\X$
with itself has a double pole.
It turns out that these singularities correspond to terms that are exact
even in the small Hilbert space\footnote{For the integrated vertex they
should be exact only after integration. In particular, singularities
multiplying total derivatives may
pop-up.}~\cite{Horowitz:1988ip,Lian:1989cy}.
Hence, one can solve the problem with these singularities by point
splitting, e.g.,
\begin{equation}
\X(z)\hat V^p(z)\rightarrow \X(z+\ep)\hat V^p(z)\,.
\end{equation}
The terms, which become singular in the $\ep\rightarrow 0$ limit, decouple
from scattering amplitudes and can simply be dropped out from the definition
of the picture-changed vertex operators. Then, the limit
$\ep\rightarrow 0$ leads to consistent local vertex operators.

\newcommand{\redcircle} {\begin{picture}(18,18)(-4,-3)
  \color{Red}\circle*{6}
  \end{picture}}
\newcommand{\greencircle} {\begin{picture}(18,18)(-4,-3)
  \color{Green}\circle*{6}
\end{picture}}
\newcommand{\bluecircle} {\begin{picture}(18,18)(-4,-3)
  \color{Blue}\circle*{6}
\end{picture}}

\newcommand{\noxi} {\phantom{\xi}}

\newcommand{\jBanticlock} {\begin{picture}(0,0)(-9,-3)
  \circle{24}
  \put(6,-18){$j_B$}
  \put(2.6,-14.4){\vector(1,1){6}}
\end{picture}}

\newcommand{\jBclockwise} {\begin{picture}(0,0)(-9,-3)
  \circle{24}
  \put(6,-18){$j_B$}
  \put(12.6,-4.4){\vector(-1,-1){6}}
\end{picture}}

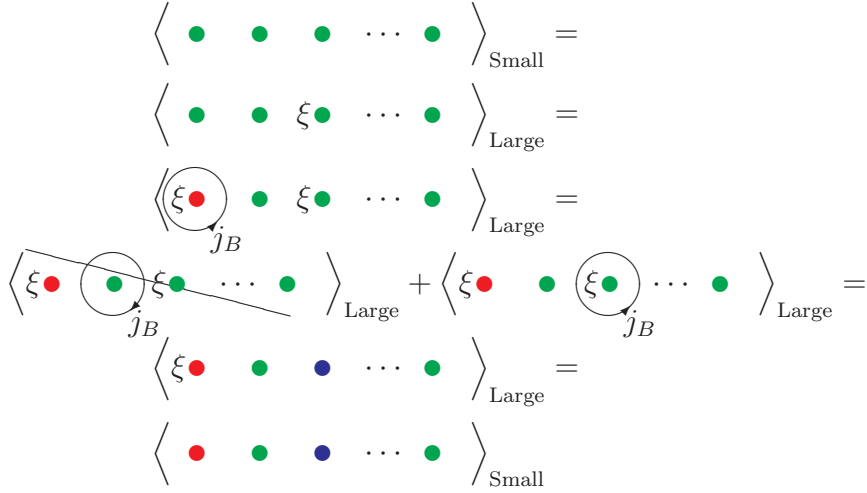
\begin{figure}
\begin{align*}
&\Big<
  \noxi\greencircle \noxi\greencircle
  \noxi\greencircle \cdots \noxi\greencircle
\Big>_{\text{Small}} = \\
&\Big<
  \noxi\greencircle \noxi\greencircle
  \xi\greencircle \cdots \noxi\greencircle
\Big>_{\text{Large}} = \\
&\Big<
  \jBanticlock\xi\redcircle
  \noxi\greencircle \xi\greencircle \cdots \noxi\greencircle
\Big>_{\text{Large}} = \\
\Big<
  \begin{picture}(0,0)(0,-16)
    \line(4,-1){100}  
  \end{picture}
  \xi\redcircle
  \jBclockwise\noxi\greencircle
&  \xi\greencircle \cdots \noxi\greencircle
\Big>_{\text{Large}} +
\Big<
  \xi\redcircle
  \noxi\greencircle
  \jBanticlock\xi\greencircle \cdots \noxi\greencircle
\Big>_{\text{Large}} = \\
&\Big<
  \xi\redcircle
  \noxi\greencircle \noxi\bluecircle \cdots \noxi\greencircle
\Big>_{\text{Large}} = \\
&\Big<
  \noxi\redcircle
  \noxi\greencircle \noxi\bluecircle \cdots \noxi\greencircle
\Big>_{\text{Small}}
\end{align*}
\caption{The equivalence of different ways to distribute picture number
         for an arbitrary scattering amplitude in six stages, from top to
         bottom:
         \newline
         1. We start from an arbitrary expectation value,
         with an arbitrary number of vertex operators (green), each one
         carrying its own arbitrary picture number.
         \newline
         2. The amplitude can also be evaluated in the large Hilbert space.
         To that end, one only has to introduce the zero mode $\xi_0$.
         This corresponds to an insertion of the field $\xi$ in an arbitrary
         place. We choose to insert it at the site of a specific vertex.
         \newline
         3. We now represent another vertex operator using the vertex at
         a picture number lower by one (red), as in~(\ref{VpicChanged}).
         Integrated vertex operators work in the same way, since the total
         derivative term in~(\ref{unVpicChanged}) does not contribute.
         \newline
         4. The integration contour of the BRST current is now deformed
         so as to circle (with opposite orientation)
         the other vertices. This results in
         many terms, such as the first one depicted, where the current
         circles around a vertex without a $\xi$ insertion. As all the
         vertices are closed (up to total derivatives), these terms
         are nullified. The only surviving term is the one where the BRST
         current circles the $\xi$ insertion. An extra ($-$) sign, coming
         from reversing the formal Grassmann ordering of $J_B$ and $\xi$,
         is represented by inverting back the orientation of the integration
         contour. In fact, there are some more minus signs coming
         from considering formal Grassmann ordering, in all the stages
         described in this figure, but they all cancel out.
         \newline
         5. The vertex with the $\xi$ insertion, surrounded by the BRST
         current is replaced by a vertex with picture number higher by one
         unit (blue).
         \newline
         6. The final result can be again evaluated in the small Hilbert
         space, by omitting the $\xi$ insertion.
         The final expression is
         identical to the initial one, except that one vertex was
         ``red-shifted'' and another one was ``blue-shifted''.
  }
\label{fig:Picture}
\end{figure}
This prescription for removing the singularities amounts essentially to
a sort of normal ordering. However, while this is always the usual
normal ordering in~(\ref{VpicChanged}) and~(\ref{unVpicChanged}),
where only the $\xi$ operator is inserted, it will not necessarily
coincide with the usual normal ordering upon usage
of~(\ref{VpicChangedX}) or~(\ref{unVpicChangedX}). Moreover, these last equations
are the ones that are easier to generalize for the purpose of lowering
the picture, as we describe next. Again, for the inverse picture changing
operator $Y$, normal
ordering and neglecting singular terms in the expansion may not coincide.

The inverse picture changing operator is defined by
\begin{equation}
\label{invPicChange}
Y=c\partial \xi e^{-2\phi}\,.
\end{equation}
This operator is the inverse of $\X$ in the sense that
\begin{equation}
\lim_{z\rightarrow w}\X(z) Y(w)=1\,.
\end{equation}
This relation implies,
\begin{equation}
\hat V^{p-1}=Y \hat V^p\,.
\end{equation}
One can check that like $\X$, the inverse picture changing operator $Y$
is closed but not exact. Thus, this
transformation maps a (closed non-exact) vertex operator into another
(closed non-exact) one, as it should\footnote{The remarks regarding the
singularities of $\X$, hold also for $Y$.}.
An interesting property of $Y$ is that it obeys,
\begin{equation}
\label{PxiY}
P=\xi Y\,.
\end{equation}
We shall use this fact in~\ref{sec:SUSYsolutions}.

For the integrated vertex we have to find,
by analogy with~(\ref{unVpicChangedX}), an operator $\Upsilon$ obeying
\begin{equation}
[Q,\Upsilon]=\partial Y\,.
\end{equation}
Then, the definition
\begin{equation}
V^{p-1}=Y V^p+\Upsilon \hat V^p\,,
\end{equation}
is consistent with~(\ref{QintegVSUSY}).
Indeed, $\Upsilon$ is easily found to be,
\begin{equation}
\Upsilon=\partial \xi e^{-2\phi}\,.
\end{equation}
A (local) primitive for $\Upsilon$ does not exist even in the large Hilbert
space. Nonetheless, one can further enlarge the Hilbert space by including
such a primitive. Inverse picture changing would then be implemented using
equations analogous to~(\ref{VpicChanged}) and~(\ref{unVpicChanged})
with $p+1\rightarrow p-1$ and the primitive of $\Upsilon$ replacing $\xi$.
The arguments presented in figure~\ref{fig:Picture} generalize
immediately for the picture lowering procedure.

The last important feature of the RNS superstring that we have to recall
is the GSO projection. This projection is based upon the operator $F$
called the ``world-sheet fermion number'', which counts the number of
occurrences at a given vertex operator of the fields
$\psi^\mu,\gamma,\beta$. In the fermionized variables the fermion
number of the superghosts is fully given in terms of the $\phi$ field,
\begin{equation}
[F,e^{l\phi}]=l e^{l\phi}\,.
\end{equation}
The $SL(2)$ vacuum of the NS sector is defined to be $F$-odd,
\begin{equation}
F\ket{0}=-\ket{0}.
\end{equation}
Since $Q$ commutes with $F$, one can consider separately the sectors
in the theory with $e^{i\pi F}$ being zero or unity. The former is
referred to as being GSO(+) while the latter is GSO($-$).
String theories contain some combinations of the NS$\pm$ and R$\pm$
sectors, subject to some consistency conditions, such as modular invariance
and closure of the OPE algebra\footnote{Closed
string theories, which we do not consider here, can have
different left and right sectors, modulo the level-matching condition.}.
Open superstring theory on a BPS D-brane has only the (+) sectors.
A non-BPS D-brane has both (+) and ($-$) sectors, while a D-\= D system
has two sectors of each kind tensored with appropriate Chan-Paton
factors.

\subsubsection{Example: The tachyon vertex operators in various pictures}
\label{sec:Example}

We illustrate the above discussion using the simplest vertex,
namely that of the tachyon field. This vertex is often ignored,
since it belongs to the GSO($-$) sector and so does not exist
on a BPS D-brane. It does exist on a non-BPS D-branes and on other systems.
Superstring field theory should be the perfect framework
to study its condensation.
At the end of this subsection we also describe the ``GSO(+)
tachyon vertex operator''.

In the natural $(-1)$ picture, the integrated tachyon vertex is
given by
\begin{equation}
V^{-1}=e^{-\phi}e^{i k\cdot X}\,.
\end{equation}
Requiring that the vertex has weight one implies that
\begin{equation}
\label{tachyonK}
k^2=1\,.
\end{equation}
Hence, this vertex describes a tachyon, as stated.
Given~(\ref{tachyonK}), one obtains
\begin{equation}
\hat V^{-1}=c e^{-\phi}e^{i k\cdot X}=c V^{-1}\,,
\end{equation}
as in the bosonic case.

Using~(\ref{unVpicChanged}) the picture is increased by one unit,
\begin{equation}
V^0=-k\cdot \psi e^{i k\cdot X}\,,
\end{equation}
while for the unintegrated vertex one gets,
\begin{equation}
\label{tachyonPic0}
\hat V^0=-(c k\cdot \psi +\eta e^\phi)e^{i k\cdot X}\,.
\end{equation}
We see that
\begin{equation}
\hat V^0\neq c V^0\,.
\end{equation}
Moreover, the second term in~(\ref{tachyonPic0}) is what one
intuitively associates with the tachyon vertex, since
\begin{equation}
\label{gammaTachyon}
\eta e^\phi(0)\ket{k}=\gamma(0)\ket{k}=\gamma_{\frac{1}{2}}\ket{k}.
\end{equation}

Continuing this way, we can evaluate $V^1$. In this case one gets
among several regular terms, also a singular one,
\begin{equation}
\label{V1tachyon}
V^1_{\text{sing}}=\frac{1}{\ep}\partial(-e^\phi e^{i k\cdot X})\,.
\end{equation}
However, as this term is a total derivative it can be safely neglected.
For the unintegrated vertex one gets again a singular
term, which equals $Q$ acting on the expression inside the parentheses
in~(\ref{V1tachyon}), in accord with~(\ref{QintegVSUSY}).
A singular term is also produced upon going in the other direction
(using inverse picture changing),
\begin{equation}
V^{-2}_{\text{sing}}=\frac{1}{\ep}\Big(
  \partial(c\partial \xi e^{-3\phi} e^{i k\cdot X})-
  Q(\partial \xi e^{-3\phi} e^{i k\cdot X})\Big)\,.
\end{equation}
For the singular part of $\hat V^{-2}$ one gets $Q$ acting on the
expression inside the first parentheses. Again, the expressions are
consistent and the singular terms can be safely dropped.

Examining~(\ref{gammaTachyon}) one can easily think of a
``more tachyonic tachyon'' living in the GSO(+) sector, namely
\begin{equation}
\label{cTach}
c(0)\ket{k}=c_1\ket{k}.
\end{equation}
For this operator to have zero conformal weight, as is appropriate
for an unintegrated vertex operator, one has to demand $k^2=2$.
However, even for these values of
the momenta the operator fails to be closed.
Thus, as is well known, there is no GSO(+) tachyon.
The operator~(\ref{cTach}) clearly carries a zero picture.
Changing the picture of an operator is a well-defined procedure
only for vertex operators. Still, one may ignore that and try to
find the form of this operator in the natural picture, by acting
on it with the inverse picture changing operator.
The result is zero and there is no other operator with the same
quantum numbers that can be added to it in order to remedy this result.
This is not a problem, since this is not a genuine vertex
operator. One may consider this operator
as a peculiarity of working with zero picture number.
Despite the above, this operator emerged in some recent
developments, described in~\ref{sec:SUSYsolutions}.

\subsection{Introducing superstring field theories}
\label{sec:SUSYtheories}

The original proposal for open superstring field theory by
Witten~\cite{Witten:1986qs} is almost a straightforward
generalization of the construction for the bosonic case.
The main difference is the appearance of picture number,
which implies that the definitions of integration
and star product should be modified, in order to saturate the $-2$
picture number in~(\ref{SmallSpaceCFT}). Also, the string field should be
assigned a fixed picture number\footnote{One could imagine
constructing a string field theory without restricting the picture number.
However, as vertices with different picture numbers are equivalent,
a gauge symmetry should be introduced in order to avoid multiple
counting. Another related option could be to start with a fixed picture
number and add other picture numbers as part of the gauge fixing
procedure, analogously to the way different ghost numbers appear
in the bosonic case. \ATFV the first of these ideas was realised
leading to the ``democratic string field theory''~\cite{Kroyter:2009rn}.
It was later shown that a partial gauge fixing of the gauge freedom
related to the picture number can be used to reduce the democratic
theory to the modified string field theory and to the non-polynomial
string field theory described below~\cite{Kroyter:2010rk}.
Furthermore, the democratic theory includes also the Ramond sector.}.
Witten suggested to modify the bosonic action to
\begin{equation}
S=-\int\Big(\frac{1}{2} \Psi\star Q \Psi+
  \X \frac{1}{3} \Psi\star \Psi\star\Psi \Big),
\end{equation}
where the picture changing operator $\X$ is inserted at the common string
mid-point.
The string fields in this scheme carry picture number $n_p=-1$.

This version of superstring field theory has some problems. The
picture changing operators appear in scattering amplitudes
and may be inserted at the same point. This, however,
produces singularities~\cite{Wendt:1987zh,Arefeva:1988nn}
(starting already at tree level for the four point function)
that render the theory erroneous.
A related issue is the collision of picture changing operators in
the derivation of the $g_o^2$-order gauge transformation of the
string field.
One can try to solve these divergences by introducing counter-terms
into the theory.
Wendt~\cite{Wendt:1987zh} gave the form of the fourth order counter-term
that resolves both problems.
Nevertheless, as also noted by him, new divergences arise
at the next order. It may be possible that an infinite set of
counter-terms exists that can regularize the theory (at least classically,
i.e., considering only tree level interactions) to all orders.
To the best of our knowledge, this avenue was never pursued.
Also, as mentioned in section~\ref{sec:LevelTrunc},
Witten's theory fails to reproduce the expected results for the tachyon
potential, at least for the first few levels in the level truncation
scheme (without the introduction of counter-terms).

A modified version of cubic superstring field theory was constructed,
using the double-step inverse picture changing operator
$Y_{-2}$~\cite{Preitschopf:1989fc,Arefeva:1989cp,Arefeva:1989cm}.
This operator is required to be closed, non-exact and obey
\begin{equation}
\lim_{z\rightarrow w}\X(z) Y_{-2}(w)=Y(w)\,.
\end{equation}
All its quantum numbers, other than the picture number, should vanish.
One can define $Y_{-2}$ either as a chiral or as a non-chiral (non-local)
operator,
\begin{align}
Y_{-2}^{chir}(z)&=-e^{-2\phi(z)}
   -\frac{i}{5}c \partial \xi e^{-3\phi}\psi_\mu \partial X^\mu (z)\,,\\
Y_{-2}^{non}(z,\bar z)&=Y(z)Y(\bar z)\,,
\end{align}
where the doubling trick is used.
The non-local version is singular on the boundary, where vertex
operators and picture changing operators are usually inserted in the
world-sheet description. It is, nonetheless,
useful for superstring field theory, in which the operator is inserted
at the string mid-point, $z=i$.
The $Y_{-2}$ insertion saturates the required picture number. Hence,
there is no need for further picture insertions and the string field
should be assigned a zero picture number. The action reads,
\begin{equation}
S=-\int Y_{-2}\Big(\frac{1}{2} \Psi\star Q \Psi+
  \frac{1}{3} \Psi\star \Psi\star\Psi \Big)\,.
\end{equation}
The $Y_{-2}$ operator can be absorbed in a redefinition of the integral.
The equation of motion derived from this action is
\begin{equation}
\label{modEOM}
Y_{-2}\Big(Q \Psi+\Psi\star\Psi \Big)=0\,.
\end{equation}
This modified form of the action does not suffer from the contact term
problems. For the gauge symmetry it is clear, since the gauge parameters
carry now zero picture number and picture changing operators do not
occur in the gauge transformation. In the expressions for scattering
amplitudes $Y_{-2}$ will still appear. However, while each vertex now
carries a factor of $Y_{-2}$, the propagator carries a factor of
``$\frac{1}{Y_{-2}}$''. These factors cancel out (at least for tree
amplitudes) leading to the expected results~\cite{Arefeva:2001ps}.

Several objections have been raised to this modified
action. One objection is that
picture changing operators have non-trivial kernels.
Thus, in might be the case that the equation of motion~(\ref{modEOM})
is not equivalent to the expected one, in which the $Y_{-2}$ factor is
absent.
However, these kernels are of a somewhat exotic nature,
containing only states which are localized at the string mid-point.
Thus, it is not clear if this is really a problem\footnote{\ATFV a
``non-minimal'' variant of the cubic theory appeared, which has no
kernel~\cite{Berkovits:2009gi,Kroyter:2009zj}. It is not clear to us whether
this formalism has any advantage over the standard one.}.
Also, as mentioned in footnote~\ref{foot:DP}, picture changing operators
should be modified at higher loop order. It may be hard to believe that
one can use the picture changing operators that are adequate for the disk
for describing scattering processes to all orders.

Another objection is related to the existence of two versions of $Y_{-2}$.
The two candidates for $Y_{-2}$ are not a-priori equivalent and
it is not clear, which one should be used. Some properties
of the theories for the two possible choices of $Y_{-2}$ were studied
using level truncation in~\cite{Urosevic:1990as}.
The theories were found to differ when truncated to the massless level.
It was claimed that the non-chiral operator is the more promising one,
since the theory with the chiral operator fails to reproduce the
Maxwell equations for the level zero component.
It was also found out there that supersymmetry is differently realized in the
two theories. Two versions of supersymmetry transformations were found
for both theories\footnote{The kernel of $Y_{-2}(i)$ was interpreted as
an extra gauge symmetry. The two realizations of supersymmetry differ by
picture changing operators inserted at $\pm i$ and so are presumably
gauge equivalent. However, as they all use explicit mid-point insertions over
string fields, they might lead to some further singularities.},
although for the chiral theory, one of them suffers
from singularities when the supersymmetry transformation is iterated.
Later, it was noticed that the theory with the chiral operator
does not respect the expected twist symmetry.
The above may suggest that the theory with the
non-chiral operator is the more promising one.
At any rate, the ultimate test would be to check which one,
if at all, reproduces the
correct results, namely correct (on-shell) scattering amplitudes and
the expected results regarding Sen's conjectures.
Due to the criticism on the chiral theory, it was studied much
less than the non-chiral theory, both in level truncation
(as described in~\ref{sec:LevelTrunc}) as well as in the recent
developments described below\footnote{\ATFV it was proven that regardless of
this discussion the chiral
and non-chiral versions of the theory (as well as other possible versions)
are classically equivalent. In particular, Sen's conjectures hold equally
well in all these versions~\cite{Kroyter:2009bg}. On the other hand, a new,
serious objection to the cubic theory was given in~\cite{Kroyter:2009zi},
where it was shown that the standard incorporation of the Ramond sector in
this theory leads to an inconsistent gauge structure.}.

Another form of open superstring field theory was developed
by Berkovits
\cite{Berkovits:1995ab,Berkovits:1998bt,Berkovits:2001nr,Berkovits:2001im}.
Unlike the cubic string field
theories discussed so far, this theory is
non-polynomial\footnote{This theory still uses only the cubic Witten
vertex, unlike the counter-terms mentioned above in the context of Witten's
superstring field theory or the higher vertices that are present in
closed string field theory~\cite{Zwiebach:1992ie},
open-closed string field theory~\cite{Zwiebach:1992bw,Zwiebach:1997fe}
and heterotic string field theory~\cite{Saroja:1992vw,Okawa:2004ii,Berkovits:2004xh}.}
and looks like a generalization of the WZW theory. This theory was
constructed using the language of $N=2$ and topological $N=4$
string theory~\cite{Berkovits:1994vy}.
Nonetheless, it can be presented (and used) without going into the
details of this construction.

One novel characteristic of this theory
is that it is defined in the large Hilbert space.
A new gauge symmetry takes care of the extra degrees of freedom that
emerge due to the use of this space.
Recall that in the large Hilbert space the picture number should equal
minus one for a non-trivial result. Thus, it is enough to use $\eta_0$,
which acts non-trivially in the large Hilbert space, in order to
saturate it. The string field can
be assigned zero picture number and there is no need to use picture
changing operators\footnote{From the $N=2$ point of view $\eta_0$ and
$Q$ should be treated on an equal footing, since they are the two
superconformal generators of this theory.}.

The action is given by\footnote{Here, we use a new integration symbol
in order to stress that it represents expectation values in the
large Hilbert space. One should not confuse the first integration symbol,
which represents the expectation value, with the second one, which stands
for a regular one-dimensional integral.}
\begin{equation}
S=\frac{1}{2g_o^2}\oint\Big(
e^{-\Phi} Q (e^\Phi) e^{-\Phi} \eta_0 (e^\Phi)-\int_0^1 dt\,
   e^{-t\Phi} \partial_t e^{t \Phi}
   \big[e^{-t\Phi}\eta_0 e^{t \Phi},e^{-t\Phi} Q e^{t \Phi}\big]
\Big),
\end{equation}
where $\Phi$ is the string field.
In the second term $\Phi$ is written as
$e^{-t\Phi} \partial_t e^{t \Phi}$ in order to bring the action to
a form resembling the WZW action. The equation of motion derived from
the action is
\begin{equation}
\label{NP_EOM}
\eta_0 (e^{-\Phi} Q e^\Phi)=0\,.
\end{equation}
This equation states that the expression inside the parentheses lies
in the small Hilbert space. This can be achieved trivially by taking
$\Phi$ to lie in the small Hilbert space. However, as mentioned above,
there is a new gauge symmetry that removes exactly these degrees of freedom.
The linearized gauge transformation is most easily given in terms of
the transformation of $G\equiv e^\Phi$,
\begin{equation}
\label{BerkoGauge}
\delta G= -(Q\tilde\Lambda)G + G(\eta_0\Lambda)\,.
\end{equation}
This can be exponentiated to give the finite gauge transformation,
\begin{equation}
\label{NP_GaugeTrans}
G \rightarrow e^{-Q\tilde\Lambda}G e^{\eta_0\Lambda}\,.
\end{equation}
The quantum numbers of the gauge string fields are
\begin{equation}
n_g(\tilde\Lambda)=-1\,,\qquad n_p(\tilde\Lambda)=0\,,\qquad
  n_g(\Lambda)=-1\,,\qquad n_p(\Lambda)=1\,.
\end{equation}
The first gauge string field in~(\ref{BerkoGauge})\footnote{Many
papers use a convention without the minus sign in front of
this term.} is analogous to the usual gauge string field.
The second gauge string field can be used to remove variations
of the form $\delta G=G \delta\Phi$ with $\Phi$ entirely within the
small Hilbert space, by defining $\Lambda=\xi_0 \Phi$, which results in
$\delta G=G \eta_0\Lambda$.
Introducing $\xi_0$ defines two copies of the small Hilbert space.
The new gauge symmetry identifies the small Hilbert space with zero.
The gauge symmetry associated with $Q$ of the quotient space $H_L/H_S$
is isomorphic to that of $Q$ as it act on $H_L$. Thus, the action
passes at least the preliminary test of having the correct amount
of degrees of freedom. It was further described
in~\cite{Berkovits:1995ab} how to write a superstring field theory
action that is manifestly supersymmetric after reduction to
four space-time dimensions. The action passed also the important
tests of describing correctly scattering
amplitudes~\cite{Berkovits:1999bs}.
Furthermore, as described in section~\ref{sec:LevelTruncSUSY},
it also reproduces the results expected according to Sen's
conjectures.

Finally, we turn to describe the inclusion of the GSO($-$) sector
in the string field theories described above. Since the states
contained in this sector differ from those in the GSO(+) sector,
it is clear that a new string field should be added to describe
them. The half-integer conformal weights of the GSO($-$) states
imply that
the cyclicity property~(\ref{intPsiPsi}) should be modified. Let
$A_{1,2}$ be two string fields with half-integer weights, then their
cyclicity property reads\footnote{If only one of them has
half-integer conformal weight then the integral gives zero.},
\begin{equation}
 \int A_1\star A_2=-(-1)^{A_1 A_2}\int A_2\star A_1\,.
\end{equation}
A related issue is that the GSO($-$) sector string field has the
opposite Grassmann character as compared to the string field from
the GSO(+) sector\footnote{That is, the GSO($-$) string field is odd
rather than even in Berkovits' formalism and even rather than odd
in the formalisms based on Witten's theory.}. In order to
solve both problems and treat the string fields from both sectors
collectively one should introduce the so called
``internal Chan-Paton indices''~\cite{Berkovits:2000hf}.
These are simply some Pauli matrices acting on the internal
space that consists of the two GSO sectors. The exact prescription
depends on the theory at hand\footnote{This should be expected.
While the GSO(+) string field is commuting and of ghost
number zero in Berkovits' theory, it is of ghost number one
and anti-commuting in the modified cubic theories. This
dictates different algebraic properties to be satisfied
upon the inclusion of the GSO($-$) sector.
One might expect that the representation of the physical
string fields in one theory resembles that of the gauge string fields
in the other one. This is indeed the case.}.
For Berkovits' theory (on the non-BPS D-brane, where both sectors
exist) one has to write,
\begin{equation}
\Phi=\Phi_+ \otimes \One + \Phi_- \otimes \sigma_1\,,
\end{equation}
where the subscripts $\pm$ refer to the GSO sector in which
the string field resides.
One should also redefine the operators $Q$ and $\eta_0$ according to
\begin{equation}
Q\Rightarrow Q \otimes \sigma_3\,,\qquad
\eta_0\Rightarrow \eta_0 \otimes \sigma_3\,,
\end{equation}
and redefine the integral so as to include also a trace over the
matrices of this internal space. Since $\tr(\One)=2$, the
coefficients in front of the action should be divided by two.
The gauge string fields are modified according to
\begin{equation}
\Lambda=\Lambda_+ \otimes \sigma_3 + \Lambda_- \otimes (i \sigma_2)
\,,\qquad
\tilde \Lambda=\tilde \Lambda_+ \otimes \sigma_3 +
   \tilde \Lambda_- \otimes (i \sigma_2)\,.
\end{equation}
For the cubic theories the string field is
given by~\cite{Arefeva:2002mb},
\begin{equation}
\Psi=\Psi_+ \otimes \sigma_3 + \Psi_- \otimes (i\sigma_2)\,,
\end{equation}
while the gauge string field is given by
\begin{equation}
\label{GSOgauge}
\Lambda=\Lambda_+ \otimes \One + \Lambda_- \otimes \sigma_1\,.
\end{equation}
The only other modification required is the assignment,
\begin{equation}
Y_{-2}\Rightarrow Y_{-2}\otimes \sigma_3\,.
\end{equation}

The description of general brane configurations
is almost as straightforward. We describe the D-\= D system
in the non-polynomial theory for simplicity.
In this case, one has to introduce also the proper Chan-Paton
factors, each one with the appropriate type of string field.
In the case at hand, the Chan-Paton matrices $\One,\sigma_3$
support GSO(+) fields, while $\sigma_1$ and $\sigma_2$
support GSO($-$) fields. Thus, we expand the string field as
\begin{equation}
\Phi=\Phi_+^1 \otimes \One \otimes \One +
     \Phi_+^2 \otimes \One \otimes \sigma_3+
     \Phi_-^1 \otimes \sigma_1 \otimes \sigma_1+
     \Phi_-^2 \otimes \sigma_1 \otimes \sigma_2\,.
\end{equation}
Here, the middle component in the tensor product describes
the internal Chan-Paton space, while the right component describes
the proper Chan-Paton factor.
The action is further divided by two, to account for the trace
that is now taken over external as well as internal Chan-Paton indices.

\subsection{Mapping solutions among (super)string field theories}
\label{sec:Map}

The first analytical solutions of superstring field theory were
constructed independently by Erler~\cite{Erler:2007rh} and by
Okawa~\cite{Okawa:2007ri,Okawa:2007it}. The solutions represent
regular marginal deformations within the framework of Berkovits'
string field theory. While it was recognized that the simplest
solutions found do not obey the reality condition,
gauge equivalent real solutions were also constructed.

It was found in these papers that despite
the apparent distinction between Witten's bosonic cubic theory
and Berkovits' supersymmetric non-polynomial theory, the
solutions are quite similar to the analogous bosonic
ones, found in~\cite{Kiermaier:2007ba,Schnabl:2007az}.
One may wonder whether this is merely some funny coincidence.
This is not the case.
In fact, given a bosonic solution in a formal pure-gauge form, it can be
used to canonically define solutions of the supersymmetric theories.

The mapping of formal pure-gauge solutions is straightforward in the case
of the modified
cubic superstring field theory. Since in this case the physical
string field has the same ghost and picture numbers as in the bosonic
case, one can simply use the
same\footnote{Note, that this is not a ``mapping''
in the strict mathematical sense, since $\La_{bos}$ lives in the
BCFT of the bosonic theory, while $\La_{cub}$ lives in the BCFT
of the RNS theory. Nonetheless, in many cases one can identify objects
on both spaces on physical grounds. For example, the photon field
has 26 (physical and unphysical) polarizations in the (flat) bosonic theory
and only ten polarizations in the RNS case. Nonetheless, the zero-mode
of a particular direction is a well defined object in both theories.
Hence, one can use it in both theories as a formal gauge string field for the
photon marginal deformation. An even simpler case is the one where the
gauge string field depends only on the $bc$ ghost sector, which is present in
both BCFT's. Then, one can literally map the solutions. In particular,
Schnabl's solution is mapped in such a way to Erler's solution.}
gauge string field and derive from it the solution.
This solution may differ from the bosonic one, since now
the supersymmetric BRST operator~(\ref{JbSUSY}) is used.

In particular, the solutions describing singular marginal deformations
can be trivially mapped to this theory (using their formal pure-gauge
representation).
Moreover, the resulting solutions, despite being different from the
bosonic ones remain $x_0$-independent, since the
condition~(\ref{LeftDE}) still holds and only formal properties
of $Q$ were used in the derivation of $x_0$-independence\footnote{Here,
we use again the terminology adequate
for the photon marginal deformation, but similar logic holds also
for other exactly marginal deformations.}.

In the case of Berkovits' theory the string field has the same
(zero) picture numbers, but the ghost number (zero) is different
from that of the cubic (bosonic and supersymmetric) theories.
Moreover, in the non-polynomial theory the string field has to contain
$\xi_0$ in order to be non-trivial.
Thus, for sending solutions of the cubic theories to
the non-polynomial theory, we have to consider a mapping that results in
\begin{equation}
\Phi_1 = \xi \cO (z) \Psi_1\,,
\end{equation}
for some $\cO$.
The quantum numbers of $\xi$ dictate that $\cO$ has to have zero
ghost number and conformal weight and minus one unit of picture
number. Hence, it is natural to identify it as the inverse
picture changing operator $Y$ and the map in this case
should take the form,
\begin{equation}
\label{Phi1PPsi1}
\Phi_1 = P(z) \Psi_1\,,
\end{equation}
for some $z$ and where $P$ is given by~(\ref{PxiY}).

The choice of $z$ in~(\ref{Phi1PPsi1}) is important.
In~\cite{Fuchs:2008zx} it was found
that the desirable value for $z$ is $\pm i$. This is a natural choice,
since these points are invariant under star-multiplication. The canonical
nature of these points was also used in the construction of the cubic
superstring field theories as described above.
On the other hand and for the same reasons, this choice is also potentially
singular. Hence, this mapping should be thought of as a singular limit
of some (unknown) family of regular mappings.

An important property of $P(z)$ is~(\ref{QP1}),
which implies that the cohomology of $Q$ is empty in the large Hilbert space.
In~\cite{Fuchs:2008zx} the operator $P$ was defined as a linear combination of
$P(\pm i)$, such that the normalization of~(\ref{QP1}) does not change,
\begin{equation}
P=k P(i)+(1-k)P(-i)\,.
\end{equation}
Then, the linearized map~(\ref{Phi1PPsi1}) was trivially extended to
a map of the cubic superstring field theory to the non-polynomial
one\footnote{The map~(\ref{TheMap}) is not adequate for mapping bosonic
solutions to
the non-polynomial theory. However, bosonic solutions that are given in a
pure-gauge form can be (trivially) mapped to the cubic theory and then to the
non-polynomial one.},
\begin{equation}
\label{TheMap}
\Phi = P \Psi\,.
\end{equation}
The operator $P$ is nilpotent, as can be seen from the OPE,
\begin{equation}
P(z)P(w)=-\frac{z-w}{12}
    (c c' \xi \xi' \xi'' \xi''' e^{-4\phi})(w)+\cO\big((z-w)^2\big)\,.
\end{equation}
The location of the $P$ insertion is invariant under the star product. This
fact implies that a state $\Psi$ carrying a $P$ insertion but no other
components with support at $\pm i$, is nilpotent with respect to
the star product. The above implies
\begin{equation}
e^{t \Phi}=1+t P\Psi\,.
\end{equation}

For solutions of the equation of motion, the map~(\ref{TheMap}) can be
inverted.
In fact, all the solutions of the equation of motion of the non-polynomial
theory~(\ref{NP_EOM}) can be mapped to formal pure-gauge solutions of the
cubic theory
by~\cite{Berkovits:2001im,Kling:2002vi},
\begin{equation}
\label{invMap}
\Psi=G^{-1}Q G\,,
\end{equation}
since~(\ref{NP_EOM}) implies that $\Psi$ lives in the small Hilbert space.
The ``gauge string field'' $G$ is formal, since it lives in the large Hilbert space.
Composing the two maps one gets the identity transformation,
\begin{equation}
\label{PsiofG2}
\Psi'=G^{-1}Q G=(1-P\Psi)(\Psi-P Q \Psi)=(1-P\Psi)(\Psi+P \Psi \Psi)=
    \Psi\,.
\end{equation}
Composing the maps in the other order results in a solution of the
non-polynomial theory, which is gauge equivalent to the original $\Phi$.
More generally it was proven in~\cite{Fuchs:2008zx} that gauge orbits
are mapped to gauge orbits under the action of both maps and all gauge orbits
(of solutions) are accessible. This implies that the cohomologies around
solutions agree in both theories. Moreover, it was found that the action of
solutions is invariant under the maps\footnote{To that end one has to
regularize the map~(\ref{TheMap}). Strictly speaking one should think of
this map as a specific limit of maps with no support at $\pm i$. In the
limiting maps the $\xi$ and $Y$
components of $P$ approach the limit points at a different pace.
However, defining the regularization by specifying a path for the
insertions that approaches $\pm i$, might lead to singularities, since
string fields are allowed to carry insertions inside the local coordinate
patch. One possibility would be to move the line in a way that avoids such
insertions, but this prescription is not universal. It is not clear
whether a universal regularization for the mid-point insertions exist.},
provided that the non-chiral choice is made for the $Y_{-2}$ operator.
All that seems to imply that at least on-shell the non-polynomial theory and
the cubic theory with non-chiral $Y_{-2}$ are equivalent, assuming that a
regularization indeed exists.
One can study some properties of a classical solution
in one theory and then move the other theory to study some issues
that are more transparent there.

Solutions of the cubic theory that are given in a formal pure-gauge form
(other than the one of~(\ref{invMap})) can be written in such a form
in the non-polynomial theory as well using,
\begin{equation}
\label{GaugeMap}
\Lambda = \xi\Lambda_{bos}\,,\qquad \tilde\Lambda= P \Lambda_{bos}\,,
\end{equation}
where the insertion point of the $\xi$ operator is of no importance.
The formal gauge solution of the non-polynomial theory is then given by
(recall~(\ref{NP_GaugeTrans})),
\begin{equation}
\label{NP_PureGauge}
G = e^{-Q\tilde\Lambda} e^{\eta_0\Lambda}\,.
\end{equation}
The solution one gets using~(\ref{TheMap}) is identical to the one
that is obtained by using~(\ref{GaugeMap}) and~(\ref{NP_PureGauge}).

The map~(\ref{TheMap}) is easily generalized to include the cases
of the non-BPS D-brane and of D-brane systems, provided we assign
a factor of $\sigma_3$ also to $P,\xi$ in the internal Chan-Paton
space,
\begin{equation}
\label{GSO-Map}
P\Rightarrow P \otimes \sigma_3\,,\qquad
\xi\Rightarrow \xi \otimes \sigma_3\,.
\end{equation}
The mapping of on-shell gauge orbits to gauge orbits and the equality of
the value of the action in the two theories carries over without any further
modification upon imposing~(\ref{GSO-Map}).

\subsection{Analytical solutions in superstring field theory}
\label{sec:SUSYsolutions}

Following~\cite{Erler:2007rh,Okawa:2007ri,Okawa:2007it}, more solutions were
found. Marginal deformations with singular OPE's were
studied in~\cite{Fuchs:2007gw,Kiermaier:2007ki},
within the non-polynomial theory. These works generalized the methods
of~\cite{Fuchs:2007yy,Kiermaier:2007vu} respectively.
Since in the bosonic case these deformations are given in a formal
pure-gauge form, they could be written in the cubic theory immediately
using the same gauge string fields. All that is needed to write them in the
non-polynomial theory is either~(\ref{TheMap}) or~(\ref{GaugeMap}).

Currently we have only one bosonic solution at hand, other than the ones
describing marginal deformations. This is Schnabl's solution, describing
tachyon condensation.
What happens when we consider its counterpart in superstring field
theory\footnote{It is possible to define the solution in the supersymmetric
theory, since Schnabl's solution can be given in a formal pure-gauge
form using the gauge
field~(\ref{LambdaSolution}).}~\cite{Fuchs:2007gw,Erler:2007xt}?
On the one hand, from the canonical nature of the map, one expects
that the solution of the supersymmetric theory has similar physical
content to the bosonic solution. On the other hand, Schnabl's
solution describes tachyon condensation, while the solution in the
supersymmetric theory lives, by
construction\footnote{The map does not involve operators
of half-integer conformal weight.}, in the GSO(+) sector, where no
tachyons are present. One may still conjecture that it describes the state
without the original D-brane, despite the fact that no dynamical
condensation process exists that connects these two solutions.

The most natural way to check the conjecture is by evaluating the
action and calculating the cohomology around the solution.
In~\cite{Erler:2007xt}, Erler studied the solution in the framework of
the modified cubic theory. He found out that the cohomology and action
of the solution are the expected ones, namely, the cohomology
vanishes and the action equals minus the tension of the original D-brane.
To that end, Erler used ``the bosonic gauge string field''~(\ref{LambdaSolution})
with the supersymmetric BRST charge. This results in a closed form
expression that differs from the bosonic solution~(\ref{SchPsi})
only slightly,
\begin{equation}
\Psi=\Psi_{bos}+B\gamma^2(0)\ket{0}=
   \Psi_{bos}+B \eta\partial \eta e^{2\phi}(0)\ket{0}.
\end{equation}
The usage of the cohomology argument is straightforward, since
$A$ of~(\ref{A}) contains the operator $B$ and so kills the extra piece,
which also contains the $B$ operator (and no $c$ insertions).
Thus, one can use
the argument of~\cite{Ellwood:2006ba} without any modification
to prove that the cohomology vanishes.

The evaluation of the action requires some sort of regularization of
the solution. This can be achieved either by level truncation
(usually not adequate for analytical calculations) or by introducing
the ``phantom pieces'' of the solution. 
As noted in section~\ref{sec:Schnabl}, Erler identified the amount
of extra terms needed in the general case and their form.
He found out that the regularization of the solution at hand should involve
two phantom terms (one more than in Schnabl's case),
so plugging into~(\ref{RegPsi}), we see that a properly regularize form for
the solution is ($B_1=-\frac{1}{2}$),
\begin{equation}
\Psi=\lim_{N\rightarrow \infty}\Big(\sum_{n=0}^{N-1} \psi_n'
  -\psi_N+\frac{1}{2}\psi_N'\Big)+B\gamma^2(0)\ket{0}.
\end{equation}

The evaluation of the action is quite similar to the bosonic case
and is in fact simpler. Erler showed that the solution obeys
the equation of motion in the strict sense, i.e., even when contracted
with the solution itself. Hence, one can use the equation of motion
in order to write the action of the solution as
\begin{equation}
S(\Psi)=-\frac{1}{6}\int Y_{-2} \Psi Q\Psi\,.
\end{equation}
The evaluation differs from the bosonic one in several ways.
First, there is the $Y_{-2}$ insertion in all correlators. Then,
in order to get a non-zero result, the $\phi$ momentum should equal
minus two~(\ref{SmallSpaceCFT}). Since $Y_{-2}$ has $-4$
charge, one has to take into account only terms whose momentum
equals two, such as the $\gamma^2$ term. In particular, all
the terms that contribute in the bosonic case do not contribute
now. This does not imply that the $\Psi_{bos}$ piece of the solution
does not contribute, since now we use the BRST charge of the
supersymmetric theory and $Q\Psi_{bos}$ contains a $\gamma^2$
piece. All in all, one has to reduce all the expression to
correlators of the form
\begin{equation}
\label{ErlerCor}
\left< Y_{-2} \int_{-i \infty}^{i\infty} dw\, b(w)
  c(y+z) c(z) \gamma^2(0)\right >_{C_{\pi(x+y+z)}}=
    -\frac{x+y+z}{2\pi^2}y\,,
\end{equation}
where the CFT evaluation is straightforward
and can be found in the appendix
of~\cite{Erler:2007xt}\footnote{Note, however, that he uses
slightly different conventions.}.
This result is much simpler to work with
than the bosonic one. Recall that in the bosonic case the
building blocks for the evaluation of the action contained
trigonometric functions~(\ref{uglyCor}),
which made the evaluation of the
sums somewhat tricky. Now, the summation is trivial.

Plugging~(\ref{ErlerCor}) into the expression for
the action, one sees that the bosonic piece contains terms that
are not only non-zero, but even diverge in the limit
$N\rightarrow \infty$. These divergences cancel against
each other, leaving a total zero contribution from the bosonic
piece. Similarly evaluating the contribution of the terms in the
action that involve also the new pieces of the solution
(the $\gamma^2$ piece and the $\Psi_N'$ piece) results in,
\begin{equation}
-E(\Psi)=S(\Psi)=\frac{1}{2\pi^2}\,,
\end{equation}
which is the expected result for a vanishing D-brane\footnote{The
space-time volume is normalized to unity.}.

The evaluation of the action and the results regarding the cohomology
support the physical interpretation of these solutions and imply that,
at least to some extent, open string field theory can be used
to describe solutions that are not continuously connected to the
original theory\footnote{In the case at hand the solutions are not
connected as they describe states with different $RR$ charge.
One can, nevertheless, in the case of a lower dimensional D-brane,
consider the continuous process of moving the D-brane to infinity.
In the case of a D9-brane, which cannot be sent to infinity, one may
expect the solution to have problems quantum mechanically, unless
the RR charge is somehow balanced.}.

One may criticize Erler's solution~\cite{Erler:2007xt},
as well as the analogous solution in Berkovits'
theory~\cite{Fuchs:2007gw}, as describing the condensation of
the wrong string field. To leading order the solution describes
the condensation of the non-physical GSO(+) tachyon~(\ref{cTach}).
It was mentioned in section~\ref{sec:Example} that this string
field is not closed and so does not give rise to a vertex operator.
From the perspective of string field theory, its non-closeness
means that its free equation of motion is a constraint equation implying
that it vanishes. It is the first of many auxiliary string fields
peculiar to the zero picture. This field can nevertheless play a
role within the framework of an interacting string field theory.
In fact, a precursor of Erler's solution was found
by Arefeva et. al. almost twenty years ago~\cite{Arefeva:1990ei}.
There, the modified chiral superstring field theory was truncated
to level zero and a condensation of the GSO(+) tachyon was obtained
as solution\footnote{A few years ago, Ohmori studied this solution
using level truncation up to level three. He found out that a
solution indeed exists for the chiral theory, but he did not
find the solution for the, presumably more reliable,
non-chiral version of the theory~\cite{Ohmori:2003vq}.
This is probably an artifact of level truncation, since
Erler constructed his analytical solution just for this
non-chiral theory.}.
It was claimed that the solution describes a
supersymmetry-breaking vacuum. This interpretation was based on
the observation that around this solution the bosonic modes
become heavier, while nothing of this sort was observed for
the Ramond sector states. One should take this observation with
a grain of salt, due to the approximate nature of the level zero
truncation and due to the inconsistency of the Ramond sector
of the modified theory.

The understanding that a solution describing the
GSO($-$) tachyon should also exist motivated
Arefeva et. al.~\cite{Aref'eva:2008ad} to look for a
generalization of Erler's solution.
On the non-BPS D-brane Erler's solution is given by setting
in~(\ref{GSOgauge}),
\begin{equation}
\Lambda_+ = B c(0)\ket{0},\qquad
\Lambda_- = 0\,.
\end{equation}
One can generalize that by allowing a non-zero $\Lambda_-$.
In the general case the solution is
\begin{equation}
\begin{aligned}
\label{PsiGSOm}
\Psi&=(Q\otimes \sigma_3)
   (\Lambda_+ \otimes \One+\Lambda_- \otimes \sigma_1)
 \frac{1}{1-(\Lambda_+ \otimes \One+\Lambda_- \otimes \sigma_1)}\\
&=(Q\Lambda_+ \Xi_+ +Q \Lambda_- \Xi_-)\otimes \sigma_3+
 (Q\Lambda_- \Xi_+ +Q \Lambda_+ \Xi_-)\otimes (i\sigma_2)\,,
\end{aligned}
\end{equation}
where we defined\footnote{The expressions
$\Lambda_+ \pm \Lambda_-$ may seem awkward, as they mix the two
GSO sectors. The resulting $\Xi_\pm$ are, however, standard.},
\begin{equation}
\Xi_\pm = \frac{1}{2}\Big(\frac{1}{1-(\Lambda_+ + \Lambda_-)} \pm
                     \frac{1}{1-(\Lambda_+ - \Lambda_-)} \Big)\,.
\end{equation}
Two simple special cases are when either of $\Lambda_\pm$
equals zero. For $\Lambda_- = 0$ one gets just the usual GSO(+)
solution, while for $\Lambda_+ = 0$ one gets
\begin{equation}
\Psi=Q \Lambda_- \frac{\Lambda_-}{1-\Lambda_-^2}\otimes \sigma_3
 +Q\Lambda_- \frac{1}{1-\Lambda_-^2}\otimes (i\sigma_2)\,.
\end{equation}
In this case one has a non-zero expression in both of the
GSO sectors, but the GSO(+) sector starts at a higher order
with respect to $\Lambda_-$.

In~\cite{Aref'eva:2008ad} it was suggested that the solution
describing tachyon condensation should contain
the physical GSO($-$) tachyon field. They advocated to
use the gauge string field,
\begin{equation}
\label{AGM}
\Lambda_+ = B c(0)\ket{0},\qquad
\Lambda_- = B \gamma(0)\ket{0}.
\end{equation}
For this specific choice one gets
\begin{equation}
\label{LambdaLambdaZero}
\Lambda_-^2=\Lambda_- \Lambda_+=0\,,
\end{equation}
which dictates that for this solution
\begin{equation}
\Xi_+=\frac{1}{1-\Lambda_+}\,,\qquad
\Xi_-=\frac{1}{1-\Lambda_+}\Lambda_- \,.
\end{equation}
Let us consider the slightly more general gauge string field~\cite{Fuchs:2008zx},
\begin{equation}
\label{AGMep}
\Lambda_+ = B c(0)\ket{0},\qquad
\Lambda_- = \ep B \gamma(0)\ket{0}.
\end{equation}
This is a one-parameter family interpolating~(\ref{AGM}) ($\ep=1$)
and Erler's solution ($\ep=0$).

The action of the whole $\ep$-family agrees with that of Erler's
solution. For the evaluation
of the action we recall that in order to get a non-zero result,
the fields should have a total power two of $\gamma$.
Since $Q$ can only increase the amount of $\gamma$'s we conclude
that terms with more than two occurrences of $\Lambda_-$ will not
contribute. Let us now consider an arbitrary $\ep$-solution as
being expanded around Erler's one. We can use the general expression
for the action around a solution~(\ref{actionAroundSol}) and the
fact that the $\ep$-solution also obeys the equation of motion,
in order to write the action of this solution as\footnote{Note
that we added to~(\ref{actionAroundSol}) the action of the
original solution (Erler's one) and also wrote down the
trace factor needed when considering the non-BPS brane.}
\begin{equation}
\label{AGMaction}
S_\ep=S_E+\frac{1}{6}\int \frac{1}{2}\tr (Y_{-2} \tilde\Psi^3)\,.
\end{equation}
Here $\tilde \Psi$ represents the deviation of the solution
from Erler's one. For $\ep=1$ the GSO($-$) part of $\tilde \Psi$,
is the term proportional to $i\sigma_2$ in~(\ref{PsiGSOm}), and the
GSO(+) part is the second summand
proportional to $\sigma_3$ in~(\ref{PsiGSOm}).
The GSO($-$) part $\tilde \Psi_-$ is proportional to $\gamma$,
while $\tilde \Psi_+$ is proportional to $\gamma^2$.
At least one factor of $\tilde \Psi_+$ is needed for a non-zero
trace.
Hence, $\tilde \Psi^3$ contributes to the action~(\ref{AGMaction})
terms with at least four $\gamma$'s. This $\gamma$-counting
works the same way for $\ep\neq 1$. This results in
\begin{equation}
S_\ep=S_E\,.
\end{equation}
It was further found in~\cite{Fuchs:2008zx} that the cohomology
of this family of solutions is empty. These results seem to
suggest that the solutions of the $\ep$-family are gauge equivalent.

Writing the gauge transformation that sends Erler's solution to
a general $\mbox{$\ep$-solution}$ is straightforward, since the solutions
of this family are given as formal pure-gauge solutions. Composing
the gauge transformations one gets
\begin{equation}
\begin{aligned}
\label{EepGauge}
e^{\Lambda_E^{\ep}}& =(1-\Lambda_E)\frac{1}{1-\Lambda_{\ep}}=
  1\otimes\One+\ep B \gamma(0)\ket{0}\otimes\sigma_1\,,\\
e^{-\Lambda_E^{\ep}}& =(1-\Lambda_{\ep})\frac{1}{1-\Lambda_E}=
  1\otimes\One-\ep B \gamma(0)\ket{0}\otimes\sigma_1\,.
\end{aligned}
\end{equation}
These transformations form an abelian group with the simple
multiplication rule,
\begin{equation}
e^{\Lambda_E^{\ep}}e^{\Lambda_E^{\tilde \ep}}=
  e^{\Lambda_E^{\ep+\tilde \ep}}\,.
\end{equation}
In order to verify that these gauge transformations are not
singular in some sense, one could try to calculate some invariants
as in~\cite{Ellwood:2008jh,Kawano:2008ry} and verify that the
result is $\ep$-independent. This check was not performed yet.
The known properties of the solutions and of these gauge
transformations support the idea that these transformations are
genuine gauge transformations. In fact, in previous cases, the formal
nature of the gauge string field manifested itself as a singularity
related to inverting the exponentiated gauge string field. Since in our
case~(\ref{EepGauge}) both $e^{\pm\Lambda_E^{\ep}}$ are well
defined and involve no implicit inverse string fields, we believe that no
problems could emerge and all the solutions in the $\ep$-family are proven
to be gauge equivalent.

One is therefore led to
believe that the $\ep=0$ solution can be used to describe
tachyon condensation on the non-BPS D-brane. It is then very
natural to assume that Erler's original solution
indeed manages to describe the state without the BPS D-brane despite
the fact that it supports no tachyons.

Erler's solution lives in the modified theory. Could one write
a counterpart of this solution in the non-polynomial theory?
The mapping between the two theories described above suggests that
it should be possible. However, this mapping has a singular character
and it should be assumed that it describes a limit of regularized mappings.
\ATFV it was claimed by Erler that a ``GSO(+) tachyon solution'' should
not exist in the non-polynomial theory~\cite{Erler:2010pr}.
It was further claimed there that
other solutions exist in the modified theory that have no counterparts in the
non-polynomial theory. All that either suggests that the singular nature
of the mapping cannot be resolved in all cases, or that there are
some problems with some of Erler's assumptions. This issue is also
relevant in the context of gauge fixing the newly constructed democratic
theory and it certainly deserves further study.

\section{Outlook}
\label{sec:out}

The advance in our understanding of string field theory described in this
work has the potential of turning string field theory to a practical
framework for (non-perturbative) string theory research. Obviously there is
still more work to be done. Let us mention some of the relevant issues.

The most obvious missing ingredient\footnote{As already mentioned, \aTFV
lump solutions were constructed in~\cite{Bonora:2010hi}.}
is the construction of analytical
solutions describing lump solutions, i.e., lower dimensional D-branes,
around the tachyon vacuum solution\footnote{There are also other examples of
string field theory solutions for which no analytical form is known. One
such example is that of solutions representing cosmological tachyon
models~\cite{Aref'eva:2003qu}.
Another interesting example includes marginal deformations on the
separated D-\= D pair~\cite{Bagchi:2008et}.}.
This is desirable both in its own right as well as for
addressing Sen's second conjecture that for now could not be proved. One
potential complication with this construction is that it should refer to a
specific BCFT, that of the lump, and is therefore not background independent.
Background independence, translated into the description of solutions using
general CFT methods only, played an important role in the constructions of
analytical solutions. One may hope to look for general lump solutions using
tools such as boundary changing operators as well as using explicit
background/oscillator representations.

The, somewhat surprising, possibility of generalizing Schnabl's solution
to the case of a BPS D-brane makes one wonder: which string backgrounds
are accessible as classical solutions of open string field theory around
a particular background?
This is an important question, whose answer can help us to estimate
the relevance of string field theory to other directions of string
theory research.
A particular interesting question is whether it is possible to obtain
a solution describing $N$ D-branes from string field theory in the
background of $M$ D-branes with $M<N$. The opposite case, $M>N$, where some
of the D-branes condense, was analytically solved in~\cite{Ellwood:2006ba}.
Is it possible to go to the other direction, or would one fail, since
the theory does not contain ``enough degrees of freedom''?
The search for a two D-brane system around a single one, within level
truncation, was not successful. Other than to a problem of principle,
one can ascribe this failure, either to the limit of validity of
the Siegel gauge, or to numerical problems resulting from
``climbing up the potential''. One can hope to avoid these problems
by finding an analytical solution. \ATFV it was reported at the
conference SFT2010 that such a solution was found~\cite{SchnablMurata}.

Following the treatment of solutions as formal pure-gauge solutions and
given the fact that all currently known analytical solutions can be cast in
such a form, some obvious questions arise:
\begin{itemize}
\item Is it possible to represent all string field theory solutions as
      formal pure-gauge solutions\footnote{\ATFV it was argued by Ellwood
      in~\cite{Ellwood:2009zf}, that the answer to this question is
      affirmative. While Ellwood's construction was performed around the
			tachyon vacuum, the result formally holds also around the perturbative
			vacuum, since the tachyon vacuum itself is a formal gauge
			solution and gauge transformations can be composed.}?
\item Given a solution how should one cast it in a pure-gauge form?
\item Is there a simple criterion to distinguish the ``large gauge
      transformations'' from the majority of pure-gauge ones?
\item Given a solution that is formally written in a pure-gauge form,
      is there a simple and universal prescription for regularizing it?
\end{itemize}
To understand the last two questions we recall that we regularized Schnabl's
solution and the marginal deformation solutions in quite a different way and
the manifestation of them being non-trivial was also quite
different\footnote{Note that evaluating the action is not always enough,
e.g., it is zero for solutions describing marginal deformations.}.

Another pressing issue is to recognize the physical nature of solutions,
since in some sense it is easier to construct solutions than to
interpret them. Generally speaking, one expects that a solution in string
field theory corresponds to a boundary CFT. One route for recognizing the
BCFT is to evaluate a large enough set of gauge invariant operators
describing the solution in order to reveal its nature. 
Gauge invariant quantities other than the action include the invariants
found in~\cite{Hashimoto:2001sm,Gaiotto:2001ji}. They are related to the
1-point disk scattering amplitudes of closed
strings\footnote{See~\cite{Zwiebach:1992bw} for an early incorporation of
on-shell closed strings into open string field theory. Closed string
amplitudes in open string field theory were also studied
in~\cite{Drukker:2002ct,Takahashi:2003kq}.
Some interesting speculations regarding the relation
between open and closed strings from a string field theoretical
perspective appeared in~\cite{Bonora:2006cd,Bonora:2006tm}.}.
Generalization to Berkovits' superstring field theory was given
in~\cite{Michishita:2004rx}.
Ellwood~\cite{Ellwood:2008jh} used these invariants in order to address
the above mentioned question in the bosonic as well as in the supersymmetric
theory. For specifically comparing Schnabl's solution with the perturbative
one, found in
the Siegel gauge, some of these gauge invariant expressions were calculated
in~\cite{Kawano:2008ry}, supporting the expectation that these solutions are
gauge equivalent\footnote{The best would be of course, if one could
somehow translate the solutions to Siegel gauge. We do not know how to
do that yet.}. They were further shown to coincide in a specific case
with the tadpole of an open string state on the boundary state that
corresponds to the closed string~\cite{Kawano:2008jv}. It would be
interesting to clarify whether these invariants fully characterize
a solution and whether other gauge invariant expressions can be defined
and evaluated. It is also desirable to find methods for improving
their calculability\footnote{\ATFV it was shown in~\cite{Kiermaier:2008qu} that
these invariants can be generalised in a way that leads to the boundary
state associated with the solution, which does characterize the solution.}.

Closed strings can be described naturally using the framework of closed
string field theory~\cite{Zwiebach:1992ie}. While recent advance in the
field proved adequate for open superstring field theory, the generalization
to closed string field theory was not found\footnote{There was,
nevertheless, quite an impressive advance in the study of tachyon
condensation within closed string field theory~\cite{Yang:2005rw,Yang:2005rx,Bergman:2006pd,Moeller:2006cv,Moeller:2006cw,Moeller:2007mu,Moeller:2008tb}.
The new vacuum is interpreted in this case as representing a big crunch of
space-time, since the metric goes to zero. Lump solutions were also found
in this framework and it was claimed that they represent lower dimensional
space-times. This claim is consistent with finding an approximately linear
profile of the dilaton for these solutions.
An important technical tool that enabled this progress is the improved
numerical description of higher string vertices.}.
One of the obstacles for the generalization lies in
the existence of an intricate structure of many string vertices, not simply
related to the star product. Another stumbling-block comes from the ghost
number of the string fields, which is two in the case of the closed string.
The required saturation of six ghost zero modes in the integral implies that
at the quadratic order an explicit ghost insertion is present in addition to
the BRST charge. Together with the Siegel gauge this insertion implies that
the string fields obey
\begin{equation}
b_0\Psi=\bar b_0 \Psi=0\,.
\end{equation}
Modifying the gauge choice to the Schnabl gauge, without changing the form of
the explicit expression leads to awkward looking formulas. It would be
very desirable to examine how can the explicit ghost insertion be changed
and in what way should the gauge choice be modified as compared to the open
string case.

\section*{Acknowledgments}

During the preparation of this work we benefited much from discussions with
Ido Adam, Ofer Aharony, Sudarshan Ananth,
Nathan Berkovits, Guillaume Bossard, Ian Ellwood, Stefan Fredenhagen,
Sunny Itzhaki, Alexey Koshelev, Carlo Maccaferri, Nicolas Moeller,
Yuji Okawa, Yaron Oz, Rob Potting, Samson Shatashvili, Martin Schnabl,
Ashoke Sen, Stefan Theisen and Barton Zwiebach.

M.~K. would like to thank Tel-Aviv University for hospitality.
The work of M.~K.\ was supported by a Minerva fellowship
during the initiation of this project.
This research was supported by the German-Israeli Project
cooperation (DIP H.52).

\newpage
\bibliography{bib}
\end{document}